\documentclass[a4paper,12pt,twoside]{article}
\usepackage[pdftex,colorlinks,bookmarksnumbered,bookmarksopen,citecolor=red,urlcolor=red]{hyperref}
\usepackage{graphicx}\usepackage{a4wide}
\usepackage{amssymb,amsmath,amsthm}
\usepackage{siunitx}
\usepackage{subfigure,color,colordvi,graphicx,xcolor}
\usepackage[only,llbracket,rrbracket,interleave]{stmaryrd}
\usepackage[sort,compress]{cite}
\usepackage{enumitem}
\usepackage{sidecap, caption}

\makeatletter
\newcommand*\wt[2][0.2ex]{%
        \begingroup
        \mathchoice{\wt@helper{#1}{#2}{\displaystyle}{\textfont}}
                   {\wt@helper{#1}{#2}{\textstyle}{\textfont}}
                   {\wt@helper{#1}{#2}{\scriptstyle}{\scriptfont}}
                   {\wt@helper{#1}{#2}{\scriptscriptstyle}{\scriptscriptfont}}%
        \endgroup
        #2%
}
\newcommand*\wt@helper[4]{%
        \def\currentfont{\the#41}%
        \def\currentskewchar{\char\the\skewchar\currentfont}%
        \setbox\tw@\hbox{\currentfont#2\currentskewchar}%
        \dimen@ii\wd\tw@
        \setbox\tw@\hbox{\currentfont#2{}\currentskewchar}%
        \advance\dimen@ii-\wd\tw@
        \rlap{\raisebox{-#1}{$\m@th#3\kern\dimen@ii\widetilde{\phantom{#2}}$}}%
}
\makeatother

\usepackage{booktabs,array,dcolumn}
\newcommand{\PreserveBackslash}[1]{\let\temp=\\#1\let\\=\temp}
\newcolumntype{C}[1]{>{\PreserveBackslash\centering}p{#1}}
\newcolumntype{R}[1]{>{\PreserveBackslash\raggedleft}p{#1}}
\newcolumntype{L}[1]{>{\PreserveBackslash\raggedright}p{#1}}

\newcommand{\bm}[1]{\text{\boldmath $#1$\unboldmath}}

\newcommand{\abs}[1]{\lvert#1\rvert}
\newcommand{\norm}[1]{\lVert#1\rVert}

\newcommand{\bddot}{\operatorname{\bm{:}}}

\newcommand{\vect}[1]{\mathbf{#1}}
\newcommand{\mat}[1]{\mathbf{#1}}

\newcommand{\dt}[1]{\frac{\partial{#1}}{\partial{t}}}

\newcommand{\Div}{{\bm{\nabla}{\!\cdot}}}
\newcommand{\grad}{\bm{\nabla}_{\!}}

\newcommand{\defo}{\bm{\nabla}^{\texttt{{s}}}}

\newcommand{\RR}{\mathbb{R}}

\newcommand{\Ga}[1]{\Gamma_{\!\!#1}}

\newcommand{\nsd}  {\ensuremath{\texttt{n}_{\texttt{sd}}}}
\newcommand{\numel}{\ensuremath{\texttt{n}_{\texttt{el}}}}
\newcommand{\numfa}{\ensuremath{\texttt{n}_{\texttt{fa}}}}

\newcommand{\nstg} {\ensuremath{\texttt{n}_{\texttt{s}}}}

\newcommand{\nnz}{\ensuremath{\texttt{n}_{\texttt{nz}}}}

\newcommand{\nequa}{\ensuremath{\texttt{n}_{\texttt{eq}}}}

\newcommand{\Insd}{\mat{I}_{\nsd\!}}

\newcommand{\bu}{\bm{u}}
\newcommand{\bhu}{\bm{\hat{u}}}

\newcommand{\bn}{\bm{n}}

\newcommand{\bt}{\bm{t}}
\newcommand{\bx}{\bm{x}}
\newcommand{\bL}{\bm{L}}
\newcommand{\bq}{\bm{q}}

\newcommand{\rhoE}{\rho_e}

\newcommand{\nuT}{\tilde{\nu}}

\newcommand{\bqT}{\tilde{\bm{q}}}
\newcommand{\hnu}{\hat{\nu}}

\newcommand{\bB}{\bm{B}}

\newcommand{\bhBi}{\vect{\hat{B}}_i}
\newcommand{\hbi}{\hat{\text{b}}_i}

\newcommand{\btau}{\bm{\tau}}
\newcommand{\btauLF}{\bm{\tau}_\texttt{LF}}
\newcommand{\btauRoe}{\bm{\tau}_\texttt{Roe}}
\newcommand{\btauHLL}{\bm{\tau}_\texttt{HLL}}
\newcommand{\tauT}{\tilde{\tau}}
\newcommand{\tauTLF}{\tauT_\texttt{LF}}
\newcommand{\tauTRoe}{\tauT_\texttt{Roe}}
\newcommand{\tauTHLL}{\tauT_\texttt{HLL}}

\newcommand{\vn}{v_{\bn}}
\DeclareMathOperator{\diag}{diag}
\DeclareMathOperator{\sign}{sign}

\newcommand{\bF}{\vect{F}}
\newcommand{\bA}{\mat{A}}
\newcommand{\bAn}{\bA_{\bn}}

\newcommand{\CFL}{\text{CFL}}

\newcommand{\CFLmax}{\CFL_\text{max}}

\usepackage{scalerel,stackengine}
\stackMath
\newcommand\reallywidehat[1]{%
	\savestack{\tmpbox}{\stretchto{%
			\scaleto{%
				\scalerel*[\widthof{\ensuremath{#1}}]{\kern-.6pt\bigwedge\kern-.6pt}%
				{\rule[-\textheight/2]{1ex}{\textheight}}
			}{\textheight}% 
		}{0.5ex}}%
	\stackon[1pt]{#1}{\tmpbox}%
}

\newcommand{\intE}{\int_{\Omega_e} \!\!}
\newcommand{\intB}{\int_{\partial \Omega_e} \!\!}
\newcommand{\intNotB}{\int_{\partial \Omega_e \setminus \partial \Omega} \hspace{-0.85em}}
\newcommand{\intDBC}{\int_{\partial \Omega_e \cap \Ga{D}} \hspace{-0.85em} }
\newcommand{\intNotDBC}{\int_{\partial \Omega_e \setminus \Ga{D}} \hspace{-0.85em}}

\newcommand{\extNotDBC}{\int_{\partial \Omega_e \cap \partial \Omega \setminus \Ga{D}} \hspace{-0.85em}}

\newcommand{\Aset}{\mathcal{A}_e}
\newcommand{\Bset}{\mathcal{B}_e}
\newcommand{\Dset}{\mathcal{D}_e}
\newcommand{\Iset}{\mathcal{I}_e}
\newcommand{\Cset}{\mathcal{C}_e}
\newcommand{\Eset}{\mathcal{E}_e}

\newcommand{\volE}{|\Omega_e|}
\newcommand{\areaFi}{|\Ga{e,i}|}
\newcommand{\areaFj}{|\Ga{e,j}|}

\newcommand{\bLe}{\vect{L}_e}
\newcommand{\bue}{\vect{u}_e}

\newcommand{\nuTe}{\tilde{\nu}_e}
\newcommand{\bqTe}{\tilde{\vect{q}}_e}
\newcommand{\bqte}{\vect{q}_{t,e}}

\newcommand{\bse}{\vect{s}_e}
\newcommand{\se}{\text{s}_e}

\newcommand{\bhuj}{\vect{\hat{u}}_j}
\newcommand{\bhui}{\vect{\hat{u}}_i}

\newcommand{\hnuj}{\hat{\nu}_j}
\newcommand{\hnui}{\hat{\nu}_i}

\newcommand{\R}{\text{R}}
\newcommand{\bR}{\vect{R}}

\newcommand{\bLV}{\vect{L}}
\newcommand{\buV}{\vect{u}}

\newcommand{\bqTV}{\tilde{\vect{q}}}
\newcommand{\bnuTV}{\tilde{\bm{\nu}}}
\newcommand{\bhuV}{\hat{\vect{u}}}
\newcommand{\brhoV}{\bm{\rho}}
\newcommand{\bhnuV}{\hat{\bm{\nu}}}

\newcommand{\bT}{\mat{T}}
\newcommand{\TLL}{\bT_{LL}}
\newcommand{\TLhu}{\bT_{L\hat{u}}}
\newcommand{\TuL}{\bT_{uL}}
\newcommand{\Tuu}{\bT_{uu}}
\newcommand{\Tuq}{\bT_{u\tilde{q}}}
\newcommand{\Tunu}{\bT_{u\tilde{\nu}}}
\newcommand{\Tuhu}{\bT_{u\hat{u}}}
\newcommand{\Tqq}{\bT_{\tilde{q}\tilde{q}}}
\newcommand{\Tqhnu}{\bT_{\tilde{q}\hat{\nu}}}

\newcommand{\Tnuq}{\bT_{\tilde{\nu}\tilde{q}}}
\newcommand{\Tnunu}{\bT_{\tilde{\nu}\tilde{\nu}}}

\newcommand{\Tnuhnu}{\bT_{\tilde{\nu}\hat{\nu}}}
\newcommand{\ThuL}{\bT_{\hat{u}L}}
\newcommand{\Thuu}{\bT_{\hat{u}u}}
\newcommand{\Thurho}{\bT_{\hat{u}\rho}}
\newcommand{\Thuhu}{\bT_{\hat{u}\hat{u}}}
\newcommand{\Thuhnu}{\bT_{\hat{u}\hat{\nu}}}
\newcommand{\Trhohu}{\bT_{\rho\hat{u}}}
\newcommand{\Thnuq}{\bT_{\hat{\nu}\tilde{q}}}
\newcommand{\Thnunu}{\bT_{\hat{\nu}\tilde{\nu}}}

\newcommand{\Thnuhnu}{\bT_{\hat{\nu}\hat{\nu}}}

\newcommand{\TUU}{\bT_{UU}}
\newcommand{\TULambda}{\bT_{U\Lambda}}
\newcommand{\TLambdaU}{\bT_{\Lambda U}}
\newcommand{\TLambdaLambda}{\bT_{\Lambda\Lambda}}

\newcommand{\bU}{\vect{U}}
\newcommand{\bLambda}{\bm{\Lambda}}

\newenvironment{keywords}{\begin{quote}\emph{\textbf{Keywords:}}}{\end{quote}}
\newtheorem{remark}{Remark}

\RequirePackage{algorithm}[0.1] %Following SIAM cls
\usepackage{algorithmic} %Algorithm style, could alternatively use algpseudocode 

\graphicspath{{figsPDF/}{figsPNG/}{figs/}}

\usepackage{hyphenat}

\begin{document}
%==========================================================================
\title{A face-centred finite volume method for laminar and turbulent incompressible flows}

\author{
\renewcommand{\thefootnote}{\arabic{footnote}}
			  Luan M. Vieira\footnotemark[1]\textsuperscript{ \ ,}\footnotemark[2]\textsuperscript{ \ ,}\footnotemark[3] , 
			  Matteo Giacomini\footnotemark[1]\textsuperscript{ \ ,}\footnotemark[2] , \\
\renewcommand{\thefootnote}{\arabic{footnote}}
			  Ruben Sevilla\footnotemark[3]\textsuperscript{ \ ,}* \  and
			  Antonio Huerta\footnotemark[1]\textsuperscript{ \ ,}\footnotemark[2]
}

\date{\today}
%________________________________________________________________________
\maketitle

\renewcommand{\thefootnote}{\arabic{footnote}}

\footnotetext[1]{Laboratori de C\`alcul Num\`eric (LaC\`aN), ETS de Ingenier\'ia de Caminos, Canales y Puertos, Universitat Polit\`ecnica de Catalunya, Barcelona, Spain.}
\footnotetext[2]{Centre Internacional de M\`etodes Num\`erics en Enginyeria (CIMNE), Barcelona, Spain.}
\footnotetext[3]{Zienkiewicz Centre for Computational Engineering, Faculty of Science and Engineering, Swansea University, Swansea, SA1 8EN, Wales, UK.
\vspace{5pt}\\
* Corresponding author: Ruben Sevilla \textit{E-mail:} \texttt{r.sevilla@swansea.ac.uk}
}

%________________________________________________________________________
\begin{abstract}
This work develops, for the first time, a face-centred finite volume (FCFV) solver for the simulation of laminar and turbulent viscous incompressible flows. The formulation relies on the Reynolds-averaged Navier-Stokes (RANS) equations coupled with the negative Spalart-Allmaras (SA) model and three novel convective stabilisations, inspired by Riemann solvers, are derived and compared numerically. The resulting method achieves first-order convergence of the velocity,  the velocity-gradient tensor and the pressure. FCFV accurately predicts engineering quantities of interest, such as drag and lift, on unstructured meshes and, by avoiding gradient reconstruction, the method is less sensitive to mesh quality than other FV methods, even in the presence of highly distorted and stretched cells. A monolithic and a staggered solution strategies for the RANS-SA system are derived and compared numerically. Numerical benchmarks, involving laminar and turbulent, steady and transient cases are used to assess the performance, accuracy and robustness of the proposed FCFV method.   
\end{abstract}

%________________________________________________________________________
\begin{keywords}
Finite volumes, Face-centred, Incompressible flows, Hybridisable discontinuous Galerkin, Spalart-Allmaras
\end{keywords}

%==========================================================================
\section{Introduction}                     \label{sc:Intro}
%==========================================================================

The simulation of incompressible flows is central to many areas of science and engineering, including aerodynamic design in the automotive industry, the modelling of water quality and the study of blood flow in the human circulatory system, just to name a few. The majority of computational fluid dynamics (CFD) tools --open-source, commercial, academic or industrial-- is based on either second-order cell-centred or vertex-centred finite volume (FV) techniques~\cite{MR1925043,Barth-BHO-17}. 

Despite formally second-order accurate, one of the main drawbacks of classical cell-centred or vertex-centred FV methods is the need to perform a reconstruction of the gradient of the solution. In the presence of highly stretched or distorted cells, the reconstruction often leads to an important loss of accuracy~\cite{diskin2010comparison,diskin2011comparison}.

The face-centred finite volume (FCFV) was initially proposed in~\cite{RS-SGH:2018_FCFV1} as an alternative to cell-centred or vertex-centred approaches, avoiding any reconstruction of the derivatives and results have proved that it is almost insensitive to mesh distortion. In the same way as second-order cell-centred FV methods can be seen as a low order discontinuous Galerkin (DG) with constant degree of approximation in each cell and second-order vertex-centred FV methods can be seen as a conforming piecewise linear continuous finite element method, the FCFV can be seen as the hybridisable DG (HDG) method~\cite{cockburn2004characterization,Jay-CG:05,Jay-CG:05-GAMM,Jay-CGL:09} using the lowest order approximation for cell and face variables. As such, the method is almost second order, in the sense that the gradient of the velocity converges with order one without the need of reconstructions and, therefore, with an accuracy that is less sensitive to mesh distortion. Furthermore, the method allows the use of the same degree of approximation for velocity and pressure, circumventing the so-called Ladyzhenskaya-Babu{\v s}ka-Brezzi  (LBB) condition~\cite{Donea2003}.Compared to classical second-order FV methods, the FCFV method provides the same order of accuracy in the gradient of the velocity without any reconstruction.

Since its introduction, the FCFV method has been applied to a variety of problems including solid mechanics~\cite{RS-SGH:2019_FCFV2}, heat transfer and Stokes flows~\cite{RS-VGSH:20,MG-RS-20,sevilla2023face,sevilla2024face} and viscous laminar compressible flows~\cite{vila2022non,vila2023benchmarking}. This work presents, for the first time, the development and application of the FCFV method to viscous incompressible flows to simulate steady-state and transient phenomena, in both laminar and turbulent regime, using the one-equation Spalart-Allmaras model~\cite{SA1992,SA2012modifications}. 

The FCFV formulation of the incompressible Reynolds-averaged Navier-Stokes (RANS) equations coupled with the negative Spalart-Allmaras (SA) turbulence model~\cite{SA2012modifications} is presented. The velocity-pressure formulation of the momentum equation is adopted to ensure that the vorticity, required in the SA model, can be computed from the mixed variable, which is the velocity-gradient tensor. This is in contrast with other mixed formulations for incompressible flow problems~\cite{giacomini2018superconvergent,Tutorial-GSH:2020,AlS-SKGWH:20} that use the Cauchy formulation of the momentum equation and the symmetric gradient of the velocity as the mixed variable. In addition, three novel stabilisation tensors for the convective part of the incompressible RANS equations are proposed. They are derived following the unified analysis presented in~\cite{JVP_HDG-VGSH:20,vila2022non} for the Riemann solvers in the context of hybrid discretisation methods. Extensive numerical examples are then used to assess the performance and robustness of the proposed FCFV methodology. The tests include benchmarks featuring highly distorted and stretched meshes. Finally, two approaches, one monolithic and one staggered, to solve the RANS-SA equations are discussed, implemented and compared in terms of their computational complexity, efficiency and robustness. 

The remainder of the paper is organised as follows. Section~\ref{sc:INS} briefly recalls the governing RANS and SA equations. The mixed integral form of the RANS-SA system is detailed in section~\ref{sc:FCFV} and the FCFV solution algorithm is described in section~\ref{sc:FCFV_solution}. Section~\ref{sc:examples} presents six numerical examples, encompassing laminar and turbulent, steady and transient cases, solved using triangular and quadrilateral meshes. The examples include test cases to assess the optimal convergence properties of the FCFV approximations and more complex benchmarks where the results are compared to reference solutions in the literature. The conclusions of the work are summarised in section~\ref{sc:conclusions}, whereas the new convective stabilisations are derived in~\ref{app:convectiveTau} and computational details, including a comparison between monolithic and staggered approaches, are presented in~\ref{app:FCFV_computational}.

%==========================================================================
\section{Governing equations}               \label{sc:INS}
%==========================================================================

Let us consider an open bounded domain $\Omega\subset\RR^{\nsd}$, where $\nsd$ is the number of spatial dimensions, and a time interval $(0,T)$, with $T>0$. The non-dimensional incompressible RANS equations coupled with the negative SA turbulence model~\cite{SA2012modifications} can be written as
\begin{equation} \label{eq:NS}
\left\{\begin{aligned}
	\dt{\bu} + \Div (\bu {\otimes} \bu) -\Div \left( \frac{2(1 {+} \nu_t)}{Re} \defo\bu - p \Insd \right)  & = \bm{s}       &&\text{in $\Omega{\times}(0,T]$,}
	\\
	\Div\bu &= 0  &&\text{in $\Omega{\times}(0,T]$,}	
	\\
	\dt{\nuT}+ \Div ( \nuT \bu ) - \Div \left( \frac{1{+}\nuT f_n}{\sigma Re } \grad\nuT \right)   & = s       &&\text{in $\Omega{\times}(0,T]$,}
\end{aligned}\right.
\end{equation}
where $\bu$ is the non-dimensional velocity vector field, $p$ is the non-dimensional modified pressure field, which involves the density and the turbulent kinematic energy, the strain-rate tensor is $\defo{\bu} = ( \grad \bu + \grad^T\! \bu )/2$, $Re$ denotes the Reynolds number and $\bm{s}$ is the body force. The non-dimensional turbulent viscosity $\nu_t$, introduced by the Spalart-Allmaras turbulence model, is computed in terms of the turbulent variable $\nuT$ as $\nu_t = \nuT f_{v1} $.  
The right-hand side of the SA turbulence model is given by
\begin{equation} \label{eq:sourceSA}
s {=}  \left\{
\begin{aligned}
	&c_{b1} (1{-}f_{t2})\tilde{S} \nuT {+} \frac{c_{b2}}{\sigma Re} \grad \nuT {\cdot} \grad \nuT {-} \frac{1}{Re}\bigl(c_{w1}f_w {-} \frac{c_{b1}}{\kappa^2} f_{t2} \bigr)\! \Bigl(\frac{\nuT}{d}\Bigr)^2  &&\text{if $\nuT \geq 0$,}\\[1ex]
  	&c_{b1} (1{-}c_{t3}){S} \nuT {+} \frac{c_{b2}}{\sigma Re} \grad \nuT {\cdot} \grad \nuT {+} \frac{c_{w1}}{Re}\Bigl(\frac{\nuT}{d}\Bigr)^2 &&\text{otherwise,}
\end{aligned}
\right.
\end{equation}
where the three terms correspond to production,  cross-diffusion and destruction, respectively. In the production term, the modified vorticity magnitude, $\tilde{S}$, is introduced to avoid negative values~\cite{SA2012modifications}, namely
\begin{equation} \label{eq:SA_S}
\tilde{S} = \left\{
\begin{aligned}
& S + \bar{S}       & \quad &\text{if $\bar{S} \geq -c_{v2}S$,}\\
& S + \frac{S(Sc_{v2}^2 + \bar{S}c_{v3})}{S(c_{v3} - 2c_{v2}) - \bar{S}} & &\text{otherwise,}
\end{aligned}
\right.
\end{equation}
where $S = \sqrt{2 \bm{S} \bddot \bm{S}}$ is the magnitude of the vorticity, $\bm{S} = ( \grad \bu {-} \grad^T \bu )/2$ is the vorticity tensor, $\bar{S} = \nuT f_{v2}/(Re\, \kappa^2 d^2)$ and $d$ is the minimum distance to the wall. The SA model is completed with the closure functions
\begin{equation}\label{eq:SA_closure}
\begin{aligned}
\chi &= \nuT, & \quad 
f_{v1} &= \frac{\chi^3}{\chi^3 + c_{v1}^3}, &   \quad
f_{v2} &= 1 - \frac{\chi}{1+ \chi f_{v1}}, \\
r &= \min \left(\frac{\nuT}{\widetilde{S} \kappa^2 d^2}, r_{lim} \right), &
g &= r + c_{w2}(r^6 -r), &
f_w &= g \left( \frac{1 + c_{w3}^6}{g^6 + c_{w3}^6} \right)^{1/6},\\
f_{t2} & = c_{t3} \exp(-c_{t4} \chi^2) ,& f_n &= \frac{c_{n1} + \chi^3}{c_{n1} - \chi^3},
\end{aligned}
\end{equation}
and closure constants $\sigma = 2/3$, $c_{b1}=0.1355$, $c_{b2} = 0.622$, $\kappa = 0.41$, $c_{w1} =c_{b1}/\kappa^2 + (1+c_{b2})/\sigma$, $c_{v1} = 7.1$, $c_{t3} = 1.2$, $c_{t4} = 0.5$, $c_{v2} = 0.7 $, $c_{v3} =0.9 $, $r_{lim} = 10$, $c_{w2} = 0.3$, $c_{w3} = 2$, and  $c_{n1}  = 16$. 

\begin{remark} \label{rk:velo-pressure}
The conservation of momentum, see the first equation in \eqref{eq:NS}, can be rewritten in the usual velocity-pressure formulation as
\begin{equation}\label{eq:vp}
  \dt{\bu} + \Div (\bu {\otimes} \bu) -\Div \left( \frac{1 {+} \nu_t}{Re} \grad\bu - p \Insd \right) 
  - \frac{1}{Re} \grad^T\!\bu\, \grad \nu_t  = \bm{s},
\end{equation}
where the term $(1 {+} \nu_t)/Re \Div (\grad^T\!\bu)$ vanishes due to the point-wise divergence-free condition on the velocity.
\end{remark}

The FCFV method described in this work employs the velocity-pressure formulation of the momentum equation~\eqref{eq:vp} to ensure that the vorticity $\bm{S}$ can be computed from the components of the velocity-gradient tensor $\grad\bu$. This strategy is adopted to avoid performing a reconstruction of the derivatives of the velocity, as commonly done in other FV methods, which induces a loss of accuracy when employing distorted meshes.

The strong form of the incompressible RANS equations and SA turbulence model is closed with appropriate boundary conditions, written in the abstract form
\begin{equation} \label{eq:NS_ICandBC}
\left\{\begin{aligned}
	\bB (\bu,p,\grad \bu,\nuT) & = \bm{0}      &&\text{on $\partial \Omega \ {\times}(0,T]$,}
	\\
	b(\bu,\nuT,\grad \nuT)& = 0       &&\text{on $\partial \Omega \ {\times}(0,T]$,}
\end{aligned}\right.
\end{equation}
using the boundary operators $\bB$ and $b$. In this work, Dirichlet, Neumann, outflow, and symmetry boundary conditions are considered. The portions of the boundary $\partial\Omega$ where such conditions are imposed, denoted by $\Ga{D}$, $\Ga{N}$, $\Ga{O}$, and $\Ga{S}$, respectively, are disjoint by pairs and such that $\partial\Omega=\Ga{D}\cup\Ga{N}\cup\Ga{O}\cup\Ga{S}$. The particular expression of the boundary operators $\bB$ and $b$ is 
\begin{subequations}\label{eq:RANS_SA_BC}
\begin{align} 
\bB& := 
\begin{cases}
	\bu - \bu_D
	& \text{ on  $\Ga{D}$,} \\
	\displaystyle \left( \frac{2(1 {+} \nu_t)}{Re} \defo{\bu} - p \Insd \right) \bn - \bm{g}
	& \text{ on  $\Ga{N}$,} \\
	\displaystyle \left( \frac{(1 {+} \nu_t)}{Re} \grad{\bu} - p \Insd \right) \bn
	& \text{ on  $\Ga{O}$,} \\               
	\left\{\begin{aligned} & \bt_k \cdot \left( \frac{(1 {+} \nu_t)}{Re} \grad{\bu} - p \Insd \right) \bn \\ &\bu \cdot \bn \end{aligned} \right\}
	& \text{ on  $\Ga{S}$,}                                      
\end{cases}
\end{align}
and
\begin{align} 
b & := 
\begin{cases}
	\nuT - \nuT_D
	& \text{ on  $\Ga{D}$,} \\
	\displaystyle \left(\frac{1{+}\nuT f_n}{\sigma Re } \grad \nuT \right) \cdot \bn
	& \text{ on  $\Ga{N}\cup \Ga{O} \cup \Ga{S}$},                         
\end{cases}
\end{align}
\end{subequations}
where $\bn$ is the outward unit normal and $\bt_k$ , for $k=1,\dotsc,\nsd - 1$, are unit tangential vectors to the boundary (one vector in two dimensions and two vectors in three dimensions), $\bu_D$ and $\nuT_D $ are the imposed velocity and turbulent variable on the Dirichlet boundary, and $\bm{g}$ is the imposed traction on the Neumann boundary representing a material surface. A numerical comparison of the outflow boundary condition considered here and other boundary conditions imposed on an outlet can be found in~\cite{Tutorial-GSH:2020}.

Finally, an initial condition is given for the velocity and turbulent variable, namely
\begin{equation} \label{eq:NS_IC}
\left\{\begin{aligned}
	\bu & = \bu_0      & &\text{in $\Omega{\times}\{0\}$,}
	\\
	\nuT & = \nuT_0     & &\text{in $\Omega{\times}\{0\}$.}
\end{aligned}\right.
\end{equation}

It is worth noting that the divergence-free condition in the RANS equations induces the compatibility condition on the velocity field given by
\begin{equation} \label{eq:compatibility} 
	 \int_{\Ga{D}} \bu_D \cdot \bn\, d\Ga{} + \int_{\partial\Omega\setminus\Ga{D}} \bu \cdot \bn\, d\Ga{} = 0. 
\end{equation}
In particular, if only Dirichlet boundary conditions are considered it is necessary to impose an extra constraint to eliminate the indeterminacy of the pressure field~\cite{Tutorial-GSH:2020}.

%==========================================================================
\section{Face-centred finite volume formulation}               \label{sc:FCFV}
%==========================================================================

Let us consider a partition of the domain $\Omega$ in a set of non-overlapping cells $\{\Omega_e\}_{e=1,\dotsc,\numel}$, where $\numel$ is the number of cells. The boundary of each cell, $\partial \Omega_e$, is formed by a set of faces $\{\Ga{e,j}\}_{j=1,\dotsc,\numfa^e}$, where $\numfa^e$ is the number of faces of cell $\Omega_e$. Finally the set of faces not on the boundary of the domain form the internal interface $\Ga{}$. As shown in previous works~\cite{RS-SGH:2018_FCFV1,MG-RS-20}, the FCFV allows for various cell types and hybrid meshes with cells of different types. Examples using triangular, quadrilateral, tetrahedral, hexahedral, pyramidal and prismatic cells, as well as hybrid meshes, have been previously considered in~\cite{RS-SGH:2018_FCFV1,MG-RS-20}.

Following the rationale of FCFV~\cite{RS-SGH:2018_FCFV1,RS-SGH:2019_FCFV2,MG-RS-20,RS-VGSH:20,vila2022non} and HDG~\cite{cockburn2009derivation,Nguyen-NPC:10,Nguyen-CNP:10,Nguyen-NPC:11} methods, the incompressible RANS equations together with the SA turbulence model are written as a system of first-order equations after introducing the mixed variables $\bL = -\grad\bu$ and $\bqT = -\grad \nuT $. This mixed formulation %in~\eqref{eq:NS_Mixed} 
is solved in two steps. First,  $\numel$ \emph{local problems} are defined, one per cell,  to express the primal variables, $\bu$, $p$ and $\nuT$, and the mixed variables, $\bL$ and $\bqT$, as a function of new, independent, face variables, $\bhu$ and $\hnu$. In addition, Dirichlet boundary conditions are enforced in the local problems, for the cells touching the portion $\Ga{D}$ of the boundary $\partial\Omega$. Second, a \emph{global problem} is defined to impose the inter-cell continuity of the velocity, the  turbulent variable and the normal fluxes of the RANS and SA equations, across the internal interface $\Ga{}$. In addition, non-Dirichlet boundary conditions on $\partial\Omega\setminus\Ga{D}$ are also accounted for in the global problem. 

\begin{figure}[!tb]
	\centering
	\includegraphics[width=0.35\textwidth]{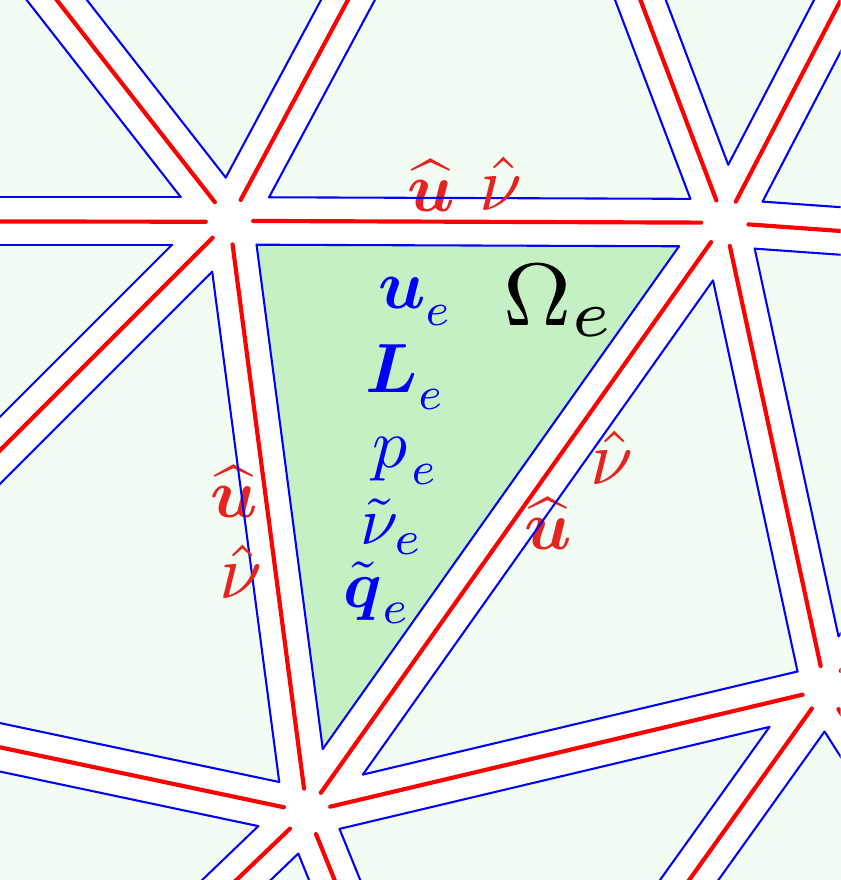}
	\caption{Detail of a triangular mesh highlighting a cell $\Omega_e$ where the velocity $\bu_e$, velocity gradient $\bL_e$, mean pressure $p_e$, turbulent variable $\nuT_e$ and gradient of the turbulent variable $\bqT_e$ are defined and internal edges where the hybrid velocity $\bhu$ and hybrid turbulent variable $\hnu$ are defined.}
	\label{fig:FCFVvariables}
\end{figure}

Figure~\ref{fig:FCFVvariables} shows a sketch of a triangular mesh with the variables defined on each cell and on each edge. As it will be shown in this section, the global system of equations to be solved only involves the variables on the edges and the cell variables can then be recovered cell-by-cell.

%==========================================================================
\subsection{Integral form of the local problems}      \label{sc:FCFV_weakLocal}
%==========================================================================

The local problems enforce, in each cell, the kinematic equations defining the mixed variables $\bL$ and $\bqT$, the RANS momentum equation and the SA equation. Thus, conservation of mass and momentum is imposed cell-by-cell. The integral form of the local problems for each cell is
\begin{equation} \label{eq:localWeakNumFluxes}
	\left\{
	\begin{aligned}
		-\intE \bL d\Omega & =  \intDBC \bu_D{\otimes}\bn\, d\Ga{}  + \intNotDBC   \bhu{\otimes}\bn\,  d\Ga{}, 
		\\
		\intE \dt{\bu} d\Omega & + \intB  \left(\reallywidehat{\bu \otimes\bu} \right) \bn  d\Ga{}  + \intE  \Div \left( \frac{1{+}\nu_t}{Re} \bL {+} p \Insd \right)  d\Omega 
		\\[-0.5ex]
		&\hspace{1em} + \intB  \left[\left( \reallywidehat{\dfrac{1{+}\nu_t}{Re} \bL {+} p \Insd} \right) {-} \left( \frac{1{+}\nu_t}{Re}\bL {+} p \Insd \right) \right] \bn d\Ga{} 
		\\[-0.5ex]
		&\hspace{12em} - \intE \frac{1}{Re} \bL^T \bq_t d\Omega   = \intE  \bm{s} d\Omega,
		\\
		\intE p\, d\Omega{} & = \abs{\Omega_e}\,\rhoE,
		\\
		- \intE \bqT  d\Omega &  =  \intDBC \nuT_D \bn d\Ga{}  + \intNotDBC  \hnu \bn  d\Ga{},
		\\
		\intE \dt{\nuT} d\Omega & + \intB  (\reallywidehat{\nuT \bu})  \cdot \bn d\Ga{} + \intE \Div \left( \frac{1{+}\nuT f_n}{\sigma Re }\, \bqT \right)  d\Omega, 
		\\[-0.5ex]
		&\hspace{4em}  + \intB \left(  \reallywidehat{\dfrac{1{+}\nuT f_n}{\sigma Re }\, \bqT} {-}  \frac{1{+}\nuT f_n}{\sigma Re }\, \bqT \right)\cdot \bn  d\Ga{} = \intE s  d\Omega ,
	\end{aligned}
	\right.
\end{equation}
where $\bhu$ and $\hnu$, known as \emph{hybrid} variables, denote the face velocity and turbulent variable on $\Ga{} \cup \partial \Omega \setminus \Ga{D}$, $\rhoE$ is the mean pressure in the cell and $\bq_t := - \grad \nu_t = (4 - 3f_{v1})f_{v1} \bqT$.  Moreover, the right-hand side of the SA model, introduced in~\eqref{eq:sourceSA}, is also written in terms of the mixed variable $\bqT$ as
\begin{equation} \label{eq:sourceSAmixed}
	s =  \left\{
	\begin{aligned}
		&c_{b1} (1{-}f_{t2})\tilde{S} \nuT {+} \frac{c_{b2}}{\sigma Re} \bqT {\cdot} \bqT {-} \frac{1}{Re}\bigl(c_{w1}f_w {-} \frac{c_{b1}}{\kappa^2} f_{t2} \bigr)\! \Bigl(\frac{\nuT}{d}\Bigr)^2  &&\text{if $\nuT \geq 0$,}\\[1ex]
		&c_{b1} (1{-}c_{t3}){S} \nuT {+} \frac{c_{b2}}{\sigma Re} \bqT {\cdot} \bqT {+} \frac{c_{w1}}{Re}\Bigl(\frac{\nuT}{d}\Bigr)^2 &&\text{otherwise.}
	\end{aligned}
	\right.
\end{equation}

Note that the second equation in~\eqref{eq:localWeakNumFluxes} uses the rewriting of the momentum equation given by Remark~\ref{rk:velo-pressure}. 

In the local problems, the divergence theorem is applied twice to the diffusive terms of both the momentum and SA equations. This implies that the boundary integrals feature the difference between physical and numerical fluxes. This is simply done to guarantee the symmetry of the global system in the case of creeping flows~\cite{RS-SH:16,Tutorial-GSH:2020}, i.e., for Stokes problems.

As the local problems only contain Dirichlet boundary conditions, it is necessary to introduce an extra constraint to eliminate the indeterminacy of the pressure field. In this work, the constraint imposes the mean value of the pressure in each cell,  as reported in the third equation in~\eqref{eq:localWeakNumFluxes}. 

Therefore, the local problems provide an expression to compute $(\bL,\bu,p,\bqT,\nuT)$, in each cell, in terms of the global unknowns $(\bhu,\brhoV,\bhnuV)$, where $\brhoV=(\rho_1\dotsc,\rho_{\numel})^T$.

%==========================================================================
\subsection{Integral form of the global problem}      \label{sc:FCFV_weakGlobal}
%==========================================================================

The global problem imposes the continuity of the inter-cell fluxes of the RANS and SA equations and the non-Dirichlet boundary conditions. The integral form of the global problem is
\begin{equation} \label{eq:globalWeakNumFluxes}
	\left\{
	\begin{aligned}
		%	\\
		& \sum_{e=1}^{\numel} \Bigg\{ \intNotB (\widehat{\bu{\otimes}\bu})\bn\, d\Ga{} 
		+ \intNotB \;\: \Bigl(  \reallywidehat{\dfrac{1{+}\nu_t}{Re} \bL {+} p \Insd } \Bigr) \bn\,  d\Ga{} \Bigg.
		\\[-0.5ex] &\hspace{14em}
		\Bigg. +\extNotDBC\: \bB(\bu,p,\bL,\nuT) d\Ga{} \Bigg\}
		= \!\bm{0},
		\\
		& \int_{\partial\Omega_e \setminus \Ga{D}} \bhu \cdot \bn\, d\Ga{} = - \int_{\partial\Omega_e \cap \Ga{D}} \bu_D \cdot \bn\, d\Ga{}   \qquad \text{ for } e=1, \dotsc,\numel,
		\\
		& \sum_{e=1}^{\numel} \Bigg\{  \intNotB (\reallywidehat{\nuT \bu} ) \cdot \bn\, d\Ga{} + \intNotB  \;\: \Bigl( \reallywidehat{\dfrac{1{+}\nuT f_n}{\sigma Re }\, \bqT} \Bigr) \cdot \bn\, d\Ga{} \Bigg.
		\\[-0.5ex] &\hspace{14em}
		\Bigg. + \extNotDBC  \; b(\bu,\nuT,\bqT)   d\Ga{}\Bigg\}  = 0,
	\end{aligned}
	\right.
\end{equation}
where the second equation represents the compatibility condition enforcing the divergence-free condition on the velocity in each cell.

The continuity of the velocity and the turbulent variable across the internal interface $\Ga{}$ is not explicitly imposed in the global problem because it is automatically satisfied due to the Dirichlet boundary conditions $\bu=\bhu$ and $\nuT=\hnu$ on $\partial\Omega_e\setminus\Ga{D}$ in the local problems and the unique value of the hybrid variables $\bhu$ and $\hnu$ on each face.

%==========================================================================
\subsection{Numerical fluxes and stabilisation}      \label{sc:stabilisation}
%==========================================================================

In the local and global problems, the numerical trace of the convective and diffusive fluxes in the RANS equations is given by
\begin{subequations}\label{eq:numFluxRANS}
	\begin{equation} \label{eq:numFluxRANSadvection}
		(\widehat{\bu{\otimes}\bu})\bn :=
		\begin{cases}
			(\bu_D{\otimes}\bu_D\!)\bn + \btau^a(\bu-\bu_D\!) & \text{on $\partial\Omega_e\cap\Ga{D}$,} 
			\\
			( \bhu{\otimes}\bhu)\bn + \btau^a(\bu-\bhu) & \text{elsewhere,}  
		\end{cases}
	\end{equation}
	and
	\begin{equation} \label{eq:numFluxRANSdiffusion}
		\hspace{-1.1em}
		\Bigl( \reallywidehat{\dfrac{1 {+} \nu_t}{Re} \bL {+} p \Insd}\! \Bigr) \bn\! :=\!\!
		\begin{cases}
			\!\!
			\Bigl(\dfrac{1 {+} \nuT_{\! D} f_{\! v1\!}(\nuT_{\! D}) }{Re} \bL {+} p \Insd\! \Bigr) \bn {+} \btau^d (\bu {-} \bu_D\!) &\text{on$\,\partial\Omega_e{\cap}\Ga{D}$,} 
			\\[0.5ex]
			\!\!\Bigl(\dfrac{1 {+} \hnu f_{\! v1\!}(\hnu) }{Re} \bL {+} p \Insd \Bigr) \bn {+} \btau^d (\bu {-} \bhu)
			&\text{elsewhere,}  
		\end{cases}
	\end{equation}	
\end{subequations}
respectively. Similarly, the numerical trace of the convective and diffusive fluxes in the SA equation is given by
\begin{subequations}\label{eq:numFluxSA}
	\begin{equation} \label{eq:numFluxSAadvection}
		(\reallywidehat{\nuT \bu} ) \cdot \bn := 
		\begin{cases}
			(\nuT_D \bu_D\!) \cdot \bn  + \tauT^a(\nuT- \nuT_D\!) & \text{on $\partial\Omega_e\cap\Gamma_D$,} 
			\\
			(\hnu \bhu) \cdot \bn  + \tauT^a(\nuT- \hnu ) & \text{elsewhere,}  
		\end{cases}
	\end{equation}
	and
	\begin{equation} \label{eq:numFluxSAdiffusion}
		\hspace{-1em}
		\Bigl( \reallywidehat{\dfrac{1{+}\nuT f_n}{\sigma Re }\, \bqT} \Bigr) {\cdot} \bn :=
		\begin{cases}
			\!\!
			\Bigl(\dfrac{1 {+} \nuT_D f_n(\nuT_D)\!}{\sigma Re }\,  \bqT \Bigr) {\cdot} \bn {+} \tauT^d (\nuT {-} \nuT_D\!) &\text{on $\partial\Omega_e\cap\Ga{D}$,} 
			\\[0.5ex]
			\!\!
			\Bigl(\dfrac{1 {+} \hnu f_n(\hnu)}{\sigma Re }\, \bqT \Bigr) {\cdot} \bn {+} \tauT^d (\nuT {-} \hnu) &\text{elsewhere,}  
		\end{cases}
	\end{equation}		
\end{subequations}
respectively.

An appropriate stabilisation, given by $\btau^a$, $\btau^d$, $\tauT^a$ and $\tauT^d$, is required in order to ensure that the numerical scheme is well posed. The effect of the stabilisation has been extensively studied in the literature for HDG~\cite{Jay-CGL:09,Cockburn-CDG:08,Nguyen-NPC:10,HDG-Elasticity2018,giacomini2018superconvergent,JVP_HDG-VGSH:20} and FCFV~\cite{RS-SGH:2018_FCFV1,RS-SGH:2019_FCFV2,RS-VGSH:20,MG-RS-20,vila2022non}. The diffusive stabilisation is taken as
\begin{equation} \label{eq:stabilisationDiffusion}
	\btau^d := \frac{\beta \ (1 + \hnu f_{\! v1\!}(\hnu) )}{Re} \Insd
	\quad \text{ and }  \quad 
	\tauT^d := \frac{\beta \ (1 + \hnu f_n(\hnu))}{\sigma Re},
\end{equation}
where $\beta$ is a numerical parameter, taken as 10 in all the examples presented in this work.

For the convective stabilisation, new definitions are proposed in this work. The derivation of three different convective stabilisations, based on the Lax-Friedrichs (LF), Roe and Harten-Lax-van Leer (HLL) Riemann solvers, is detailed in~\ref{app:convectiveTau} following the ideas presented in~\cite{JVP_HDG-VGSH:20,vila2022non,vila2023benchmarking} for compressible flows.

%==========================================================================
\section{Face-centred finite volume solution}               \label{sc:FCFV_solution}
%==========================================================================

To simplify the presentation, the set of indices for the faces of a cell is denoted by $\Aset := \{1, \dotsc, \numfa^e \}$. In addition, the set of indices for the faces of a cell on the Dirichlet boundary, interior to the domain and on the boundary of the domain are denoted by $\Dset := \{j \in \Aset \; | \; \Gamma_{e,j} \cap \Gamma_D \neq \emptyset \}$, $\Iset := \{j \in \Aset \; | \; \Gamma_{e,j} \cap \partial \Omega = \emptyset \}$ and $\Eset := \{j \in \Aset \; | \; \Gamma_{e,j} \cap \partial \Omega \neq \emptyset \}$, respectively. Given the imposition of Dirichlet boundary conditions in the local problems, it is also convenient to denote by $\Bset := \Aset \setminus \Dset$ and $\Cset := \Eset \setminus \Dset$ the set of indices for the faces of a cell not on a Dirichlet boundary and for the external faces not on a Dirichlet boundary. Finally, it is useful to introduce the indicator function of a set $\mathcal{S}$, i.e.
\begin{equation}
	\chi_{\mathcal{S}}(i) = 
	\begin{cases}
		1 & \text{ if } \ i\in\mathcal{S}, \\
		0 & \text{ otherwise}.                                                                 
	\end{cases}
\end{equation}

%==========================================================================
\subsection{Spatial and temporal discretisation}               \label{sc:FCFV_discretisation}
%==========================================================================

The FCFV method introduces a constant approximation of the primal, $\bu$, $p$, and $\nuT$, and mixed, $\bL$ and $\bqT$, variables in each cell as well as a constant approximation of the hybrid variables, $\bhu$ and $\hnu$ on each cell face. The value of the primal and mixed variables in each cell is denoted by $\bue$, $\rhoE$, $\nuTe$, $\bLe$ and $\bqTe$ respectively, whereas the value of the hybrid variables on the $j$-th face of a cell is denoted by $\bhuj$ and $\hnuj$.

The time integration is performed in this work using implicit multi-step backward differentiation formulae (BDF)~\cite{ascher1998computer}. The first-order time derivative in the local problems is approximated as
\begin{equation}\label{eq:BDF}
	\dt{y}\bigg\rvert^{n} \approx \sum_{s=0}^{\nstg} a_s y^{n-s},
\end{equation}
where $y^r(\bx):=y(\bx, t^r)$, $\nstg$ is the number of stages and, to simplify the notation, the coefficients $a_s$ include the dependence on the time step $\Delta t$.

For steady-state computations, the first-oder BDF method (BDF1) is employed as a pseudo-time marching scheme, which corresponds to $\nstg=1$ with $a_0=1/\Delta t$ and $a_1 = -1/\Delta t$. For transient simulations the second-oder BDF method (BDF2) is employed, which corresponds to $\nstg=2$ with $a_0=3/(2\Delta t)$, $a_1 = -2/\Delta t$ and $a_2=1/(2\Delta t)$.

Introducing the expression of the numerical fluxes given by equations~\eqref{eq:numFluxRANS} and \eqref{eq:numFluxSA} into the local problem~\eqref{eq:localWeakNumFluxes} and the approximation of the time derivative defined in \eqref{eq:BDF}, the fully discrete residuals of the local problems are
\begin{equation} \label{eq:localDiscrete}
	\left\{
	\begin{aligned}
		\bR^{n}_{e,L}\! &:=   \volE \bLe^n + \sum_{j \in \Dset}\! \areaFj \bu_{D,j}^n {\otimes} \bn_j  
		+ \sum_{j \in \Bset}\!  \areaFj \bhuj^n {\otimes} \bn_j,
		\\
		\bR^{n}_{e,u}\! &:=    \volE \sum_{s=0}^{\nstg}\! a_s \bue^{n-s} 
		{+} \sum_{j \in \Aset}\! \areaFj \btau_j^{n-1} \bue^n 
		\\
		& {-} \frac{\volE}{Re} (\bLe^n{+}[\bLe^n]^T) \bqte(\nuTe^n,\bqTe^n) {-} \volE \bse^n
		\\
		& {-} \sum_{j \in \Dset}\!  \areaFj  \bigl( \btau_j^{n-1} {-} (\bu_{D,j}^n {\cdot} \bn_j)\Insd \bigr) \bu_{D,j}^n  
		\\
		& {-} \sum_{j \in \Bset}\!  \areaFj  \bigl( \btau_j^{n-1} {-} (\bhuj^n       {\cdot} \bn_j)\Insd \bigr) \bhuj^n ,
		\\
		\bR^{n}_{e,\tilde{q}}\! &:=   \volE \bqTe^n + \sum_{j \in \Dset}\! \areaFj \nuT_{D,j}^n \bn_j  + \sum_{j \in \Bset}\!  \areaFj  \hnu_j^n \bn_j,
		\\
		\R^{n}_{e,\nuT}\! &:=   \volE \sum_{s=0}^{\nstg}\! a_s \nuTe^{n-s}  {+} \sum_{j \in \Aset}\! \areaFj \tauT_j^{n-1} \nuTe^n  
		\\
		& {-}  \frac{\volE}{\sigma Re } \bqTe^n {\cdot} \bqTe^n  f_{n,e}(\nuTe^n)  {-}  \frac{\volE}{\sigma Re } \bqTe^n {\cdot} \bqTe^n   \nuTe^n \frac{\partial f_{n,e}(\nuTe^n)}{\partial \nuTe^n}{-} \volE \se(\nuTe^n,\bqTe^n,\bhuV^n)  
		\\
		& {-}  \sum_{j \in \Dset}\!  \areaFj  ( \tauT_j^{n-1} {-} \bu_{D,j}^n {\cdot} \bn_j) \nuT_{D,j}^n  {-} \sum_{j \in \Bset}\!  \areaFj  ( \tauT_j^{n-1} {-} \bhuj^n {\cdot} \bn_j) \hnu_j^n,
	\end{aligned}
	\right.
\end{equation}
where $\btau = \btau^a + \btau^d$, $\tauT = \tauT^a + \tauT^d$, $\volE$ is the area/volume of the cell $\Omega_e$ and $\areaFj$ is the length/area of the edge/face $\Gamma_{e,j}$.

\begin{remark} \label{rk:vorticity}
	The right-hand side term $s_e$ in the last equation of the local problem\ \eqref{eq:localDiscrete} requires the vorticity magnitude $S$. The vorticity can be computed from the mixed variable matrix $\mat{L}$ as
	\begin{equation} \label{eq:vorticityL}
		\mat{S}_e   = (\bLe^T - \bLe )/ 2,
	\end{equation}
	or in terms of the hybrid velocity $\bhu$ as
	\begin{equation} \label{eq:vorticityUhat}
		\mat{S}_e   = \frac{1}{\volE} \Bigl(  \sum_{j \in \Dset}\! \areaFj(\bu_{D,j} {\otimes} \bn_j {-} \bn_j {\otimes} 
		\bu_{D,j} ) {+}\!\! \sum_{j \in \Bset}\! \areaFj ( \bhuj {\otimes} \bn_j {-} \bn_j {\otimes} \bhuj )    \Bigr).
	\end{equation}
	The two expressions are equivalent and numerical tests have shown to provide an approximation that converges linearly as the mesh is refined. In all the numerical examples involving turbulent flows, the simpler expression given by equation~\eqref{eq:vorticityL} is employed.
\end{remark}

The residuals of the global problem, denoted by $\bR_{\hat{u}}$, $\bR_{\rho}$ and $\bR_{\hat{\nu}}$, are obtained by assembling the contributions 
\begin{equation} \label{eq:globalDiscrete}
	\left\{
	\begin{aligned}
		\bR^n_{e,i,\hat{u}} := &  \areaFi \Bigl\{ \Bigl(\! \btau_i^{n-1} \bue^n   + \frac{1 {+} \hnui^n f_{\! v1\!}(\hnui^n) }{Re} \bLe^n \bn_i + \rhoE^n \bn_i - \btau_i^{n-1} \bhui^n \Bigr)\!  \chi_{\Iset}(i)  \Bigr.\\
		& \qquad\quad \Bigl. +  \bhBi(\bue^n,\bhui^n,\bLe^n,\rhoE^n,\hnui^n,\btau_i^{n-1}) \chi_{\Cset}(i)\! \Bigr\},
		\\
		\R^n_{e,\rho} := & \sum_{j \in \Bset}  \areaFj \bhuj^n \cdot \bn_j + \sum_{j \in \Dset} \areaFj \bu_{D,j}^n \cdot \bn_j, 
		\\
		\R^n_{e,i,\hat{\nu}} := &  \areaFi  \Bigl\{ \Bigl(\! \tauT_i^{n-1} \nuTe^n  + \frac{1{+}\hnui^n f_n(\hnui^n)}{\sigma Re }\,  \bqTe^n \cdot \bn_i  - \tauT_i^{n-1} \hnui^n  \Bigr)\!  \chi_{\Iset}(i)   \Bigr.\\
		& \qquad\quad\Bigl. {+}   \hbi(\nuTe^n,\hnui^n,\bqTe^n,\tauT_i^{n-1}) \chi_{\Cset}(i) \Bigr\}.
	\end{aligned}
	\right.
\end{equation}
for all $i \in \Bset$, where the FCFV version of the boundary operators is given by
\begin{subequations}\label{eq:FCFV_BC}
  \begin{align} 
    \mat{\hat{B}} & := 
      \begin{cases}
        \Bigl( \dfrac{1 {+} \hnu f_{\! v1\!}(\hnu) }{Re} (\mat{L}{+}\mat{L}^T) {+} \rho \Insd \Bigr) \vect{n} {+} \btau (\vect{u} -\vect{\hat{u}})  {+} \vect{g} 
			& \text{ on  $\Ga{N}$,} \\[1ex]
			\Bigl( \dfrac{1 {+} \hnu f_{\! v1\!}(\hnu) }{Re} \mat{L}  {+} \rho \Insd \Bigr) \vect{n} {+} \btau^d (\vect{u} -\vect{\hat{u}}) 
			& \text{ on  $\Ga{O}$,} \\[1ex]              
			\left\{\begin{aligned} & \vect{t}_k \cdot \Bigl[ \Bigl( \frac{1 {+} \hnu f_{\! v1\!}(\hnu) }{Re} \mat{L}  {+} \rho \Insd \Bigr) \vect{n} {+} \btau^d (\vect{u} -\vect{\hat{u}}) \Bigr]\\ 
			                                 &\vect{\hat{u}} \cdot \vect{n} \end{aligned} \right\} 
			& \text{ on  $\Ga{S}$,}                                         
		\end{cases}
	\end{align}
	and
	\begin{align} 
		\hat{\text{b}} & := 
		\begin{cases}
		\Bigl(  \dfrac{1{+}\hnu f_n(\hnu)}{\sigma Re }\, \vect{\tilde{q}}  \Bigr)\cdot \vect{n} + \tauT(\nuT - \hnu) 
		& \text{ on  $\Ga{N}$,}   \\
		\Bigl(  \dfrac{1{+}\hnu f_n(\hnu)}{\sigma Re }\, \vect{\tilde{q}}  \Bigr)\cdot \vect{n} + \tauT^d(\nuT - \hnu) 
		& \text{ on  $\Ga{O} \cup \Ga{S}$.}   
		\end{cases}
	\end{align}
\end{subequations}
As the Dirichlet boundary conditions are accounted for in the local problems, the FCFV boundary operators are only required for Neumann, outflow and symmetry boundaries.

The third equation in~\eqref{eq:localWeakNumFluxes} is omitted in~\eqref{eq:localDiscrete} and the value of $\text{p}_e$ in the global residual $\bR_{e,i,\hat{u}}$ of equation~\eqref{eq:globalDiscrete} is directly taken as $\rhoE$.

\begin{remark} \label{rk:tauLinearisation}
The residuals in the local and global problems assume the stabilisation to be evaluated at time $t^{n-1}$ to avoid the need to linearise the stabilisation tensors and scalars for the RANS and SA equations, respectively. This choice significantly facilitates the linearisation and, as it will be shown in the numerical examples, it does not have a noticeable impact in the stability of the time marching scheme. 
\end{remark}

The resulting FCFV discrete problem consists of solving the nonlinear system of equations
\begin{equation} \label{eq:nonLinearR}
	\bR(\bLe^n,\bue^n,\dotsc,\bue^{n-\nstg},\bqTe^n,\nuTe^n,\dotsc,\nuTe^{n-\nstg},\bhuV^n,\bhuV^{n-1},\brhoV^n,\bm{\hnu}^n,\bm{\hnu}^{n-1}) = \bm{0},
\end{equation}
at each time step, where the residual is obtained by assembling the cell contributions
\begin{equation} \label{eq:nonLinearRElem}
	\bR_{e,i} := 
	\begin{Bmatrix}	
		\bR_{e,L}(\bLe^n,\bhuV^n)
		\\
		\bR_{e,u}(\bLe^n,\bue^n,\dotsc,\bue^{n-\nstg},\bqTe^n,\nuTe^n,\bhuV^n,\bhuV^{n-1},\bm{\hnu}^{n-1})
		\\
		\bR_{e,\tilde{q}}(\bqTe^n,\bm{\hnu}^n)
		\\
		\R_{e,\nuT}(\bqTe^n,\nuTe^n,\dotsc,\nuTe^{n-\nstg},\bhuV^n,\bhuV^{n-1},\bm{\hnu}^n,\bm{\hnu}^{n-1})
		\\
		\bR_{e,i,\hat{u}}(\bLe^n,\bue^n,\rhoE^n,\bhuV^n,\bhuV^{n-1},\bm{\hnu}^n,\bm{\hnu}^{n-1})
		\\
		\R_{e,\rho}(\bhuV^n)
		\\
		\bR_{e,i,\hat{\nu}}(\bqTe^n,\nuTe^n,\bhuV^{n-1},\bm{\hnu}^n,\bm{\hnu}^{n-1})
	\end{Bmatrix},
\end{equation}
for $e = 1,\dotsc,\numel$ and $i \in \Bset$. 

Following Remark~\ref{rk:tauLinearisation}, the dependence of the residuals on $\bhuV^{n-1}$ and $\bm{\hnu}^{n-1}$, used to compute the stabilisation of the RANS and SA equations, is explicitly stated.

\begin{remark} \label{rk:vectorisationL}
To simplify the presentation, an abuse of notation is introduced in equation~\eqref{eq:nonLinearRElem}, where $\bR_{e,L}$ and $\bLe^n$ denote the vectorised version of the matrices appearing in equations~\eqref{eq:localDiscrete} and \eqref{eq:globalDiscrete}. 
\end{remark}

%==========================================================================
\subsection{Newton-Raphson linearisation}               \label{sc:FCFV_NR}
%==========================================================================

The Newton-Raphson method is applied to linearise the residual of equation~\eqref{eq:nonLinearR}. After truncating the Taylor expansion at first order, the linear system to be solved at each Newton-Raphson iteration, $m$, can be written as
\begin{equation} \label{eq:NRiteration}
	\begin{bmatrix}
		\TUU   & \TULambda  \\
		\TLambdaU   & \TLambdaLambda\\
	\end{bmatrix}^{n,m}
	\begin{Bmatrix}
		\Delta \bU \\
		\Delta \bLambda 
	\end{Bmatrix}^{n,m}
	= - 
	\begin{Bmatrix}
		\bR_U \\
		\bR_{\Lambda}
	\end{Bmatrix}^{n,m} ,
\end{equation}
where 
\begin{equation} \label{eq:NRiterationDefs}
	\begin{aligned}
		\TUU &= \begin{bmatrix}
			\TLL   & \bm{0} & \bm{0} & \bm{0}   \\
			\TuL   & \Tuu   & \Tuq   & \Tunu    \\
			\bm{0} & \bm{0} & \Tqq   & \bm{0}   \\
			\bm{0}  & \bm{0} & \Tnuq  & \Tnunu 
		\end{bmatrix},
		&\TULambda &= \begin{bmatrix}
			\TLhu  & \bm{0} & \bm{0}  \\
			\Tuhu  & \bm{0} & \bm{0}  \\
			\bm{0} & \bm{0} & \Tqhnu  \\
			\bm{0} & \bm{0} & \Tnuhnu 
		\end{bmatrix},
		\\
		\TLambdaU &= \begin{bmatrix}
			\ThuL  & \Thuu  & \bm{0} & \bm{0}  \\
			\bm{0} & \bm{0} & \bm{0} & \bm{0}  \\
			\bm{0} & \bm{0} & \Thnuq & \Thnunu 
		\end{bmatrix},
		&\TLambdaLambda &= \begin{bmatrix}
			\Thuhu & \Thurho& \Thuhnu \\
			\Trhohu& \bm{0} & \bm{0}  \\
			\bm{0} & \bm{0} & \Thnuhnu
		\end{bmatrix},
		\\
		\bU &= \begin{Bmatrix}
			\bLV \\
			\buV \\
			\bqTV \\
			\bnuTV 
		\end{Bmatrix},
		\qquad
		\bLambda = \begin{Bmatrix}
			\bhuV \\
			\brhoV \\
			\bhnuV 
		\end{Bmatrix},
		\quad
		& \bR_U &= \begin{Bmatrix}
			\bR_L \\
			\bR_u \\
			\bR_{\tilde{q}} \\
			\bR_{\nuT}
		\end{Bmatrix},
		\qquad
		\bR_{\Lambda} = \begin{Bmatrix}
			\bR_{\hat{u}} \\
			\bR_\rho \\
			\bR_{\hat{\nu}} 
		\end{Bmatrix},
	\end{aligned}
\end{equation}
with $\bT_{ab}$ denoting the tangent matrix obtained by assembling the contributions $(\bT_{ab})_{e,i} := \partial \bR_{e,i,a}/\partial b$ and $\Delta \circledcirc^{n,m} = \circledcirc^{n,m+1} - \circledcirc^{n,m}$ the solution increment from the Newton-Raphson iteration $m$ to $m+1$.

Given the block-diagonal structure of the matrix $\TUU$, the linear sytem~\eqref{eq:NRiteration}, to be solved at each Newton-Raphson iteration, can be reduced to involve only the unknowns $\bLambda$, namely
\begin{equation} \label{eq:NRiterationReduced}
	\mathbb{K}^{n,m} \Delta \bLambda^{n,m} = \mathbb{F}^{n,m},
\end{equation}
where $\mathbb{K} = \TLambdaLambda - \TLambdaU \TUU^{-1} \TULambda$ and $\mathbb{F} = -\bR_{\Lambda} + \TLambdaU \TUU^{-1}\bR_U$.

After solving the reduced system~\eqref{eq:NRiterationReduced}, the local problems can be used to compute the solution in each cell as
\begin{equation} \label{eq:NRiterationLocal}
	\TUU^{n,m} \Delta \bU^{n,m} =  -\bR_U^{n,m} - \TULambda^{n,m} \Delta \bLambda^{n,m}.
\end{equation}
It is worth emphasising that the block-diagonal structure of the matrix $\TUU$ enables the solution of a small system of equations, cell-by-cell.

Computational aspects are discussed in detail in~\ref{app:FCFV_computational}, including the solution of local and global problems, the use of pseudo-time marching for steady state problems and a staggered approach to handle the turbulence, alternative to the monolithic approach presented in this section. The current implementation is done in Fortran 90. For the readers interested in the details of the FCFV method, a simpler implementation of the HDG method, for linear elliptic problems and using low and higher order approximations, in Matlab~\cite{HDGlab-GSH-20} is available.

%==========================================================================
\section{Numerical examples}               \label{sc:examples}
%==========================================================================

This section presents a set of numerical examples selected to test the optimal approximation properties of the proposed FCFV method and to compare the performance of the different convective stabilisations proposed. Examples include steady and transient cases, both in laminar and turbulent regimes.

When errors with respect to an analytical or reference solutions are reported, $E$ denotes the relative error and $\| \cdot \|_{L^2}$ denotes the $\mathcal{L}^2(\Omega)$ norm for velocity, pressure and gradient of velocity and the $\mathcal{L}^2(\Gamma)$ norm for the face velocity.

%==========================================================================
\subsection{Couette flow}               \label{sc:couette}
%==========================================================================

The first example is the so-called Couette flow, which involves the flow in an annulus with imposed angular velocity on the boundary. The inner and outer radii are $R_i=1$ and $R_o=2$, respectively, and the imposed angular velocities are $\Omega_i = 0$ and  $\Omega_o = 0.5$, respectively. This example is used to test the optimal rate of convergence of the FCFV for laminar flows under mesh refinement. In polar coordinates, the exact solution is 
\begin{equation}\label{eq:Couette_Statment}
\left\{\begin{aligned}
	u_r &= 0, \qquad u_\phi = C_1 r + C_2 \frac{1}{r}, \\
	p	&=  C_1^2 \frac{r^2}{2} + 2 C_1 C_2 \log(r) - \frac{C_2^2}{2 r^2} + C,
\end{aligned} \right.
\end{equation}
where $C_1 = (\Omega_o R_o^2 - \Omega_i R_i^2) /(R_o^2 - R_i^2)$, $C_2 = (\Omega_i - \Omega_o)R_i^2 R_o^2/(R_o^2 - R_i^2)$ and $C$ is a constant such that the pressure at the outer radius is equal to 1. As the exact solution does not depend upon the viscosity, the Reynolds number is selected as $Re = 1$.

\begin{figure}[!tb]
	\centering
	\subfigure[]{\includegraphics[width=0.24\textwidth]{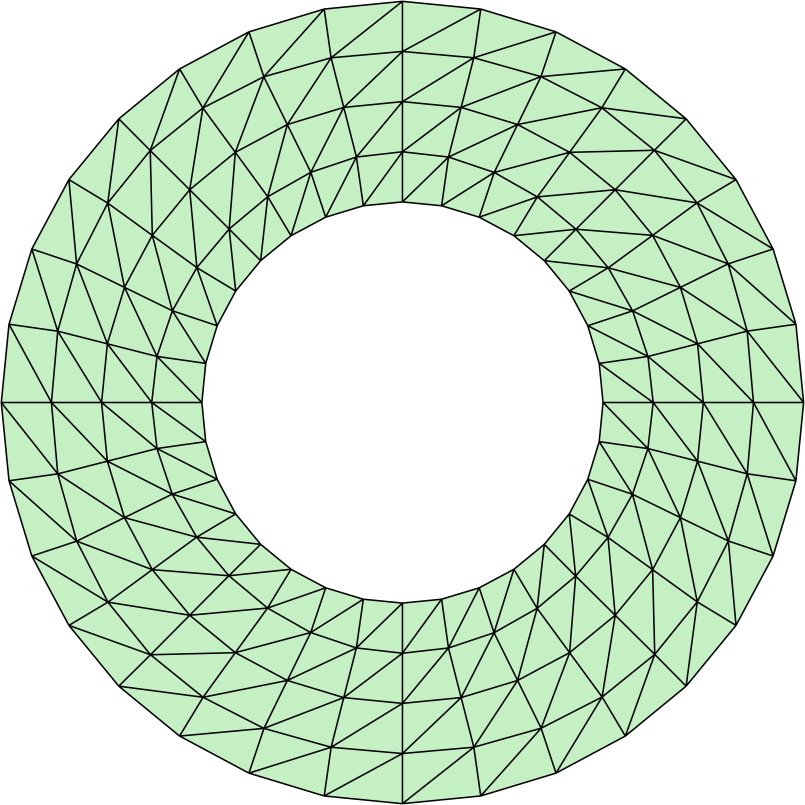}}
	\subfigure[]{\includegraphics[width=0.24\textwidth]{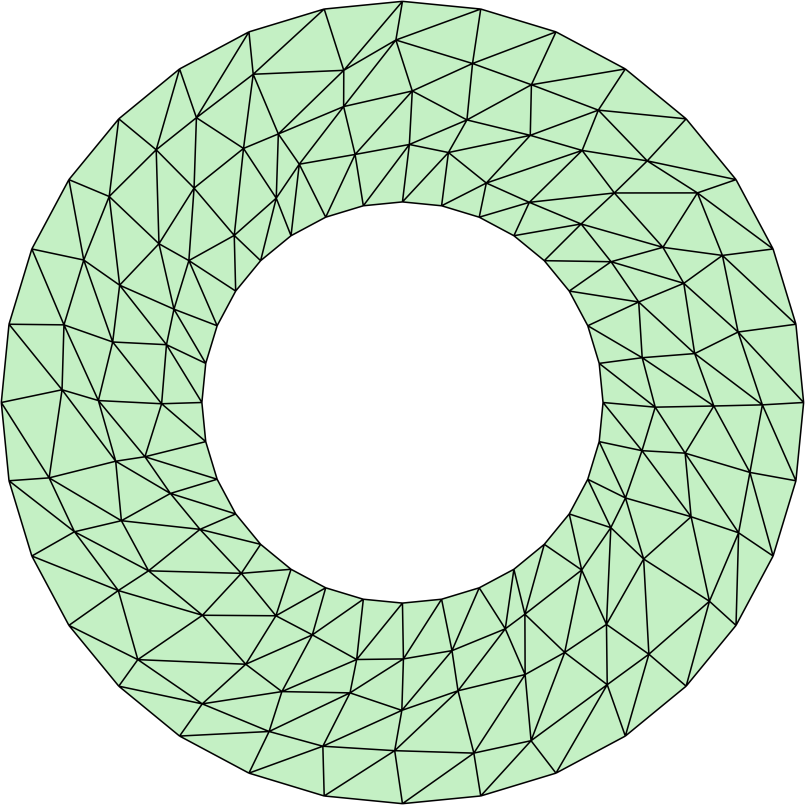}}
	\subfigure[]{\includegraphics[width=0.24\textwidth]{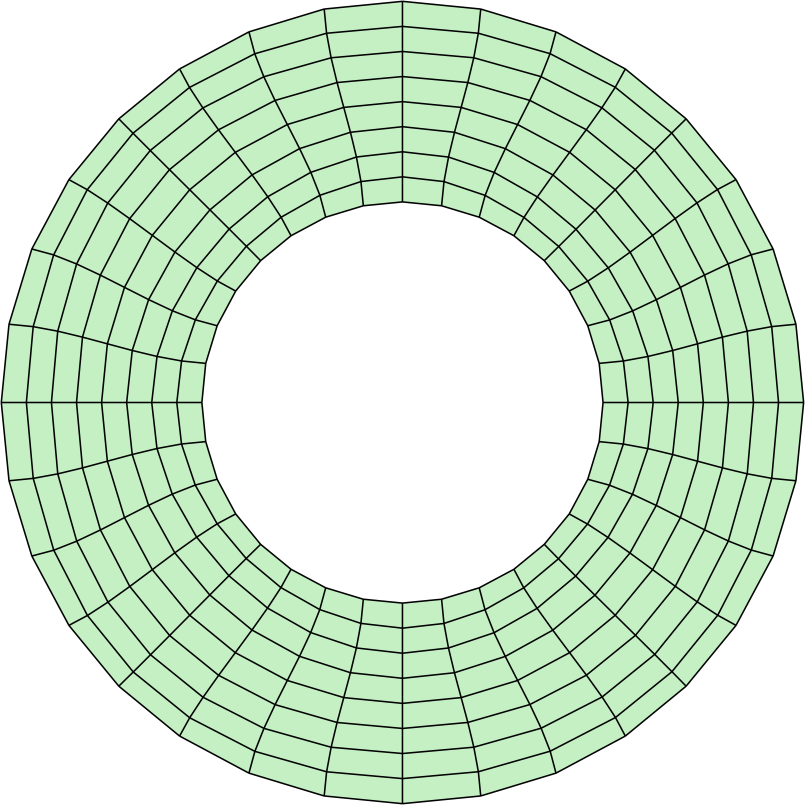}}
	\subfigure[]{\includegraphics[width=0.24\textwidth]{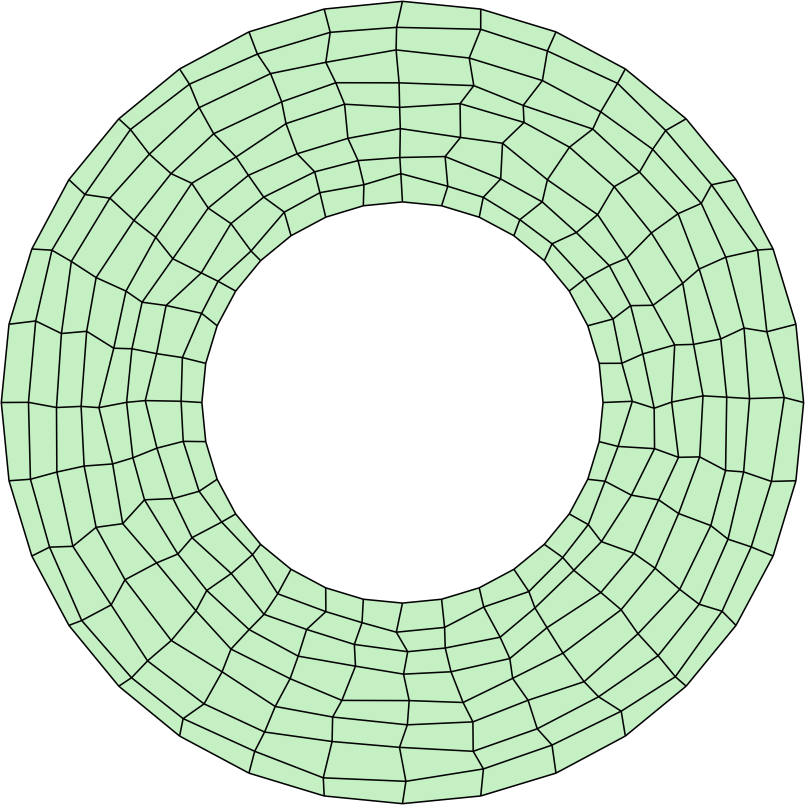}}
	\caption{Couette flow: (a,b) Triangular and (c,d) quadrilateral meshes used for the convergence study.}
	\label{fig:CouetteMeshes}
\end{figure}

Triangular and quadrilateral structured meshes are considered and the effect of cell distortion on the accuracy of the computations is evaluated. Figure~\ref{fig:CouetteMeshes} shows two structured triangular and quadrilateral meshes and the corresponding meshes where the internal nodes have been randomly moved to test the effect of the cell distortion. The strategy used to distort the meshes can be found in~\cite{RS-SGH:2018_FCFV1}.

\begin{figure}[!tb]
	\centering
	\subfigure[Triangles]{\includegraphics[width=0.49\textwidth]{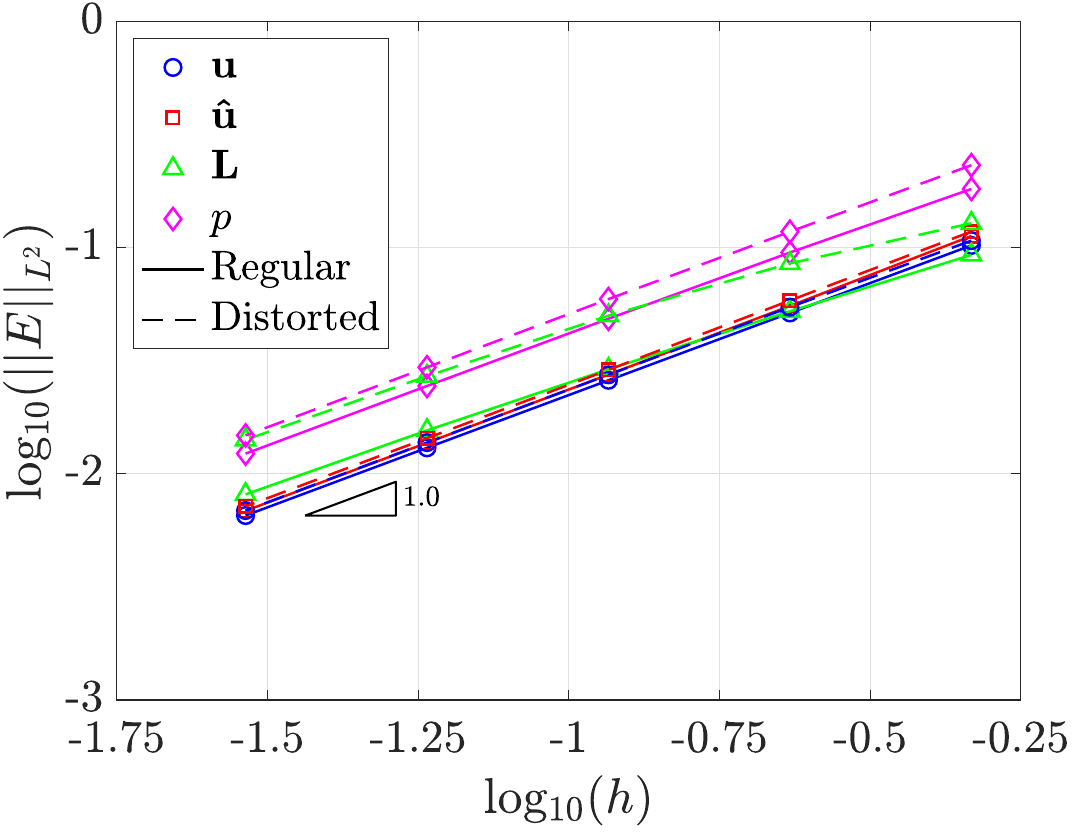}}
	\subfigure[Quadrilaterals]{\includegraphics[width=0.49\textwidth]{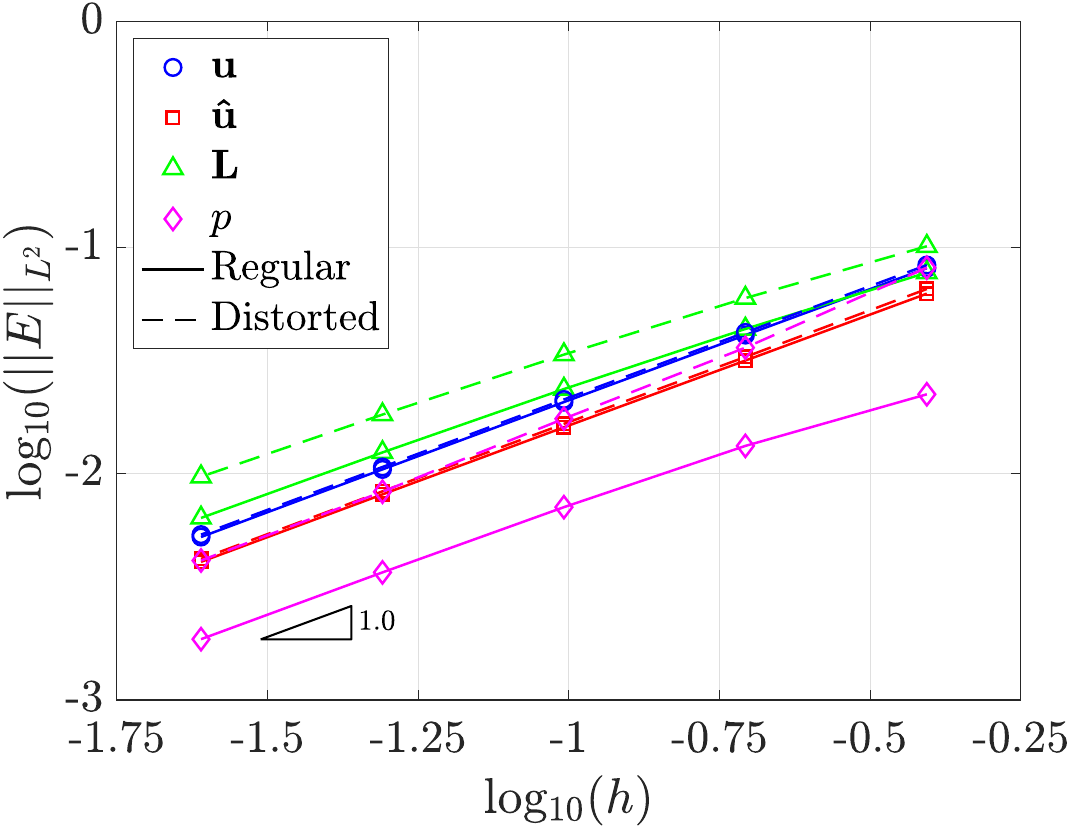}}
	\caption{Couette flow: Mesh convergence of the error of the cell velocity, face velocity, gradient of the velocity and pressure for regular and distorted triangular and quadrilateral meshes using the HLL convection stabilisation.}
	\label{fig:Couette_hConv}
\end{figure}

To perform a mesh convergence analysis, the meshes are uniformly refined and the error in all quantities, namely cell velocity, face velocity, gradient of the velocity and pressure is measured. The $i$-th triangular and quadrilateral meshes have $(8 \times 2^i) \times (8 \times 2^i)$ cells. The results, reported in figure~\ref{fig:Couette_hConv} for regular and distorted triangular and quadrilateral meshes, show the optimal convergence of the error for all quantities under mesh refinement.
The computations performed with distorted meshes show a very similar accuracy when compared to the computation on the corresponding regular mesh. It can be observed that a slightly more accurate result on the pressure and the gradient of the velocity is obtained in regular meshes, whereas the accuracy of the velocity is not influenced by the cell distortion. In addition, the effect of cell distortion on the accuracy of the pressure and the gradient of the velocity is more noticeable when using quadrilateral meshes.

The extra accuracy obtained using quadrilateral meshes in this example is simply due to the nature of the analytical solution, only dependent on the radial coordinate, and the selected meshes.

In this example, the HLL convection stabilisation is employed, but further numerical examples not reported for brevity, show that the results with the LF or the Roe stabilisation are almost identical. This is expected due to the diffusion dominated character of the solution.

Due to the simplicity of this example, the steady problem is directly solved in all cases, without marching in pseudo-time. A tolerance of $10^{-12}$ is imposed for the nonlinear Newton-Raphson iterations and quadratic convergence is observed in all cases. Details about the residuals used to check convergence and an example showing the quadratic convergence can be found in~\ref{app:NRconvergence}.

%==========================================================================
\subsection{Lid-driven cavity flow}               \label{sc:cavity}
%==========================================================================

The second problem is the lid-driven cavity flow, defined in $\Omega=[0,1]^2$. A constant horizontal velocity of magnitude one is imposed in the upper part of the domain, whereas homogeneous Dirichlet boundary conditions are imposed on the rest of the boundary. As no analytical solution is available, the accuracy of the computations is measured against a reference solution computed with the Taylor-Hood (Q2Q1) element and the streamline-upwind Petrov-Galerkin (SUPG) method~\cite{Donea2003}. The reference solution is computed on a mesh of $700 \times 700$ cells that introduces local refinement near the boundaries, with the height of the first cell being $9.6 \times 10^{-5}$. It is worth noting that the reference solution corresponds to the so-called leaky cavity, due to the strong imposition of the incompatible boundary conditions at the top left and right corners. With the FCFV approach, the incompatible boundary conditions do not represent an issue because the velocity is imposed at the barycentre of the faces. To account for this discrepancy between the FCFV and reference solutions, the $\mathcal{L}^2$ norm of the error, for all variables, is measured excluding the regions $[0,0.05]\times[0.95,1]$ and $[0.95,1]\times[0.95,1]$.

This example aims at comparing the accuracy of the different convection stabilisations derived in~\ref{app:convectiveTau} and to assess the influence of the mesh distortion for convection dominated problems. To this end, the Reynolds number is taken as $Re=1,000$ and regular and distorted triangular meshes are employed. As in the previous example, the steady problem is directly solved, without marching in pseudo-time, with a tolerance of $10^{-12}$.

\begin{figure}[!tb]
	\centering
	\subfigure[]{\includegraphics[width=0.35\textwidth]{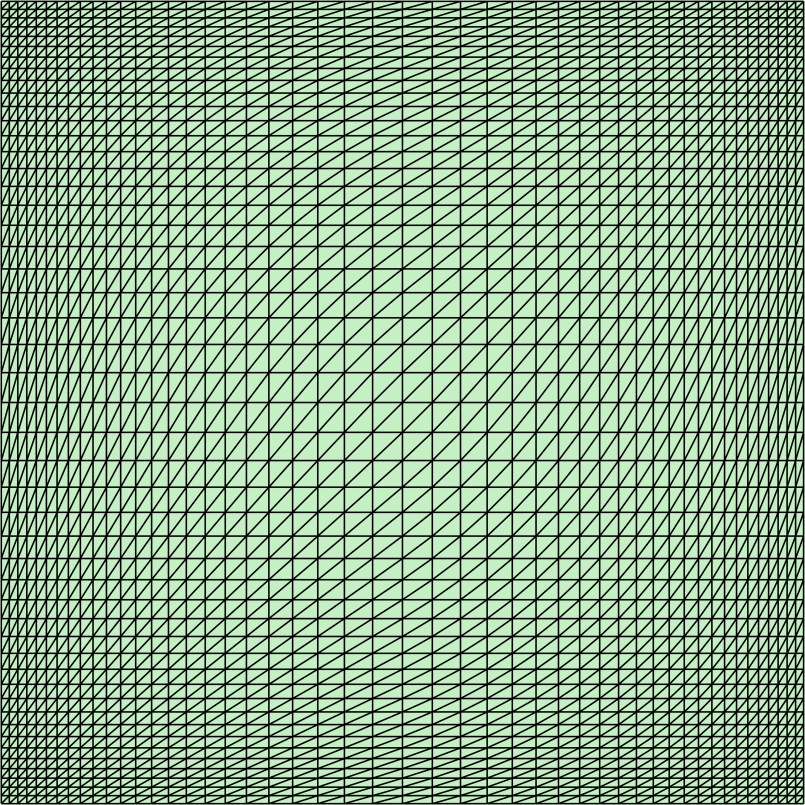}}
	\subfigure[]{\includegraphics[width=0.35\textwidth]{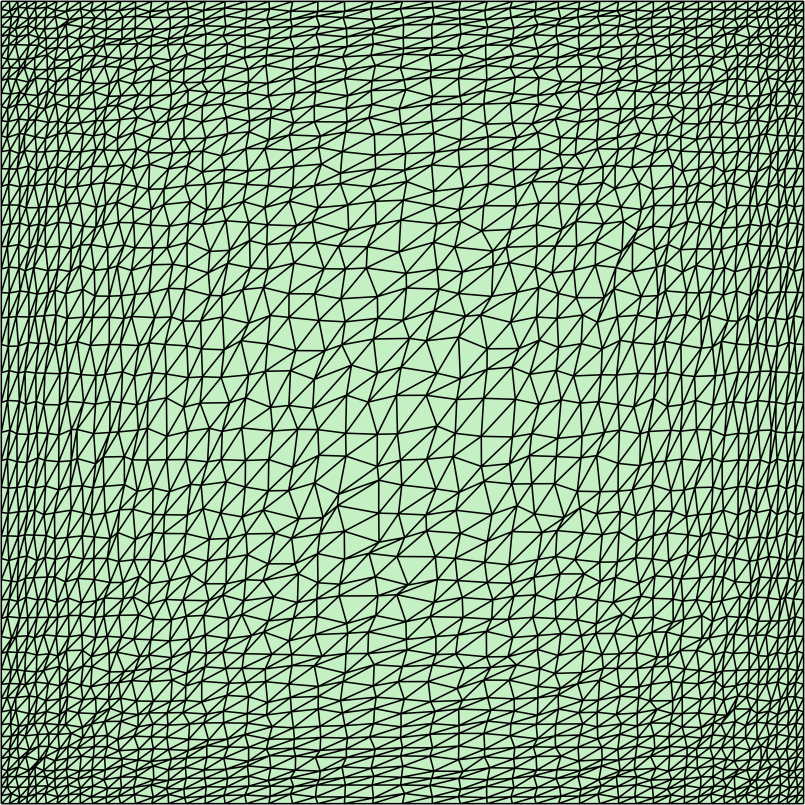}}
	\caption{Lid-driven cavity flow: (a) Regular and (b) distorted triangular meshes used for the convergence study.}
	\label{fig:CavityMeshes}
\end{figure}

Five meshes are considered for the mesh convergence study. The $i$-th mesh has $(24 \times 2^i) \times (24 \times 2^i) \times 2$ cells, the height of the first layer of boundary cells is $h_0/i$, where $h_0$ is taken as $ 10^{-2}$, and the growth ratio that dictates the height of the next layer of cells varies from 1.06 in the coarse mesh to 1.003 in the finest mesh. Figure~\ref{fig:CavityMeshes} shows the first regular and  distorted triangular meshes.

\begin{figure}[!tb]
	\centering
	\subfigure[]{\includegraphics[width=0.49\textwidth]{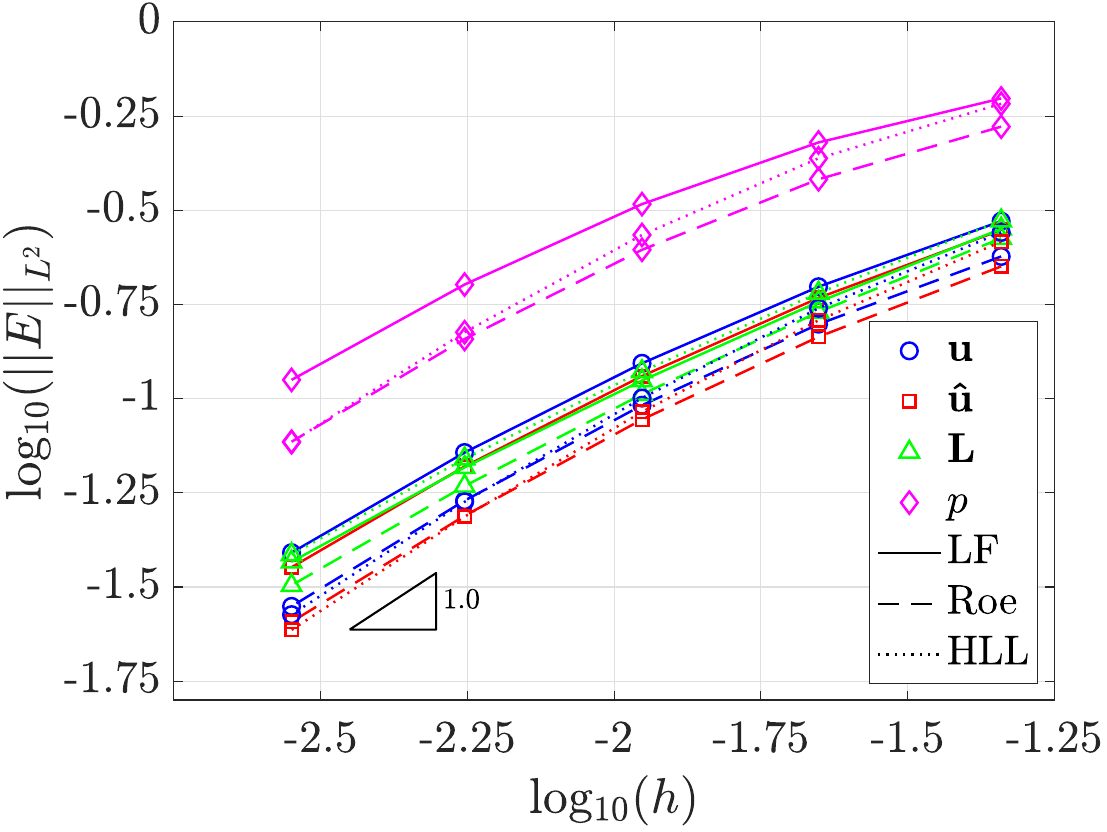}}
	\subfigure[]{\includegraphics[width=0.49\textwidth]{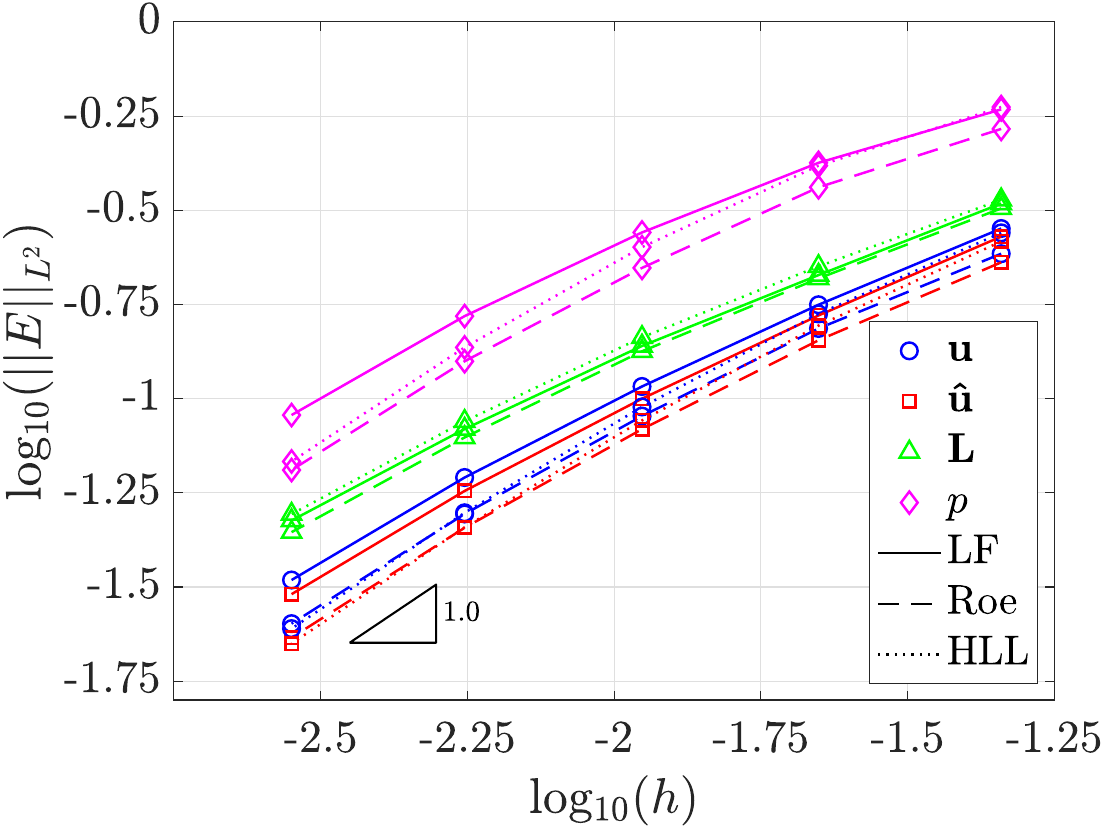}}
	\caption{Lid-driven cavity flow: Mesh convergence of the error of the cell velocity, face velocity, gradient of the velocity and pressure for (a) regular and (b) distorted triangular meshes using different convection stabilisations.}
	\label{fig:Cavity_hConv}
\end{figure}

The results of the mesh convergence study are depicted in figure~\ref{fig:Cavity_hConv} and show again the expected first-order convergence of the error for all variables, as the mesh is refined.
For this example, the Roe and HLL convective stabilisations provide very similar accuracy, whereas the LF stabilisation is less accurate. This is observed for all variables and on both regular and distorted meshes. It is also worth noting that the use of distorted meshes provides accuracy comparable to regular meshes, even for this example where resolving the boundary layers is crucial. 

\begin{figure}[!tb]
	\centering
	\subfigure[]{\includegraphics[width=0.49\textwidth]{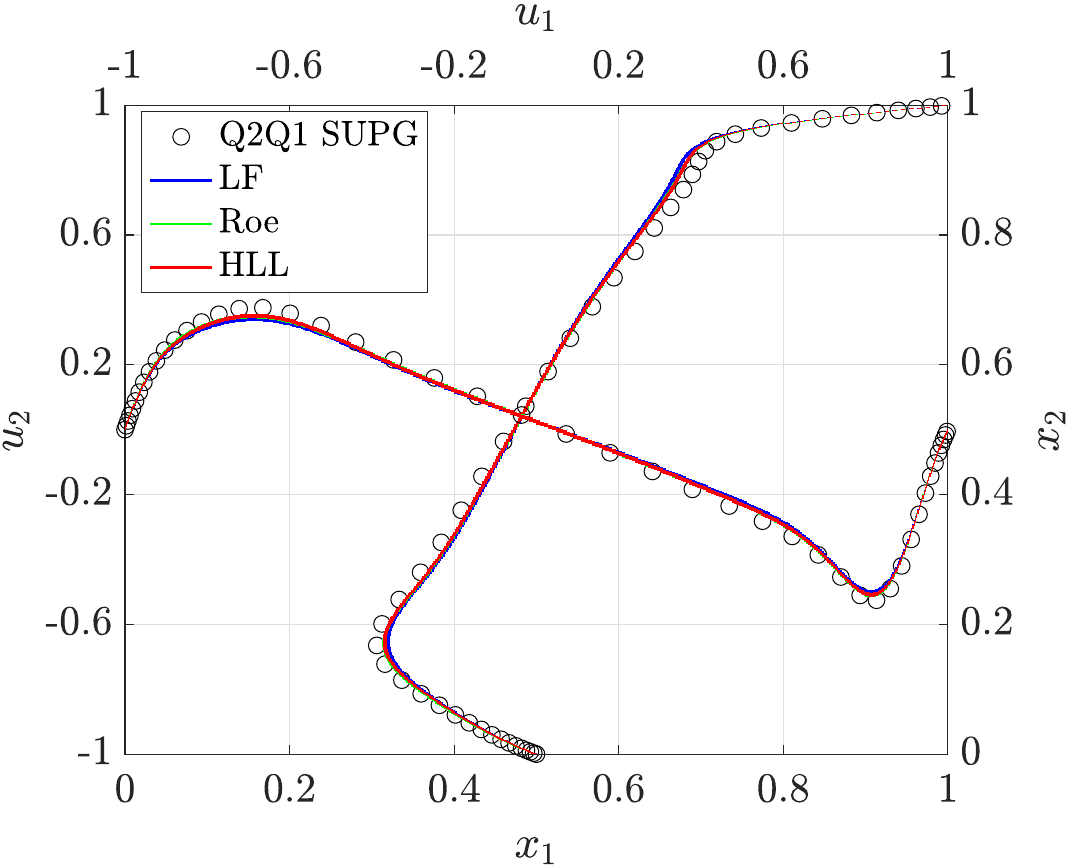}}
	\subfigure[]{\includegraphics[width=0.49\textwidth]{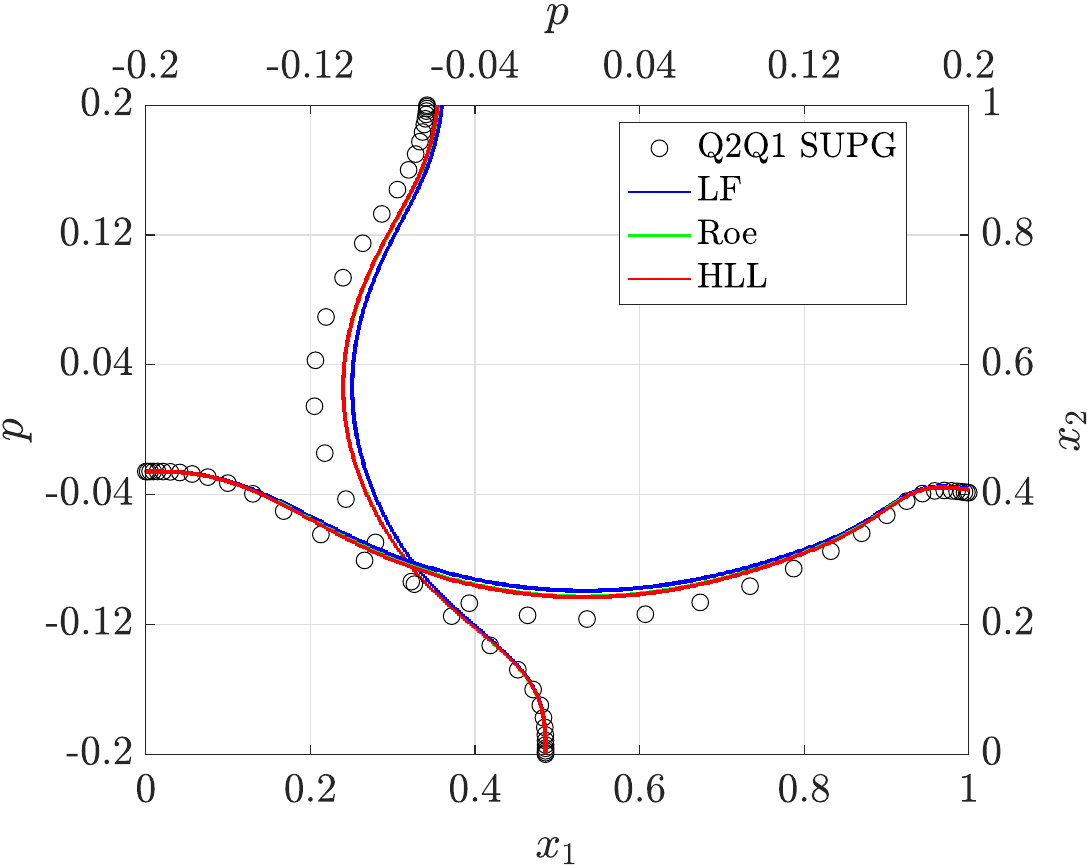}}
	\caption{Lid-driven cavity flow: Profiles of (a) velocity and (b) pressure along the centrelines, using the fourth regular mesh and different convection stabilisations.}
	\label{fig:CavitySections}
\end{figure}

To further assess the accuracy of the results, figure~\ref{fig:CavitySections} displays the profiles of the velocity and pressure fields along the centrelines, using the fourth regular mesh and different convection stabilisations. 
The results of the proposed FCFV formulation are compared to the reference solution and illustrate the extra accuracy obtained by using the Roe and HLL stabilisations, when compared to the LF stabilisation.

%==========================================================================
\subsection{Assessment of the convective stabilisation}               \label{sc:convectiveSta}
%==========================================================================

The next example, taken from~\cite{cheung2015staggered}, is used to evaluate the dissipative effect of the three convective stabilisations proposed in~\ref{app:convectiveTau}. 

The computational domain is $\Omega = (0,1)^2$ and the manufactured solution is given by
\begin{equation}\label{eq:manufactured}
	\left\{\begin{aligned}
		u_1 & = t^4 \sin(2\pi x_2) \big( 1 - \cos(2\pi x_1)\big), \\
		u_2 & = -t^4 \sin(2\pi x_1) \big( 1 - \cos(2\pi x_2)\big), \\
		p	&=  t^4 \big( \cos(\pi x_1) + \cos(\pi x_2) \big),
	\end{aligned} \right.
\end{equation}
and the Reynolds number is $Re=10^5$. For such a high Reynolds number, the stabilisation of the FCFV is dominated by the convective stabilisation because the diffusive stabilisation is of the order of $1/Re$, as detailed in~\eqref{eq:stabilisationDiffusion}.

The solution is advanced until a final time $T=1$ using the BDF2 time integrator scheme and $\Delta t = 0.001$. The time step has been selected to ensure that all the error is dominated by the spatial discretisation and a clear comparison of the three different stabilisations can be performed. To assess the dissipative effect of the stabilisation, the quantity
\begin{equation}\label{eq:energy}
E(T) = \int_{\Omega} \bu(\bx,T) \cdot \bu(\bx,T) d \Omega,
\end{equation}
proportional to the kinetic energy, is evaluated and compared to the exact value, which is 1.5.

Three meshes are considered to perform the study. The $i$-th quadrilateral mesh is a structured mesh of  $(32 \times 2^i) \times (32 \times 2^i)$ cells, whereas the $i$-th triangular mesh is obtained after splitting in four triangles each quadrilateral of a mesh with $(16 \times 2^i) \times (16 \times 2^i)$ cells. Therefore, the $i$-th triangular and quadrilateral meshes contain the same number of cells and a further comparison of the dissipative effect of the stabilisation for different element shape can be performed.

Table~\ref{tab:errorE} reports the relative error on $E(1)$ for three different triangular and quadrilateral meshes and using the three different convective stabilisations.
\begin{table}[!tb]
	\centering
	\begin{tabular}{|c|c|c|c|c|c|c|} 
		\hline
		& \multicolumn{3}{|c|}{Triangles} & \multicolumn{3}{|c|}{Quadrilaterals} \\
		\hline
	    Mesh & LF & Roe & HLL &  LF & Roe & HLL \\
		\hline
		1 & $8.5\!\times\!10^{-2}$ & $5.6\!\times\!10^{-2}$ & $2.8\!\times\!10^{-2}$ & $9.4\!\times\!10^{-2}$ & $6.6\!\times\!10^{-2}$ & $3.3\!\times\!10^{-2}$ \\
		2 & $4.6\!\times\!10^{-2}$ & $3.0\!\times\!10^{-2}$ & $1.5\!\times\!10^{-2}$ & $5.2\!\times\!10^{-2}$ & $3.7\!\times\!10^{-2}$ & $1.8\!\times\!10^{-2}$ \\
		3 & $2.4\!\times\!10^{-2}$ & $1.5\!\times\!10^{-2}$ & $7.7\!\times\!10^{-3}$ & $2.7\!\times\!10^{-2}$ & $1.9\!\times\!10^{-2}$ & $9.8\!\times\!10^{-3}$ \\
		\hline		
	\end{tabular}
	\caption{Relative error on $E(1)$ for three different triangular and quadrilateral meshes and using the three different convective stabilisations.}
	\label{tab:errorE}
\end{table}
The results show that the HLL stabilisation consistently produces the most accurate results, whereas the LF stabilisation always produces the least accurate results. This is observed for all meshes and both element types.

More precisely the error using Roe stabilisation induce an error 1.5 times lower than the error using LF. When using the HLL stabilisation the error is two times lower than the error obtained with the Roe stabilisation. This conclusion is observed for both triangular and quadrilateral cells.

Comparing the element types marginal differences are observed, with the triangles providing a slightly better accuracy for all meshes and all three stabilisations. 

%==========================================================================
\subsection{Unsteady laminar flow past a circular cylinder}               \label{sc:laminarCyl}
%==========================================================================

\begin{figure}[!tb]
	\centering
	\includegraphics[width=0.70\textwidth]{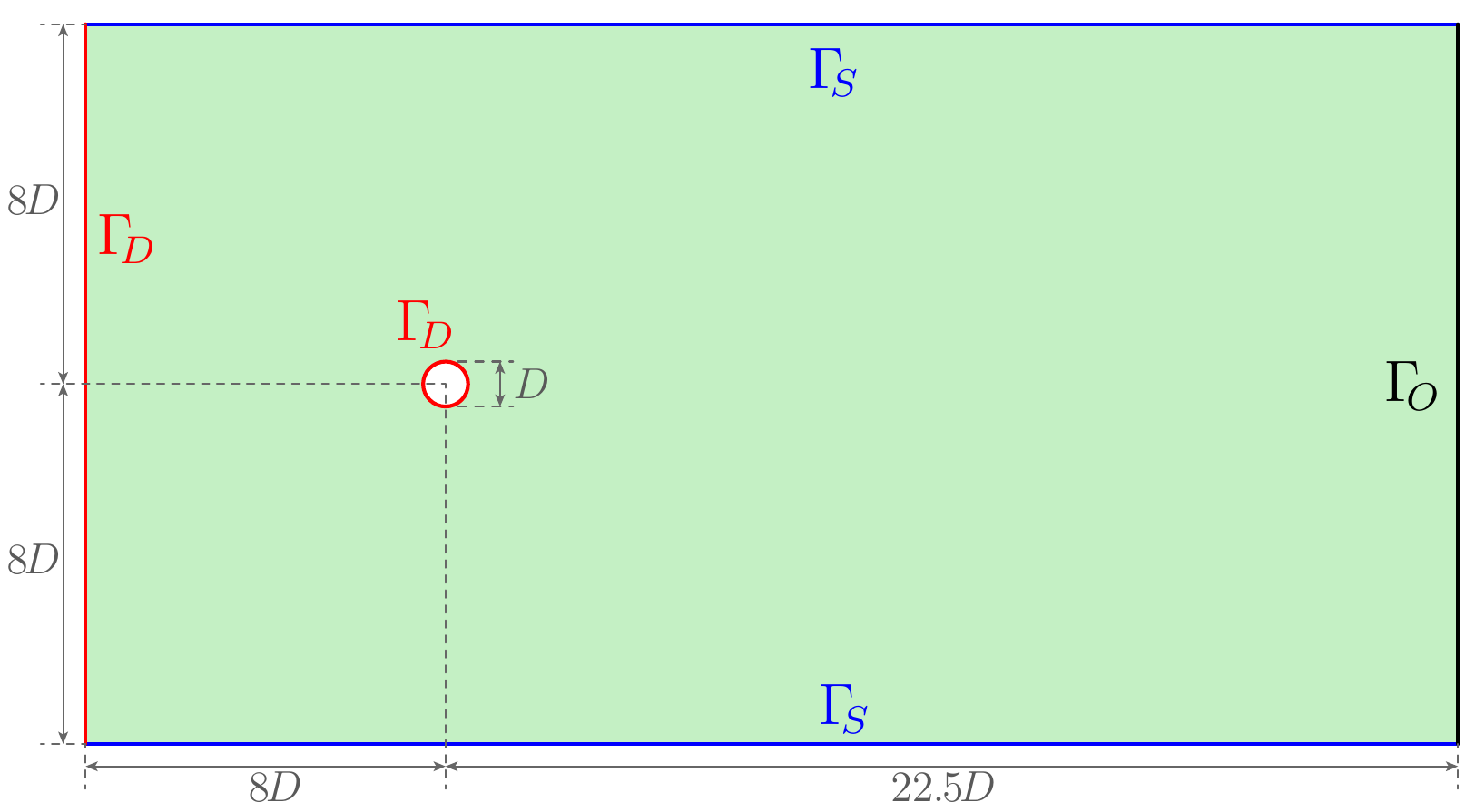}
	\caption{Unsteady laminar flow past a circular cylinder: Problem setup.}
	\label{fig:laminarCyl_Setup}
\end{figure}

The next example considers the unsteady flow past a circular cylinder of diameter $D$ at $Re=100$. This classical benchmark is used to assess the performance of the proposed method with different convective stabilisations for a transient laminar flow. The setup of the problem is depicted in figure~\ref{fig:laminarCyl_Setup}, including the relevant dimensions and the boundary conditions.
The Dirichlet boundary consists of the inflow boundary, where a constant horizontal velocity of magnitude one is imposed, and the cylinder, where a no-slip condition is enforced. 

\begin{figure}[!tb]
	\centering
	\subfigure[]{\includegraphics[height=0.2\textheight]{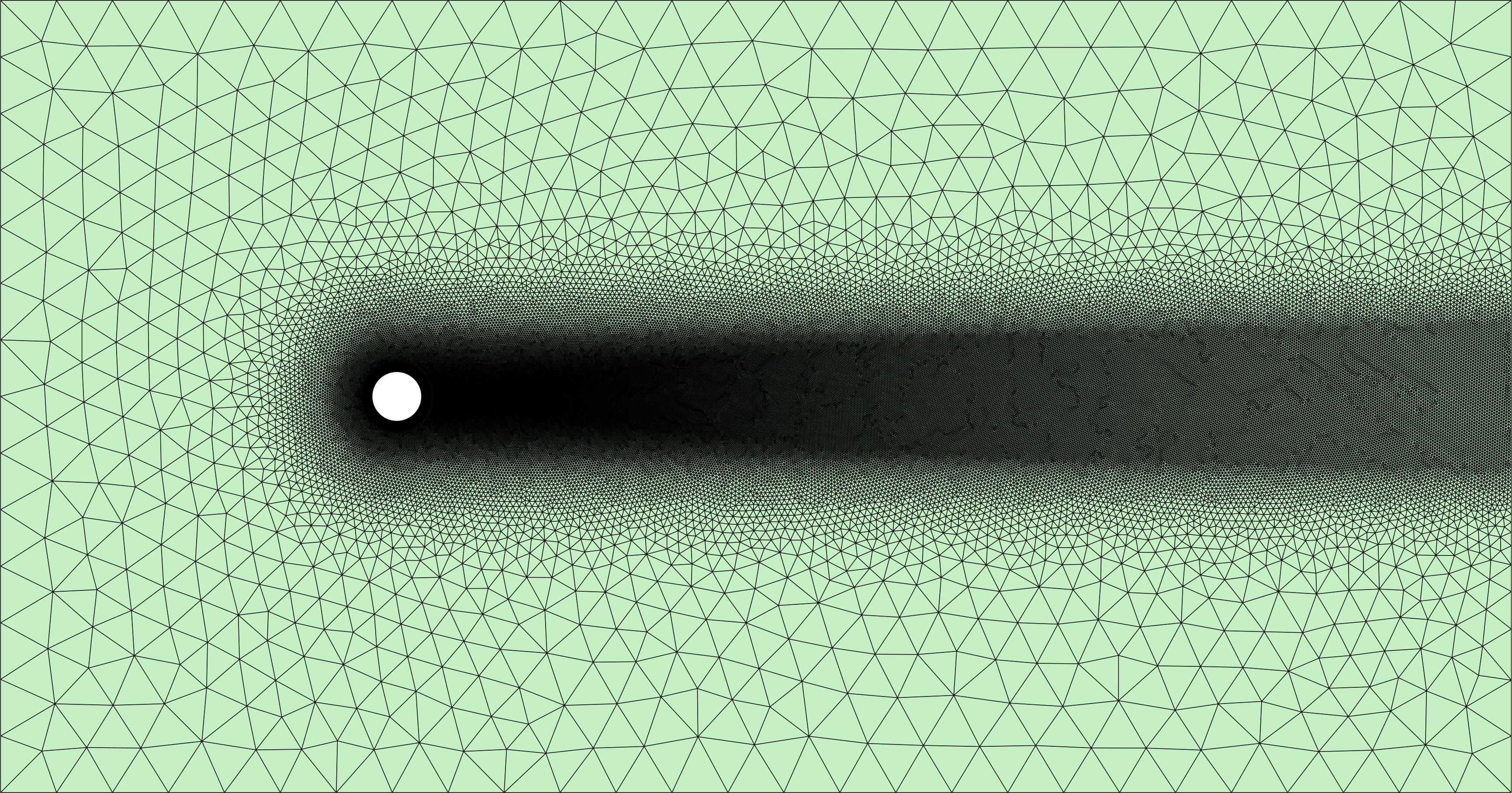}}
	\subfigure[]{\includegraphics[height=0.2\textheight]{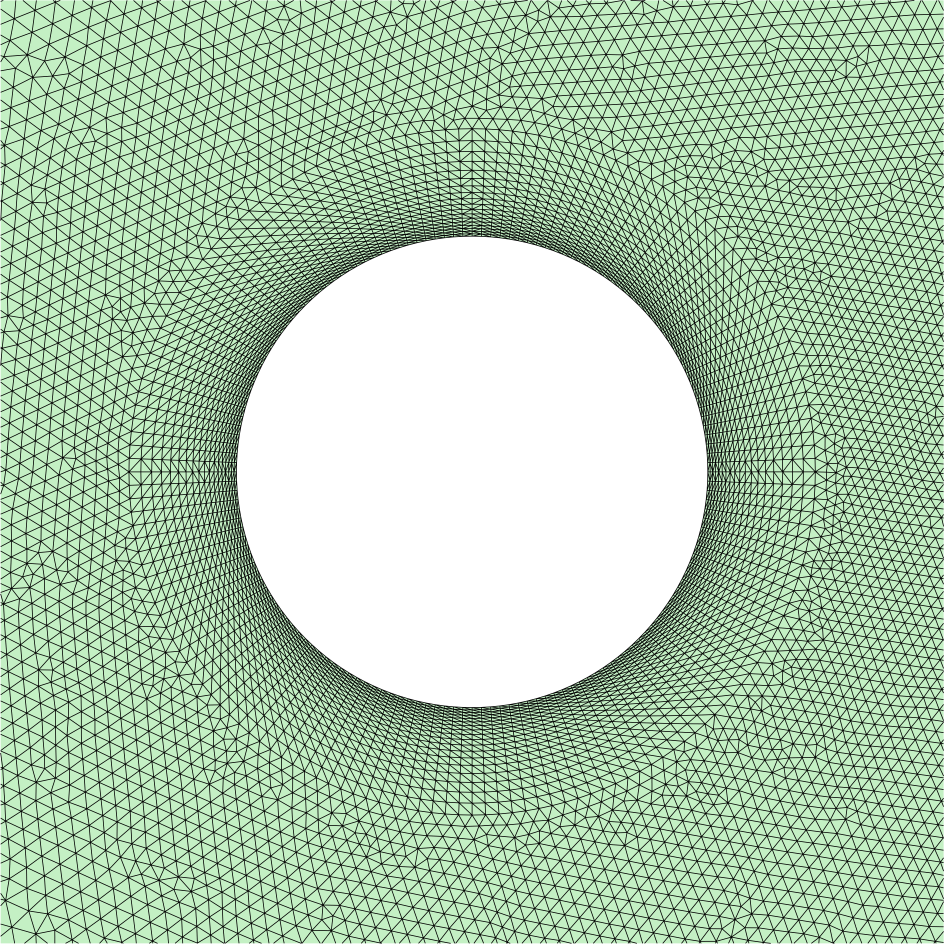}}
	\caption{Unsteady laminar flow past a circular cylinder: (a) Second mesh with 135,273 cells and (b) detail of the inflation layer.}
	\label{fig:laminarCylMesh2}
\end{figure}

Two unstructured triangular meshes are considered. For the first mesh, the inflation layer is defined using a height for the first cell on the cylinder equal to $0.01D$. A refinement behind the cylinder is introduced to capture the wake by imposing a cell size of $0.03D$ near the cylinder and $0.075D$ near the outflow boundary. For the second mesh, the height of the first cell in the inflation layer is $0.005D$ and the sizes to locally refine the region of the wake are $0.02D$ near the cylinder and $0.05D$ near the outflow. In both cases the growth ratio within the inflation layer is 1.1 and the maximum cell stretching near the cylinder is 10. The first mesh has 73,251 cells whereas the second mesh, shown in figure~\ref{fig:laminarCylMesh2}, has 134,801 cells.

For the time-integration, the BDF2 scheme is employed with a time-step $\Delta t = 0.1$, and the initial condition is taken as the steady-state solution with $Re = 10$. 

\begin{figure}[!tb]
	\centering
	\subfigure[]{\includegraphics[width=0.49\textwidth]{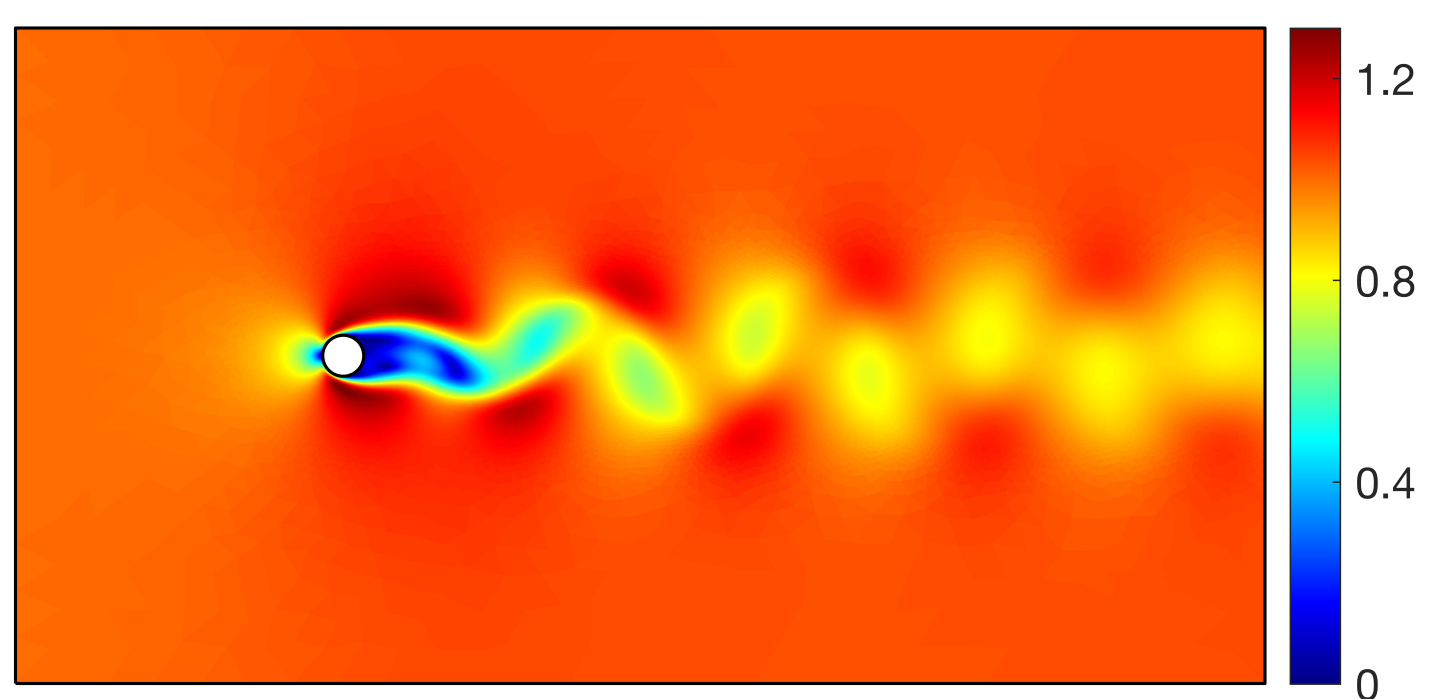}}
	\subfigure[]{\includegraphics[width=0.49\textwidth]{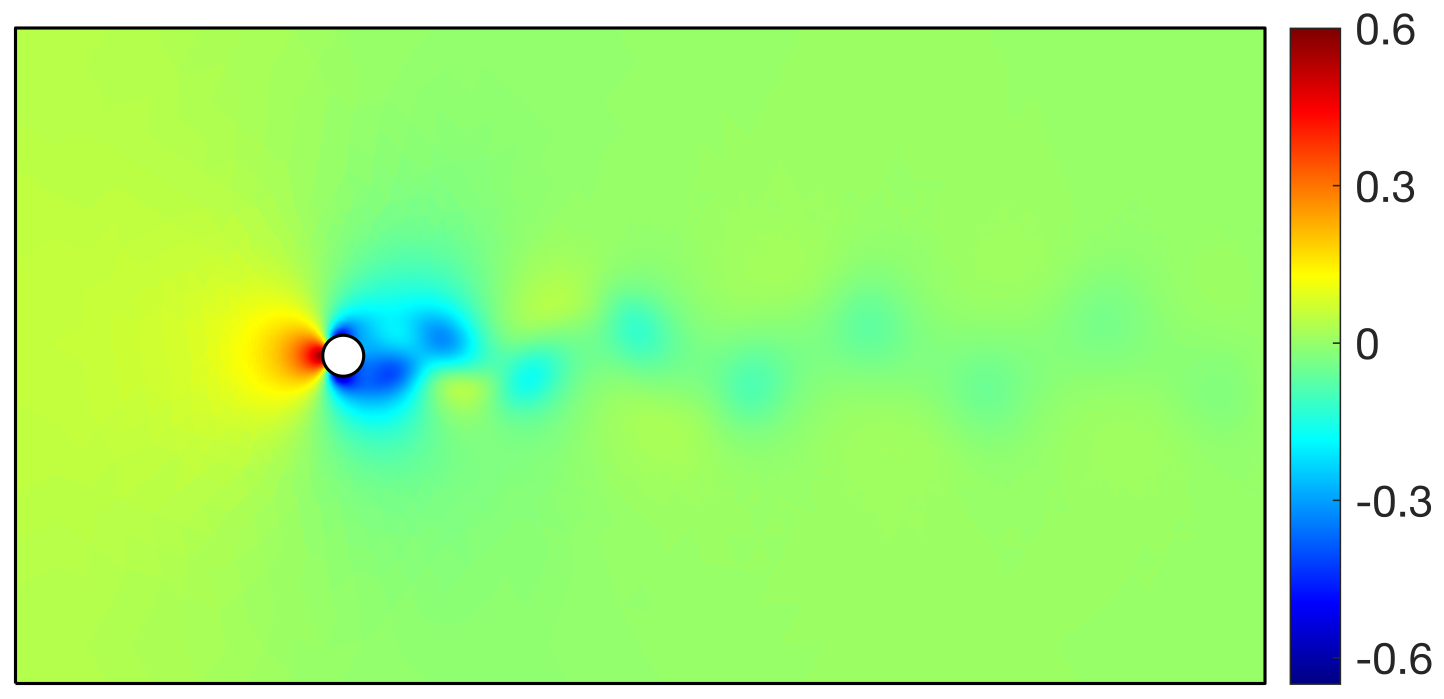}}
	\caption{Unsteady laminar flow past a circular cylinder: Snapshots of the (a) magnitude of the velocity and (b) pressure at $t=65$.}
	\label{fig:laminarCyl_Snapshots}
\end{figure}

Figure~\ref{fig:laminarCyl_Snapshots} shows a snapshot of the magnitude of the velocity and pressure fields at time $t=65$, when the solution has reached a periodic state, illustrating the ability of the proposed method to capture the von Kármán vortex street.
The displayed simulation corresponds to the FCFV solution in the second mesh using the HLL stabilisation.

A tolerance of $10^{-6}$ is used for the nonlinear problems and the average number of Newton-Raphson iterations across all time steps is two for the HLL stabilisation and one for the Roe and LF stabilisations.  

\begin{figure}[!tb]
	\centering
	\subfigure[]{\includegraphics[width=0.49\textwidth]{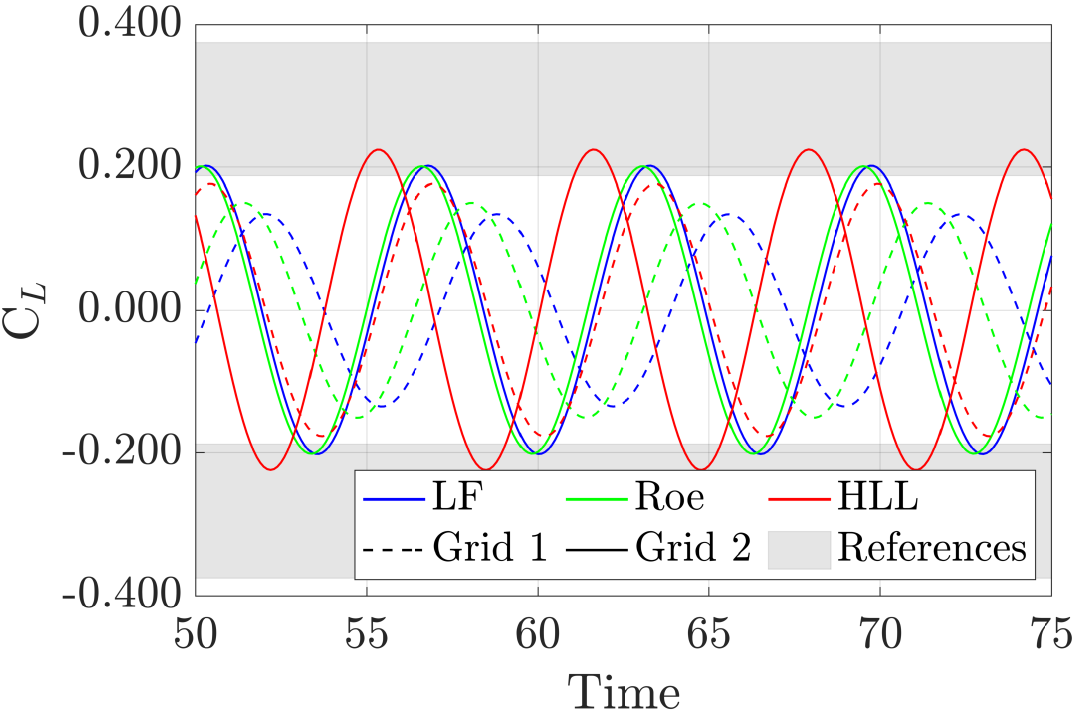}}
	\subfigure[]{\includegraphics[width=0.49\textwidth]{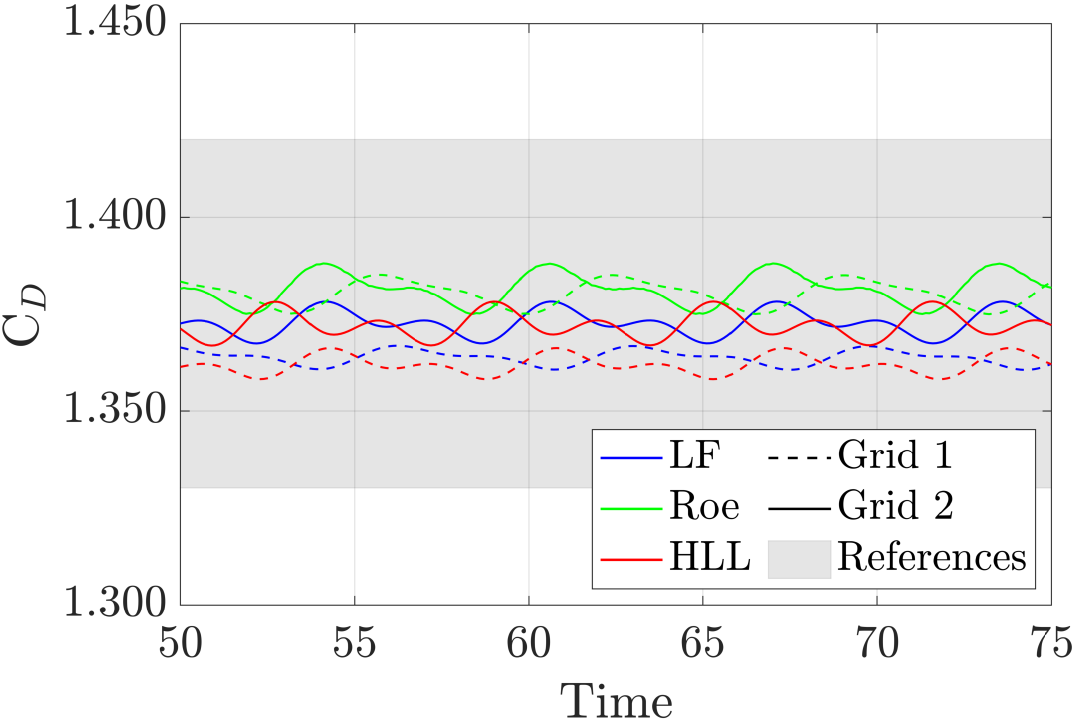}}
	\caption{Unsteady laminar flow past a circular cylinder: (a) Lift and (b) drag coefficients as a function of time for the two meshes and the three convective stabilisations.}
	\label{fig:laminarCyl_ClCd}
\end{figure}

To assess the accuracy of the simulations, the computed lift ($C_L$) and drag ($C_D$) coefficients and the Strouhal number ($S_t$) are compared against a number of results found in the literature. Figure~\ref{fig:laminarCyl_ClCd} shows the lift and drag coefficients as a function of time for the two meshes and the three convective stabilisations, where the shadowed area represents the range of values reported in the literature~\cite{Tezduyar1991,Tezduyar1992,Kjellgren1997,Rajani2009,Kadapa2015,Kadapa2020}.

For the references considered the number of elements varies between 5,000 and 18,000, but it is important to note that in these references linear, quadratic and higher order elements are considered for the simulations. The results reported in the literature are obtained for a variety of numerical schemes including finite volumes, stabilised finite elements and immersed methods with B-Splines.

A detailed comparison is provided in table~\ref{tab:laminarCyl}, showing the amplitude of the lift coefficient, the mean value of the drag and the Strouhal number using both meshes and the three different stabilisations. 
\begin{table}[!tb]
	\centering
	\begin{tabular}{|c|c|c|c|c|c|c|}
		\hline
		\multicolumn{7}{|c|}{FCFV results} \\  
		\hline
		& \multicolumn{2}{|c|}{$C_L$ amplitude} & \multicolumn{2}{|c|}{Mean $C_D$} & \multicolumn{2}{|c|}{$S_t$} \\
		\hline
		Stabilisation & Mesh 1 & Mesh 2 & Mesh 1 & Mesh 2 & Mesh 1 & Mesh 2\\
		\hline
		LF  & 0.142 & 0.202 & 1.365 & 1.373 & 0.148 & 0.154 \\
		Roe & 0.159 & 0.201 & 1.381 & 1.382 & 0.150 & 0.155 \\
		HLL & 0.177 & 0.225 & 1.362 & 1.372 & 0.154 & 0.159 \\
		\hline	
		\multicolumn{7}{|c|}{Literature results} \\  
		\hline
		Reference & \multicolumn{2}{|c|}{$C_L$ amplitude} & \multicolumn{2}{|c|}{Mean $C_D$} & \multicolumn{2}{|c|}{$S_t$} \\
		\hline
		\cite{Tezduyar1991} & \multicolumn{2}{|c|}{[0.188, 0.375]} & \multicolumn{2}{|c|}{[1.35, 1.41]} & \multicolumn{2}{|c|}{[0.156, 0.171]} \\
		\cite{Tezduyar1992} & \multicolumn{2}{|c|}{[0.370, 0.375]} & \multicolumn{2}{|c|}{[1.38, 1.40]} & \multicolumn{2}{|c|}{[0.166, 0.170]} \\
		\cite{Kjellgren1997} & \multicolumn{2}{|c|}{[0.250, 0.330]} & \multicolumn{2}{|c|}{[1.34, 1.37]} & \multicolumn{2}{|c|}{[0.160, 0.170]} \\
		\cite{Rajani2009} & \multicolumn{2}{|c|}{0.253} & \multicolumn{2}{|c|}{1.335} & \multicolumn{2}{|c|}{0.157} \\
		\cite{Kadapa2015} & \multicolumn{2}{|c|}{[0.338, 0.362]} & \multicolumn{2}{|c|}{[1.39, 1.42]} & \multicolumn{2}{|c|}{[0.165, 0.173]} \\
		\cite{Kadapa2020} & \multicolumn{2}{|c|}{[0.293, 0.338]} & \multicolumn{2}{|c|}{--} & \multicolumn{2}{|c|}{[0.159, 0.169]} \\
		\hline		
	\end{tabular}
	\caption{Unsteady laminar flow past a circular cylinder: Amplitude of the lift coefficient, mean value of the drag and Strouhal number using both meshes and the three different stabilisations and reference results.}
	\label{tab:laminarCyl}
\end{table}
Results reported in the literature for the three quantities of interest are also included.

For this example the HLL stabilisation provides the most accurate results. With the first mesh the lift coefficient is almost within the range reported in the literature, whereas the Roe and LF stabilisations require the second grid to provide similar accuracy. Using the second mesh, the results of all convective stabilisations lie within the range reported in the literature. The extra accuracy of the HLL stabilisation is also observed when comparing the Strouhal number to reference values. The results obtained with the HLL stabilisation is almost within the range reported in the references and almost identical to the results obtained with LF and Roe in the second mesh. For the mean value of the drag, the results obtained on both meshes and with the three stabilisations lie within the range reported in the literature.

%==========================================================================
\subsection{Turbulent flow over a flat plate}               \label{sc:plate}
%==========================================================================

\begin{figure}[!tb]
	\centering
	\includegraphics[width=0.70\textwidth]{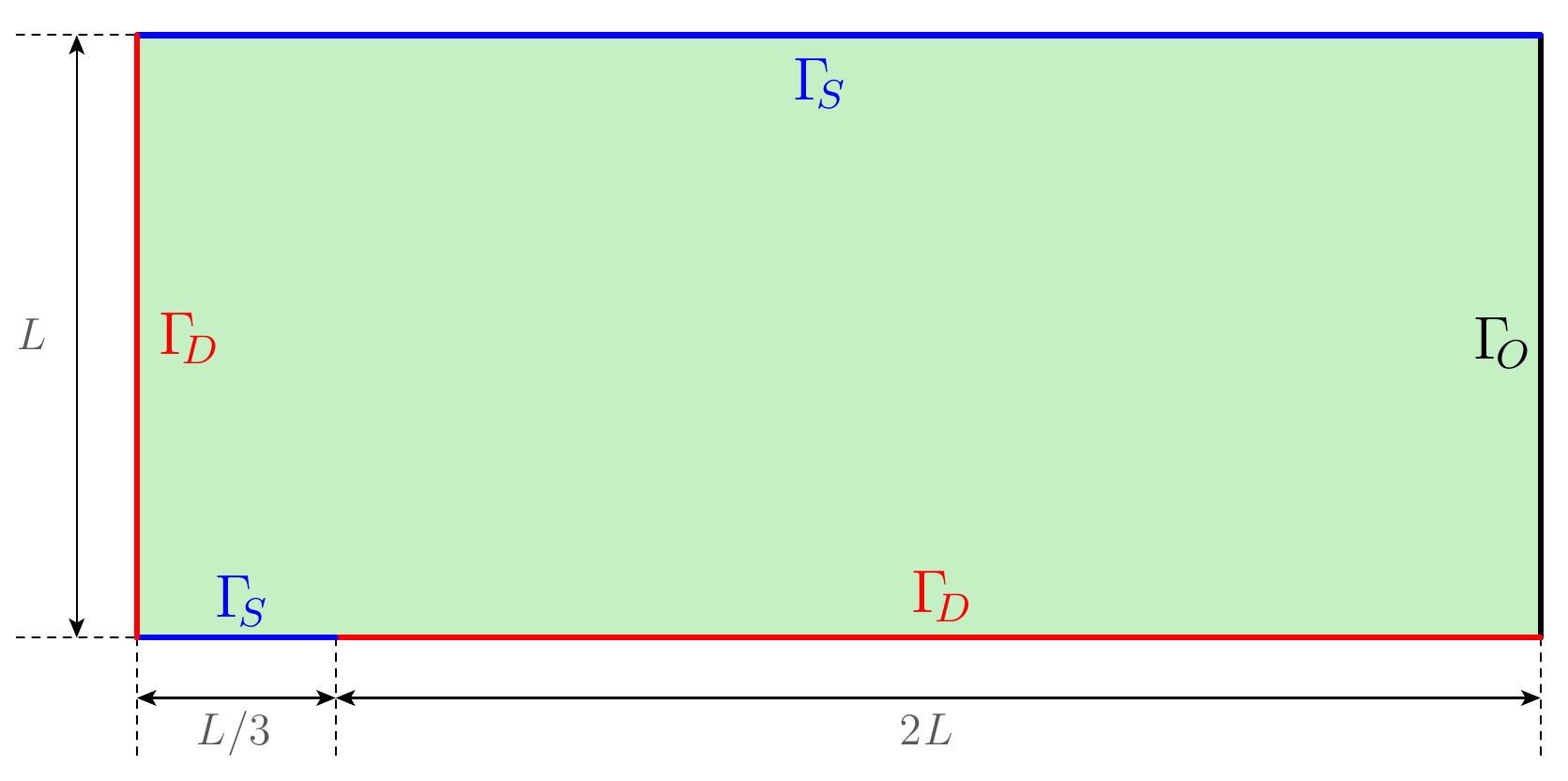}
	\caption{Turbulent flow over a flat plate: Problem setup.}
	\label{fig:plate_Setup}
\end{figure}

The next example considers the steady turbulent flow over a flat plate at $Re=5 \times 10^6$, a classical benchmark to test the accuracy and robustness of steady turbulent flow solvers. The setup of the problem is depicted in figure~\ref{fig:plate_Setup}, including the relevant dimensions and the boundary conditions.
The Dirichlet boundary consists of the inflow boundary, where a constant horizontal velocity of magnitude one is imposed, and the plate, where a no-slip boundary condition is imposed. The inflow SA viscosity is taken as $\nuT = 3$, assuming that the flow is fully developed at the inlet~\cite{SA1992,SA2012modifications}.

\begin{figure}[!tb]
	\centering
	\subfigure[]{\includegraphics[width=0.48\textwidth]{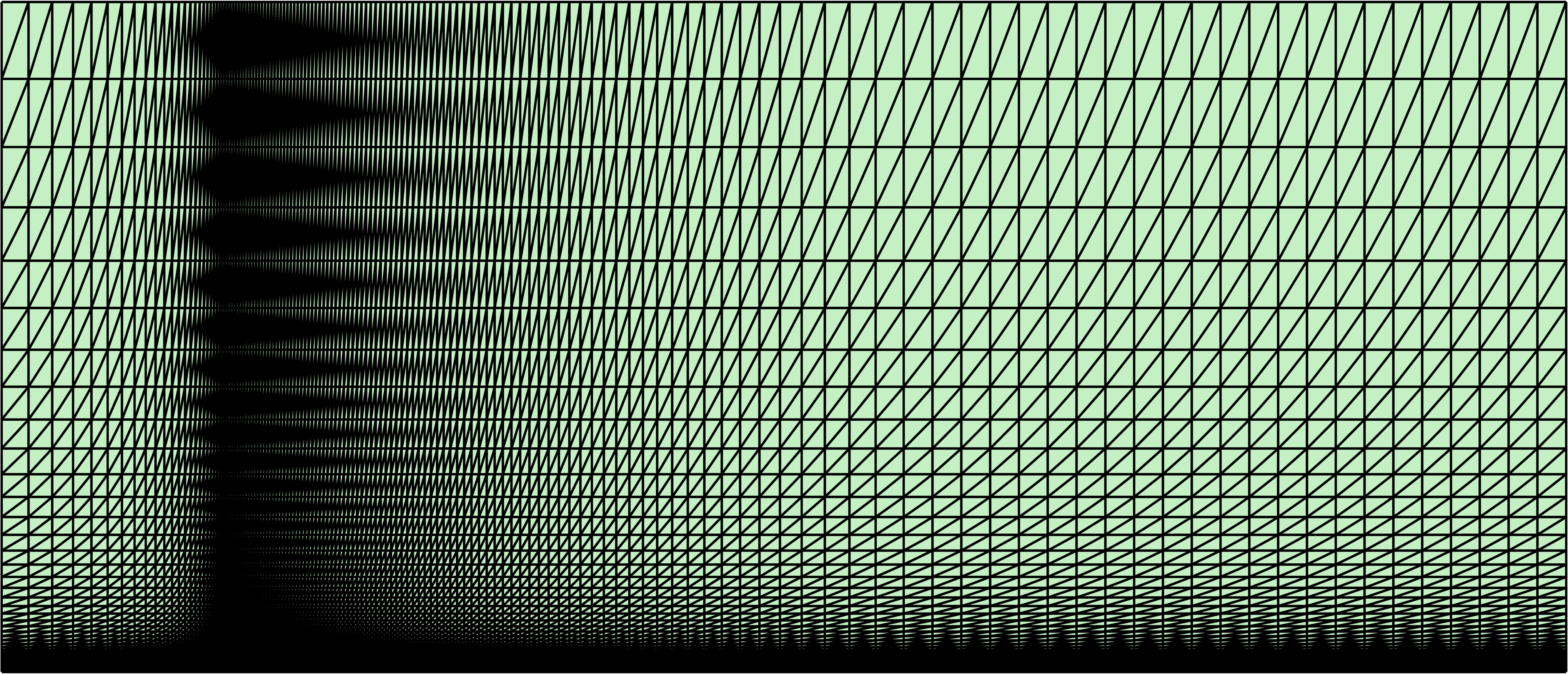}}
	\subfigure[]{\includegraphics[width=0.48\textwidth]{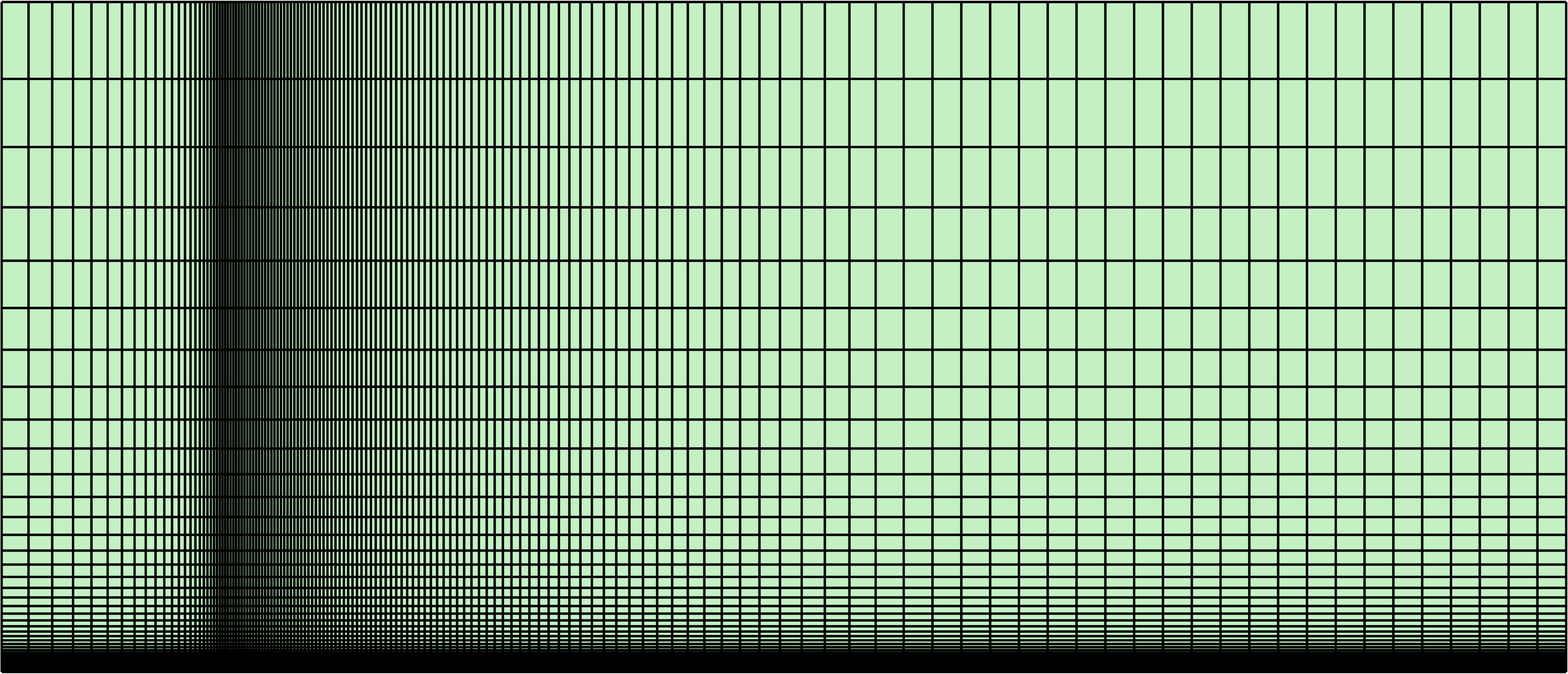}}
	\caption{Turbulent flow over a flat plate: (a) Triangular and (b) quadrilateral coarse meshes.}
	\label{fig:plate_Meshes}
\end{figure}

Three triangular and quadrilateral meshes, typically employed with NASA solvers CLF3D and FUN3D~\cite{NASAflatPlate,NASAOverflowReport}, are used to test the accuracy and robustness of the proposed FCFV. The $i$-th quadrilateral mesh has $(68\times 2^i+1) \times (48\times 2^i+1)$ cells, whereas the corresponding triangular meshes are obtained by splitting each quadrilateral into two triangles. The maximum aspect ratio of the cells near the wall in the first mesh is $2.14 \times 10^5$ and it is halved for each successive mesh refinement. The spacing is selected so that the average $y^+$ at the wall is close to 1 and 0.1 for the coarser and finest meshes, respectively. Figure~\ref{fig:plate_Meshes} displays the coarsest triangular and quadrilateral meshes.

The pseudo-time marching strategy described in~\ref{app:relaxation} is applied, using an initial condition corresponding to the free-stream conditions. The solution is advanced until the steady state is reached with a tolerance of $10^{-12}$. As no time accuracy is required, only one Newton-Raphson iteration is performed per time step.

\begin{figure}[!tb]
	\centering
	\subfigure[]{\includegraphics[width=0.48\textwidth]{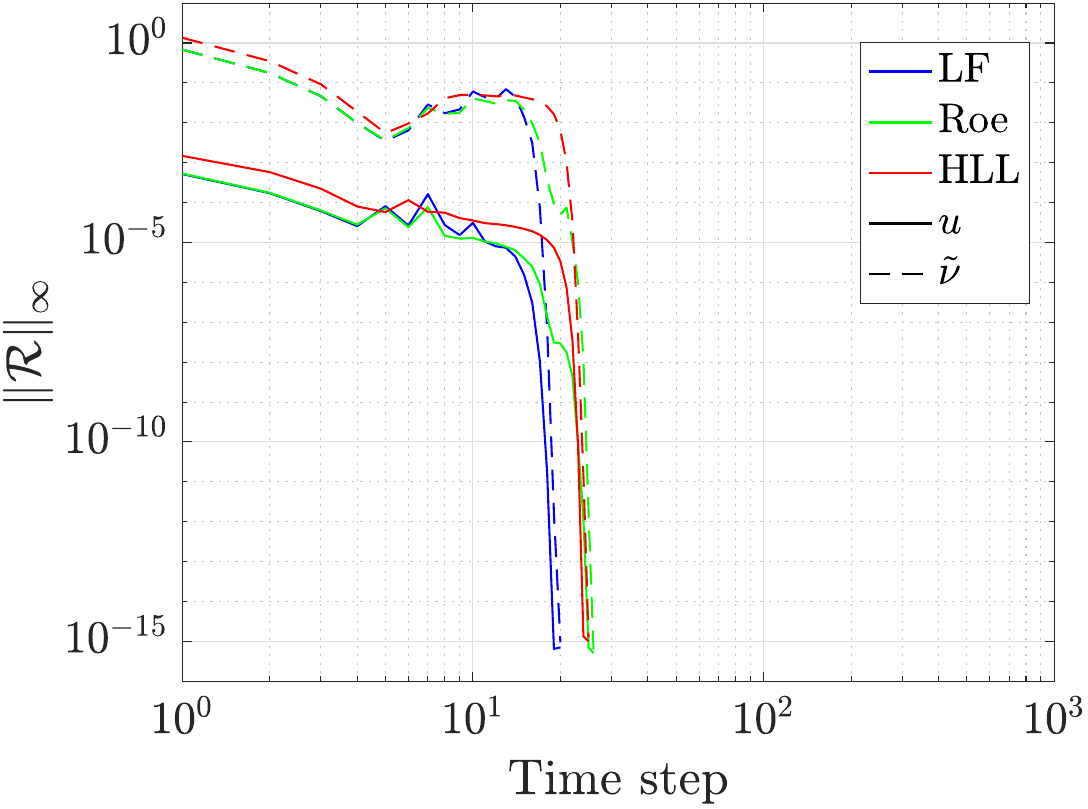}}
	\subfigure[]{\includegraphics[width=0.48\textwidth]{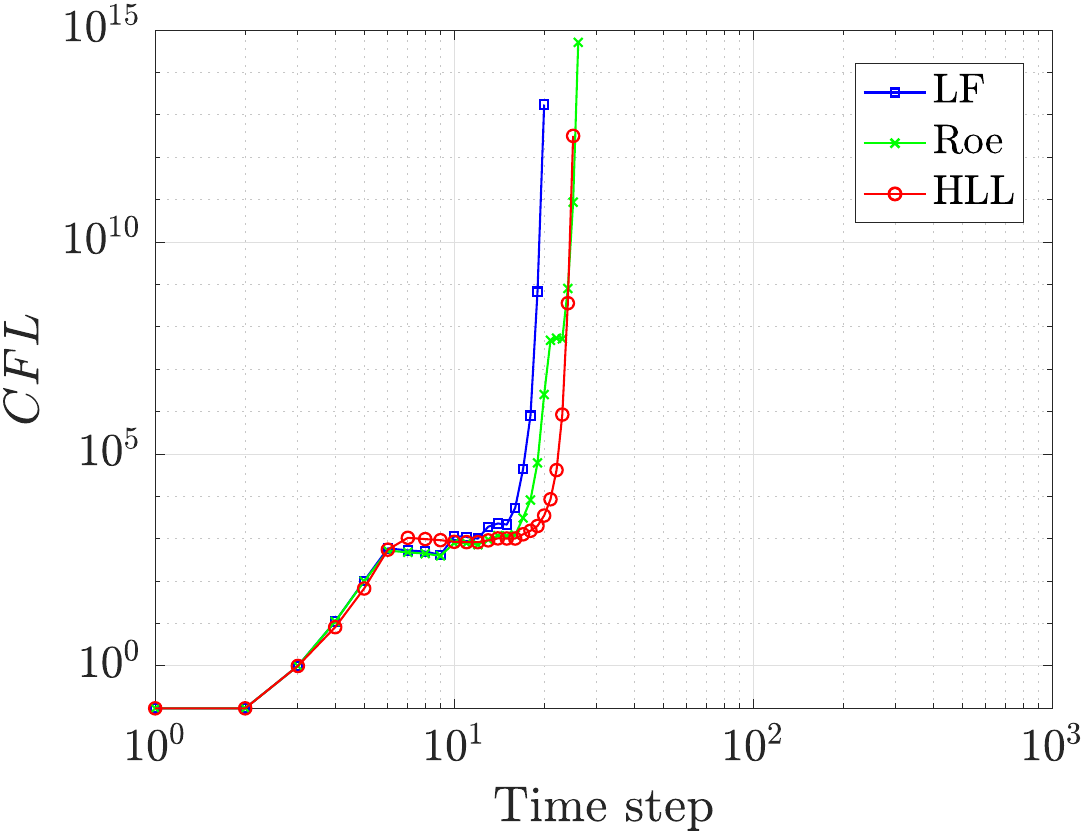}}
	\caption{Turbulent flow over a flat plate: Evolution of the (a) residuals and (b) CFL during the pseudo-time marching for the computation using the second quadrilateral mesh.}
	\label{fig:plate_ResAndCFL}
\end{figure}

Figure~\ref{fig:plate_ResAndCFL} reports the evolution of the residuals of the RANS and SA equations, measured in the maximum norm, and the CFL during the pseudo-time marching for the computation using the coarsest quadrilateral mesh and for the three different convective stabilisations.
The results show a very similar performance for the three stabilisations considered. During the first two time steps a low CFL is required due to the non-physical transient effects caused by the initial condition not satisfying the boundary conditions, but after only six time steps the CFL is already near $10^3$. The CFL stays between $10^3$ and $10^5$ until the solution develops and during the last time steps the CFL exceeds $10^{12}$, showing the robustness and stability of the proposed FCFV method. The behaviour is very similar using finer meshes or triangular cells.

\begin{figure}[!tb]
	\centering
	\subfigure[]{\includegraphics[width=0.48\textwidth]{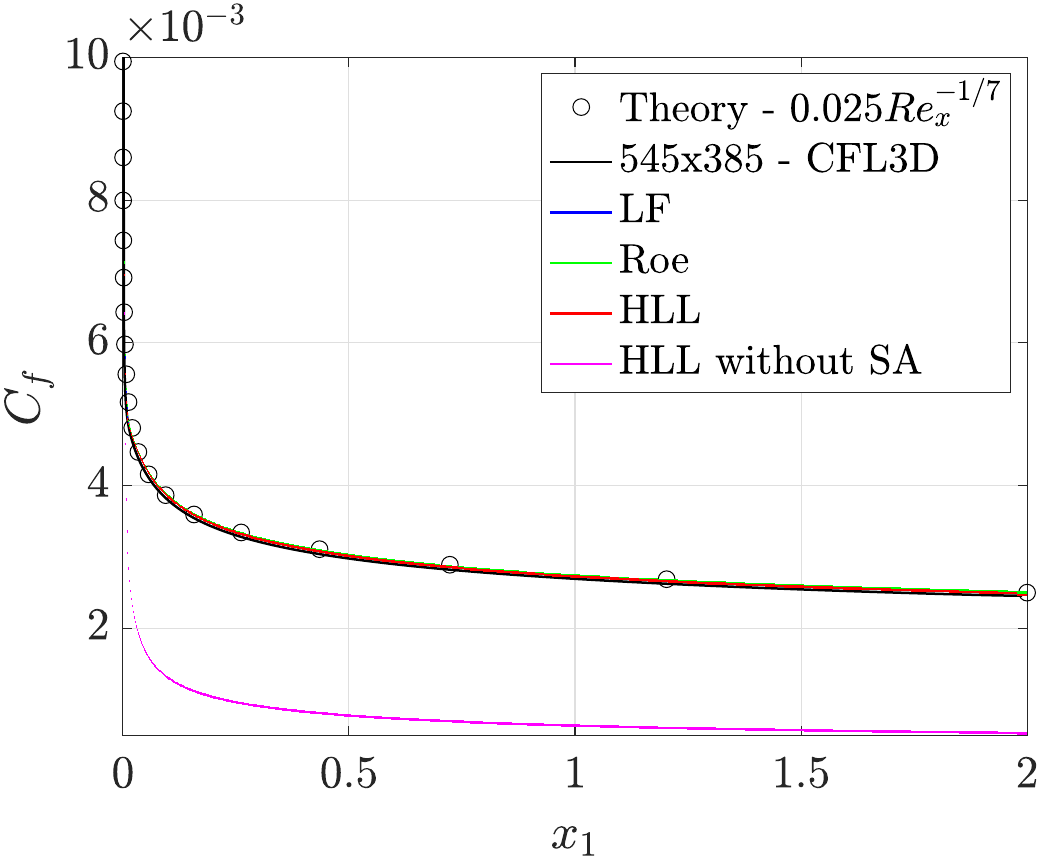}}
	\subfigure[]{\includegraphics[width=0.48\textwidth]{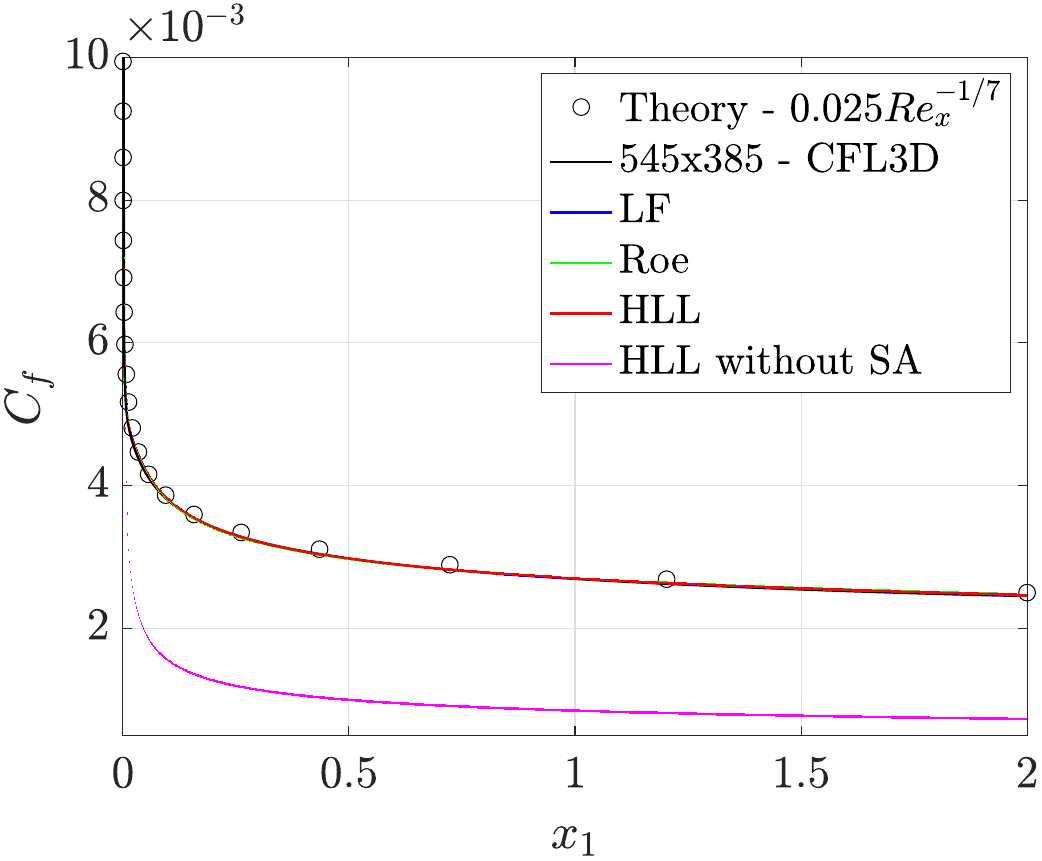}}
	\caption{Turbulent flow over a flat plate: Skin friction along the plate for (a) the second triangular mesh and (b) the second quadrilateral mesh, compared to the Blasius solution and a reference numerical solution. The solution for the HLL convective stabilisation without the SA model is also shown.}
	\label{fig:plate_Cf}
\end{figure}

To assess the accuracy of the FCFV computations, the skin friction along the plate is compared in figure~\ref{fig:plate_Cf} to two references, corresponding to the Blasius solution and the numerical solution obtained with the CFL3D solver~\cite{NASAflatPlate,NASAOverflowReport}.
The FCFV computations are performed on the second triangular and quadrilateral meshes, with $273 \times 193 \times 2$ cells and with $273 \times 193$ cells, respectively, and employing the three different convective stabilisations. The numerical reference solution is computed on the finest quadrilateral mesh, featuring $545 \times 385$ cells, with the CFL3D solver. The FCFV results show an excellent agreement with the reference behaviour. For the quadrilateral mesh, the results using different convective stabilisations are almost indistinguishable, whereas on the triangular mesh a slightly better performance of the HLL stabilisatison is observed.

\begin{figure}[!tb]
	\centering
	\subfigure[]{\includegraphics[width=0.48\textwidth]{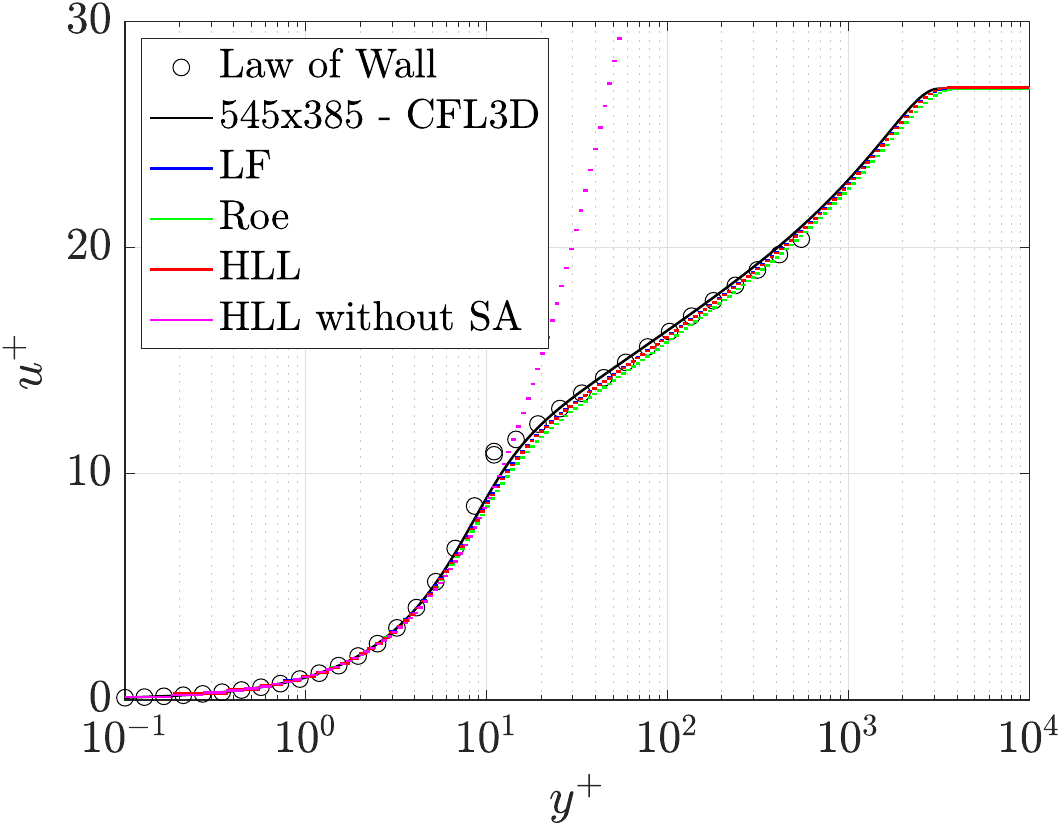}}
	\subfigure[]{\includegraphics[width=0.48\textwidth]{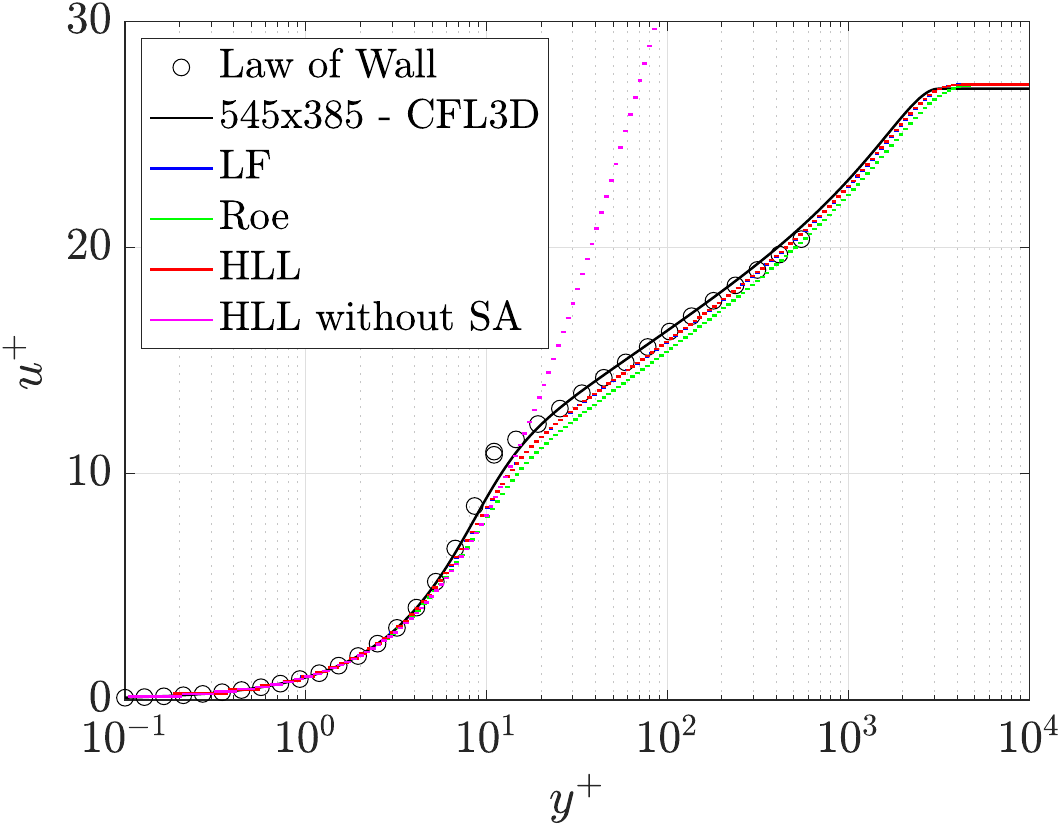}}
	\caption{Turbulent flow over a flat plate: Dimensionless value of $u^+$ as a function of $y^+$ at $x_1=0.97008$ for (a) the second triangular mesh and (b) the second quadrilateral mesh, compared to the law of wall and a reference numerical solution. The solution for the HLL convective stabilisation without the SA model is also shown.}
	\label{fig:plate_UPlus}
\end{figure}

\begin{figure}[!tb]
	\centering
	\subfigure[]{\includegraphics[width=0.48\textwidth]{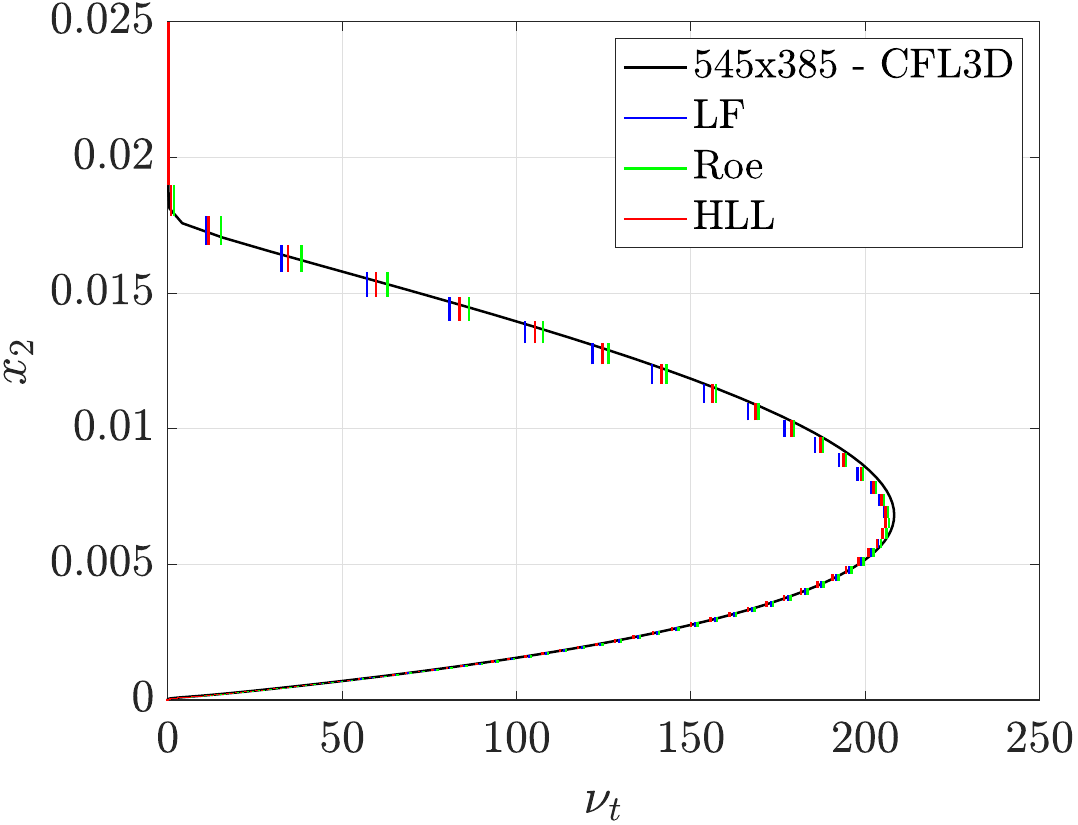}}
	\subfigure[]{\includegraphics[width=0.48\textwidth]{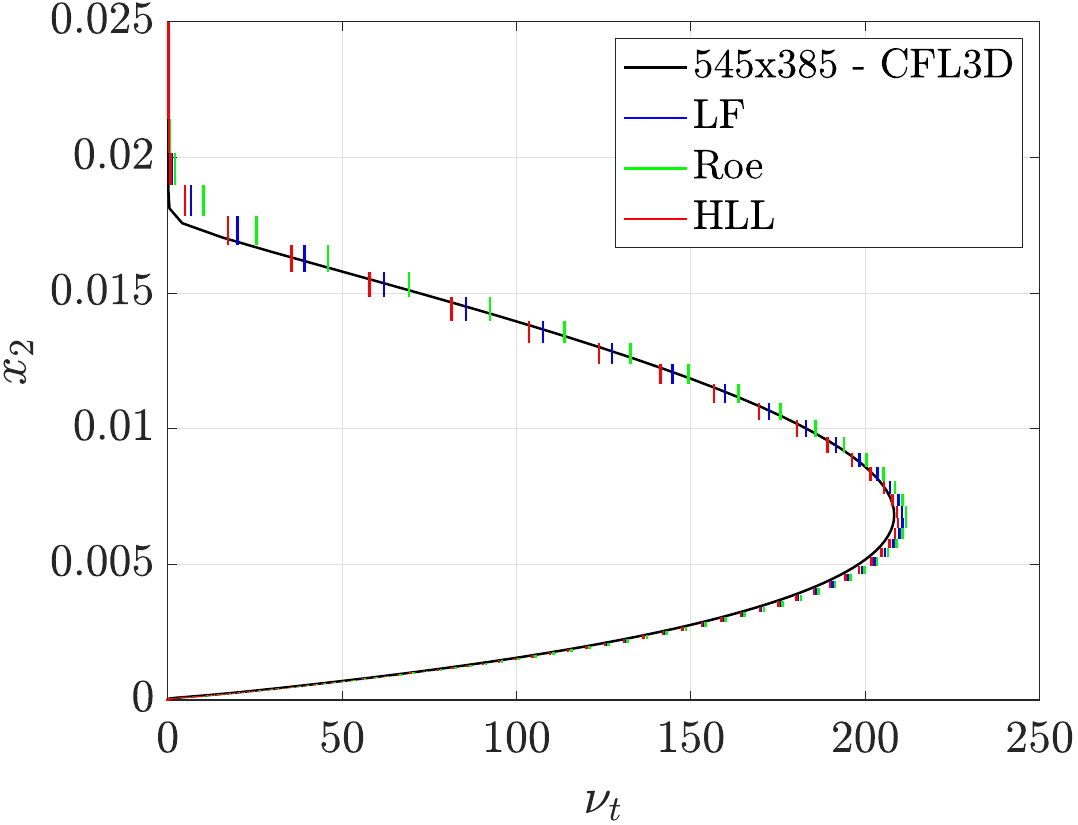}}
	\caption{Turbulent flow over a flat plate: Turbulent viscosity at $x_1=0.97008$ for (a) the second triangular mesh and (b) the second quadrilateral mesh, compared to a reference numerical solution.}
	\label{fig:plate_Nu}
\end{figure}

Similar differences between the stabilisations are obtained when comparing the dimensionless value of $u^+$ as a function of $y^+$ at $x_1=0.97008$, as shown in figure~\ref{fig:plate_UPlus}, and the turbulent viscosity at the same location, as shown in figure~\ref{fig:plate_Nu}.

Figures~\ref{fig:plate_Cf} and ~\ref{fig:plate_UPlus} also show the result obtained with the HLL stabilisation without activating the SA turbulence model. The results clearly show that a turbulent model is required to obtain accurate results, despite the extra dissipation of the first order FCFV method when compared to higher order methods.

To further analyse the accuracy of the results, table~\ref{tab:plate} reports the computed drag coefficient using the three available triangular and quadrilateral meshes and the three different convective stabilisations. 
\begin{table}[!tb]
	\centering
	\begin{tabular}{|c|c|c|c|c|c|c|}
		\hline
		\multicolumn{7}{|c|}{FCFV results} \\  
		\hline
		& \multicolumn{3}{|c|}{Triangles} & \multicolumn{3}{|c|}{Quadrilaterals}  \\
		\hline
		Stabilisation & Mesh 1 & Mesh 2 & Mesh 3 & Mesh 1 & Mesh 2 & Mesh 3\\
		\hline
		LF  & 0.00290 & 0.00290 & 0.00290 & 0.00281 & 0.00287 & 0.00289 \\
		Roe & 0.00290 & 0.00291 & 0.00291 & 0.00280 & 0.00287 & 0.00289 \\
		HLL & 0.00289 & 0.00290 & 0.00290 & 0.00281 & 0.00287 & 0.00289 \\
		HLL without SA & 0.00099 & 0.00077 & 0.00062 & 0.00129 & 0.00098 & 0.00076\\
		\hline	
		\multicolumn{7}{|c|}{Reference results} \\  
		\hline
		Reference & \multicolumn{3}{|c|}{Triangles} & \multicolumn{3}{|c|}{Quadrilaterals}  \\
		\hline
		\text{CFL3D}  & -- & -- & -- &  0.00287 & 0.00286 & 0.00286 \\
		\text{FUN3D} & -- & -- & -- &  0.00284 & 0.00285 & 0.00285 \\
		\hline		
	\end{tabular}
	\caption{Turbulent flow over a flat plate: Drag coefficient computed using triangular and quadrilateral meshes and the three different stabilisations, and reference results.}
	\label{tab:plate}
\end{table}
The results obtained using the CFL3D and FUN3D solvers~\cite{NASAflatPlate,NASAOverflowReport} are also included and show that the FCFV results are less than one drag count away from the reference results. The extra accuracy of the FCFV results using triangular meshes is only due to the extra degrees of freedom introduced when splitting the quadrilaterals into two triangles.

\begin{figure}[!tb]
	\centering
	\subfigure[]{\includegraphics[width=0.48\textwidth]{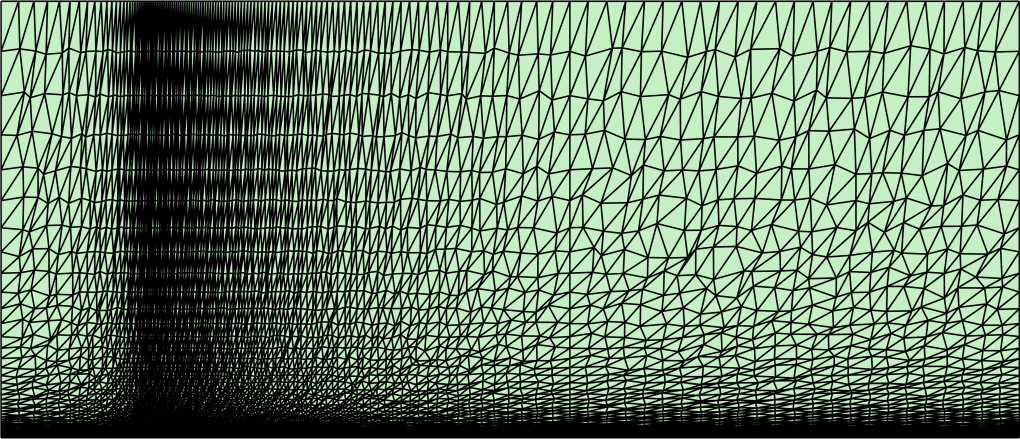}}
	\subfigure[]{\includegraphics[width=0.48\textwidth]{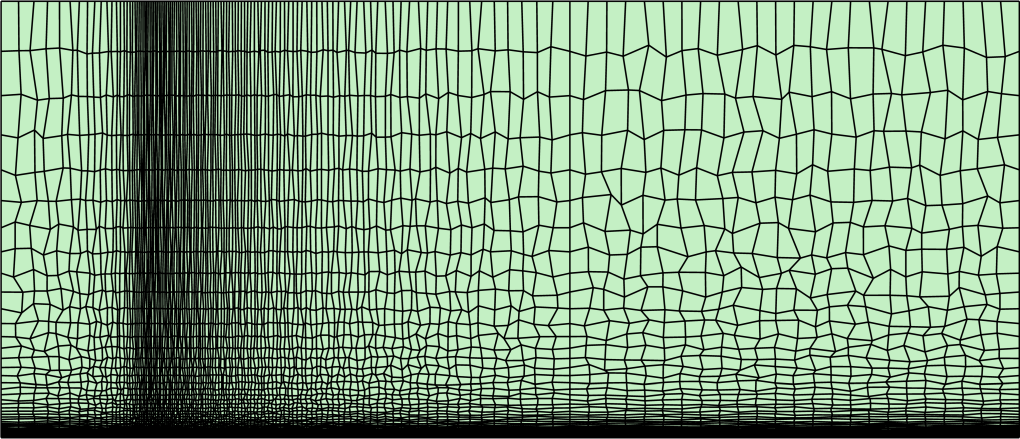}}
	\caption{Turbulent flow over a flat plate: Distorted (a) triangular and (b) quadrilateral coarse meshes.}
	\label{fig:plate_MeshesDistorted}
\end{figure}

\begin{figure}[!tb]
	\centering
	\subfigure[]{\includegraphics[width=0.48\textwidth]{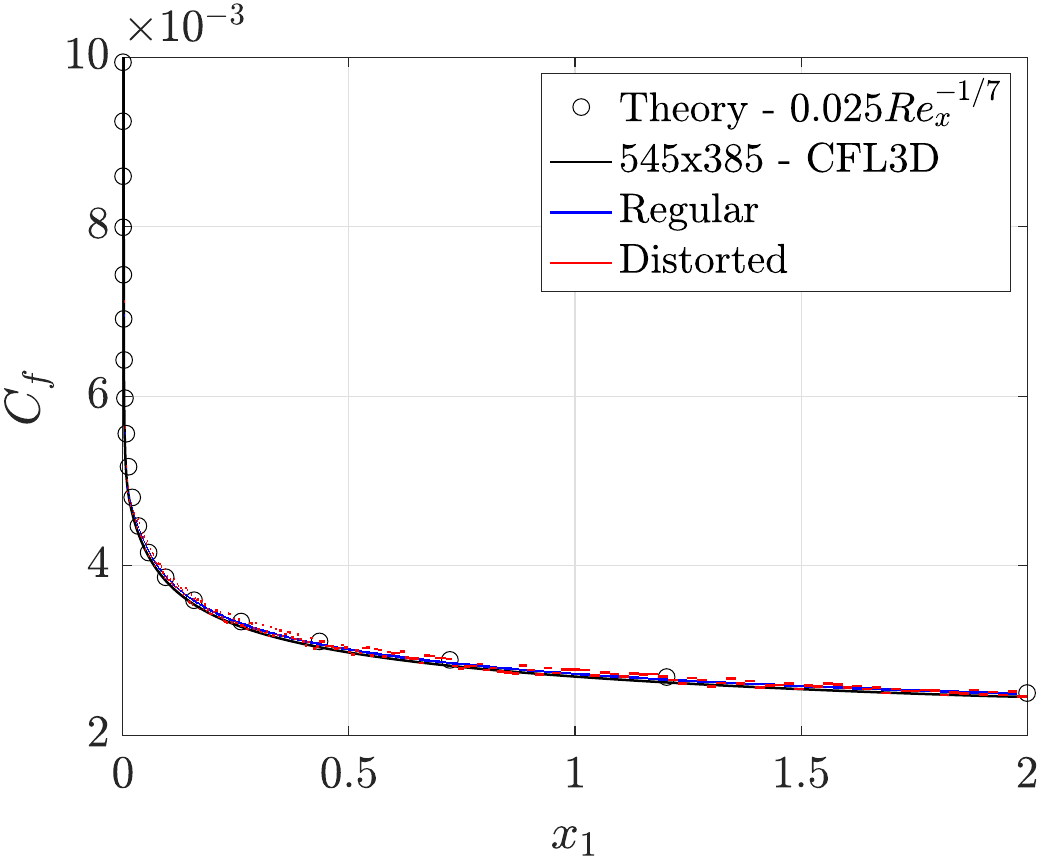}}
	\subfigure[]{\includegraphics[width=0.48\textwidth]{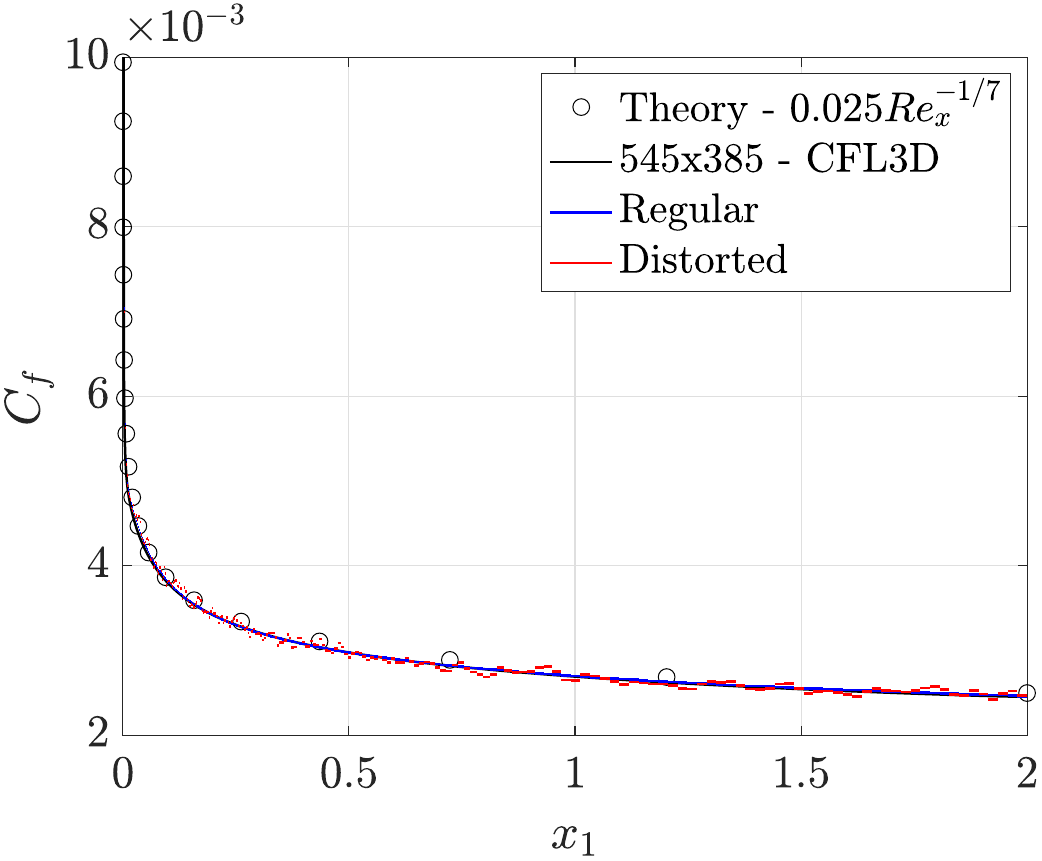}}
	\caption{Turbulent flow over a flat plate: Skin friction on the second (a) triangular and (b) quadrilateral distorted meshes compared to the computation with regular meshes, the Blasius solution and a reference numerical solution.}
	\label{fig:plate_CfDistorted}
\end{figure}

To assess the robustness of the proposed FCFV method with respect to mesh quality, a random perturbation to the interior nodes is introduced. The resulting, coarse, distorted triangular and quadrilateral meshes are shown in figure~\ref{fig:plate_MeshesDistorted}.
The skin friction using the HLL convective stabilisation on the second regular and distorted triangular and quadrilateral meshes are compared in figure~\ref{fig:plate_CfDistorted}.
The results show very little influence of the substantial mesh distortion introduced, demonstrating the insensitivity of the FCFV method to mesh quality and the overall robustness of the proposed solver, even for high Reynolds number turbulent flows.

\begin{figure}[!tb]
	\centering
	\subfigure[]{\includegraphics[width=0.48\textwidth]{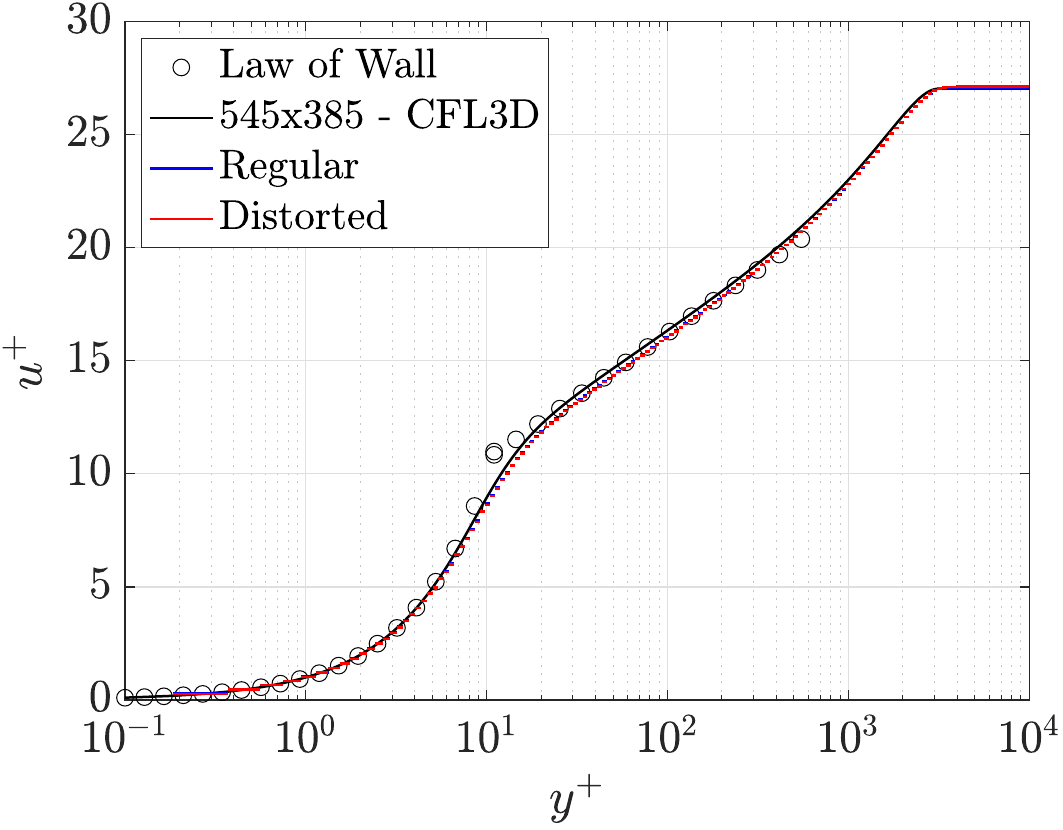}}
	\subfigure[]{\includegraphics[width=0.48\textwidth]{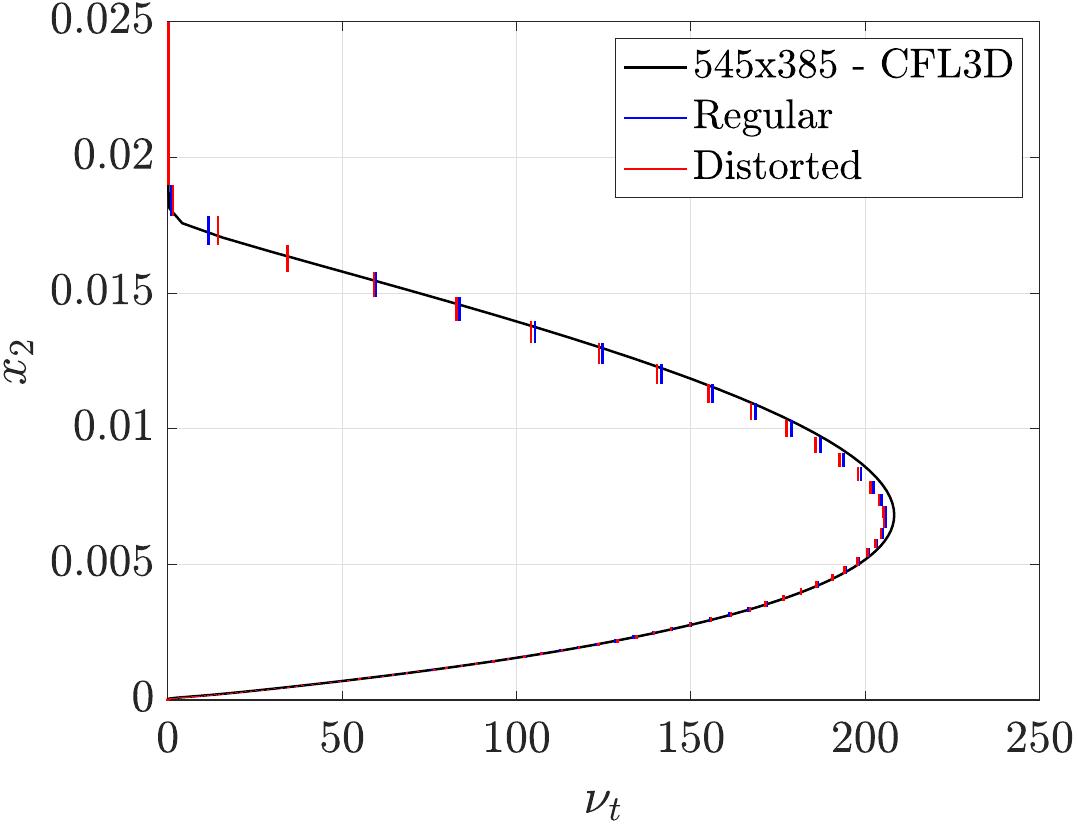}}
	\caption{Turbulent flow over a flat plate: (a) Dimensionless value of $u^+$ as a function of $y^+$ and (b) turbulent viscosity, at $x_1=0.97008$, on the second triangular distorted mesh compared to the computation with regular meshes and a reference numerical solution.}
	\label{fig:plate_UplusNuDistorted}
\end{figure}

To further illustrate the robustness of the method with respect to cell distortion, figure~\ref{fig:plate_UplusNuDistorted} compares the dimensionless value of $u^+$ as a function of $y^+$ at $x_1=0.97008$ and the turbulent viscosity at the same location, when using regular and distorted triangular meshes. 
In both cases, the HLL convective stabilisation is used and the results on distorted meshes are almost identical to the ones obtained with regular meshes. The unnoticeable difference induced by the mesh distortion is corroborated by comparing the drag coefficient computed with distorted meshes. For instance, the drag coefficient obtained using the second distorted triangular mesh is 0.00291, which is less than one drag count away from the result with the second triangular mesh, 0.00290. The same conclusions are obtained for quadrilateral meshes and other levels of mesh refinement.

All above results have been computed using a monolithic strategy to solve the RANS-SA system. A comparison of the performance and robustness of this approach with a staggered algorithm, described in~\ref{app:FCFV_computationalMonoStaggered}, can be found in~\ref{app:relaxation}.

%==========================================================================
\subsection{Unsteady turbulent flow past a circular cylinder}               \label{sc:turbulentCyl}
%==========================================================================

\begin{figure}[!tb]
	\centering
	\includegraphics[width=0.70\textwidth]{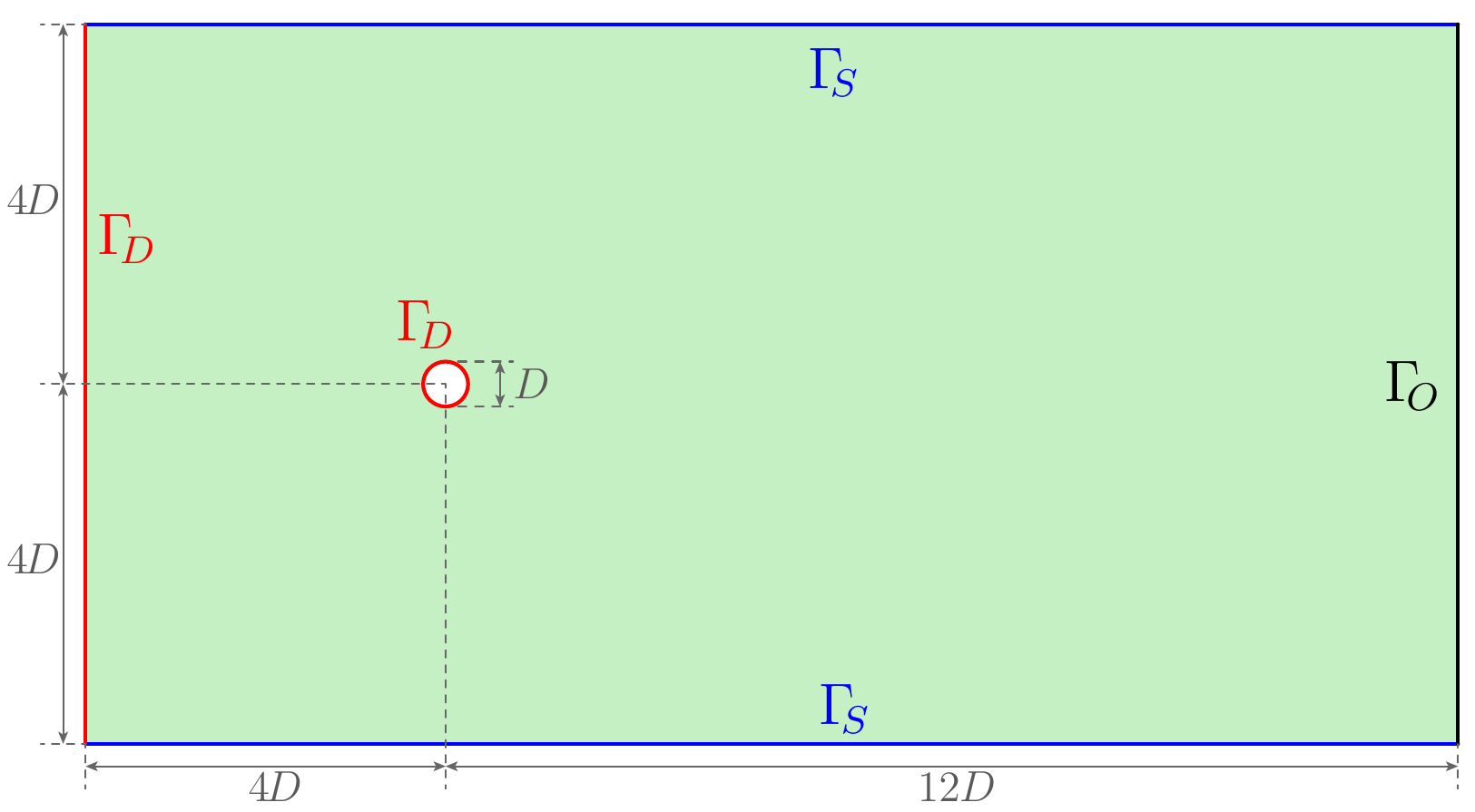}
	\caption{Unsteady turbulent flow past a circular cylinder: Problem setup.}
	\label{fig:turbulentCyl_Setup}
\end{figure}

The last example considers the unsteady turbulent flow past a circular cylinder of diameter $D$ at $Re=10^4$. This example is used to assess the applicability of the proposed method to turbulent transient cases. The setup of the problem is depicted in figure~\ref{fig:turbulentCyl_Setup}, including the relevant dimensions and the boundary conditions.

\begin{figure}[!tb]
	\centering
	\subfigure[]{\includegraphics[height=0.2\textheight]{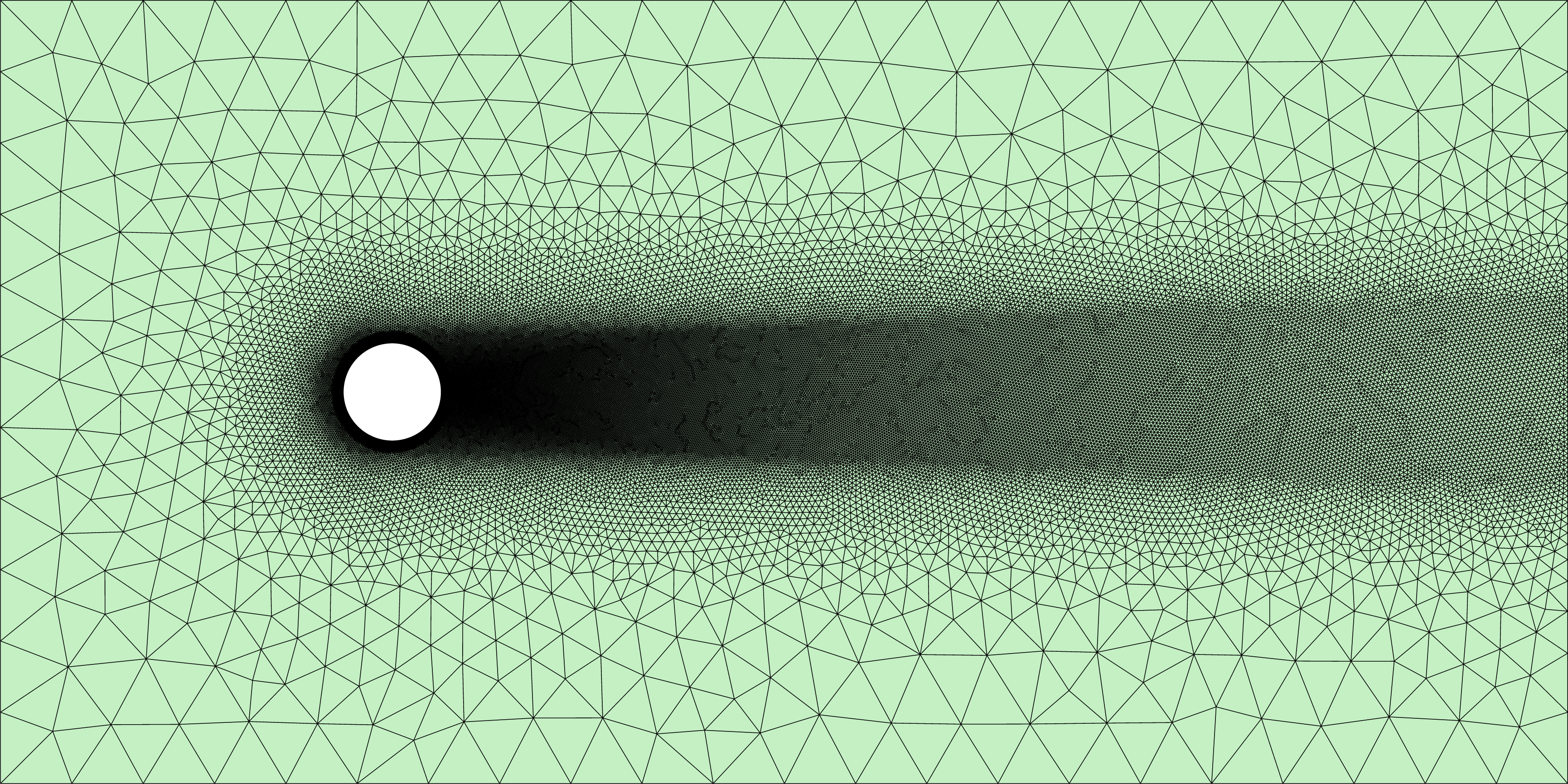}}
	\subfigure[]{\includegraphics[height=0.2\textheight]{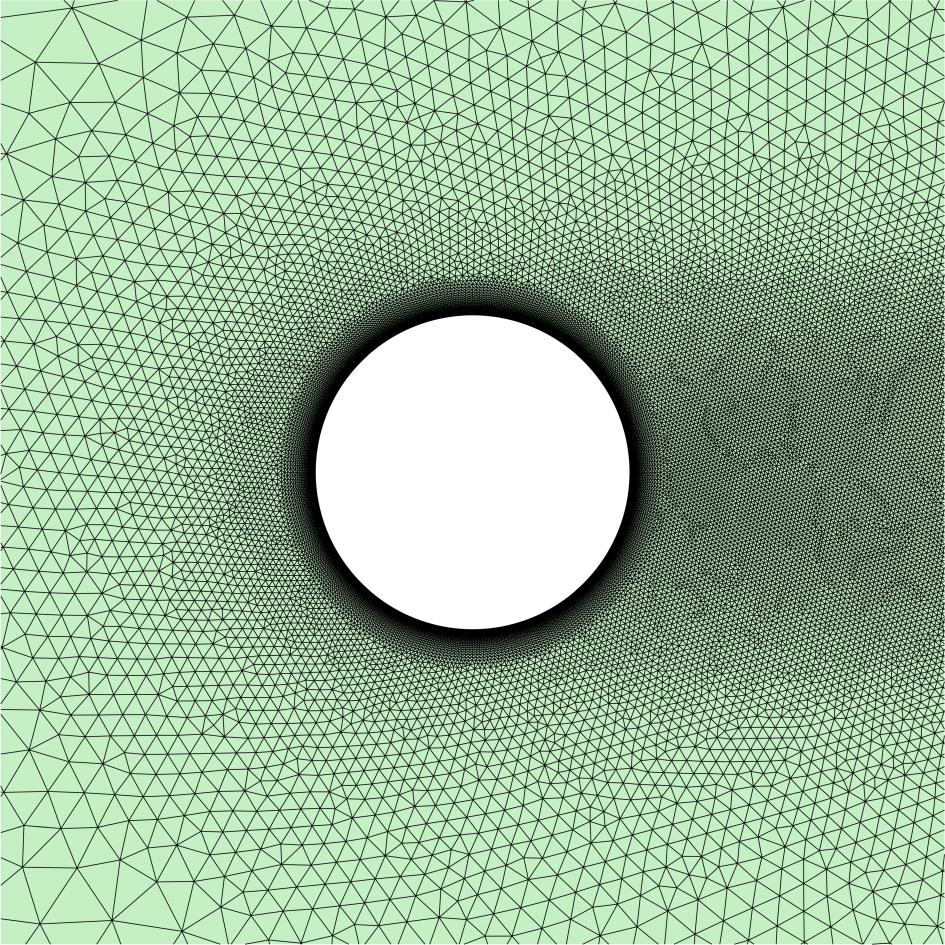}}
	\caption{Unsteady turbulent flow past a circular cylinder: (a) Second mesh with 98,193 cells and (b) detail of the inflation layer.}
	\label{fig:turbulentCylMesh2}
\end{figure}

Two unstructured triangular meshes are considered. For the first mesh, the inflation layer is defined using a height for the first cell on the cylinder equal to $0.002D$ and a growth ratio of 1.15. A refinement behind the cylinder is introduced to capture the wake by imposing a cell size of $0.015D$ near the cylinder and $0.075D$ near the outflow boundary. For the second mesh, the height of the first cell in the inflation layer is $0.001D$ and the growth ratio is set to 1.1. The sizes to locally refine the region of the wake are $0.01D$ near the cylinder and $0.05D$ near the outflow. In both cases, the maximum cell stretching near the cylinder is 7.5. The first mesh has 52,094 cells, with a maximum $y^+$ of 1.6, whereas the second mesh, displayed in figure~\ref{fig:turbulentCylMesh2}, has 98,319 cells, with a maximum $y^+$ of 0.8.

For the time-integration, the BDF2 scheme is employed with a time-step $\Delta t = 0.01$. The initial condition for the RANS equations is taken as the steady-state solution with $Re = 10$, and following~\cite{Nithiarasu2006}, the turbulent variable $\nuT$ is taken as 0.05.

\begin{figure}[!tb]
	\centering
	\subfigure[]{\includegraphics[width=0.32\textwidth]{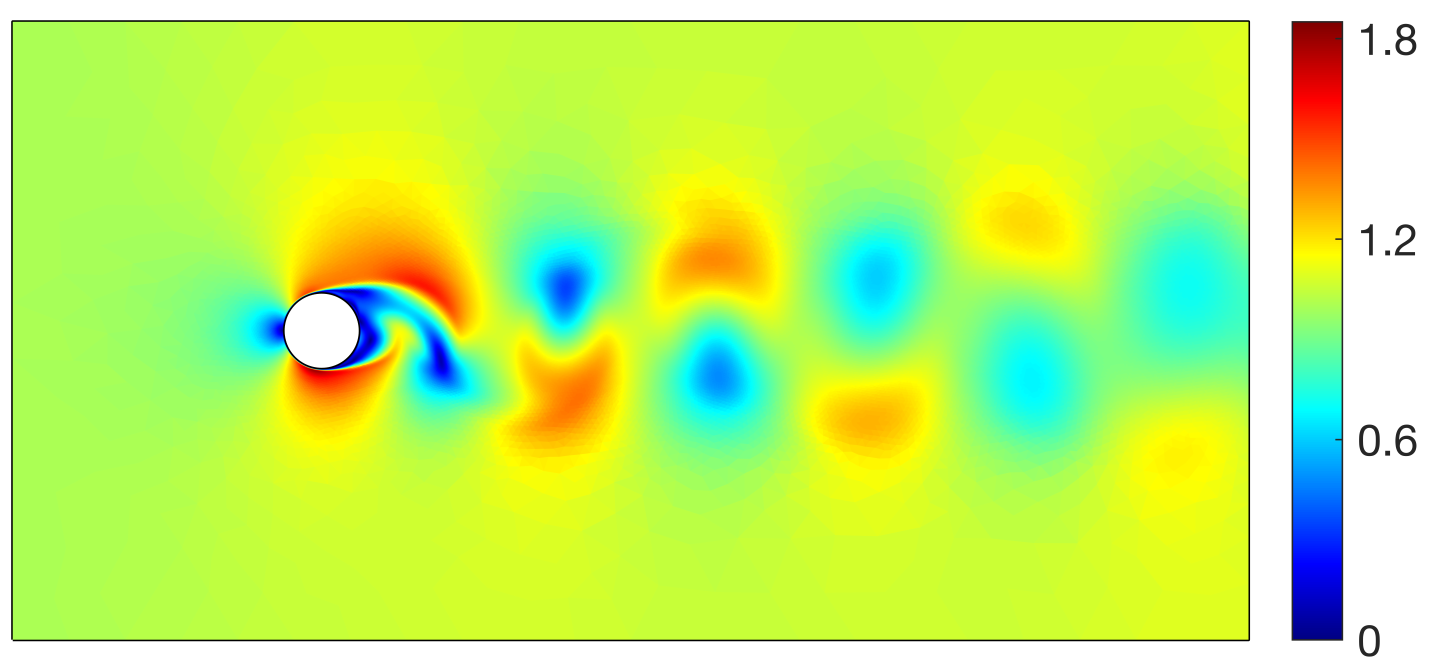}}
	\subfigure[]{\includegraphics[width=0.32\textwidth]{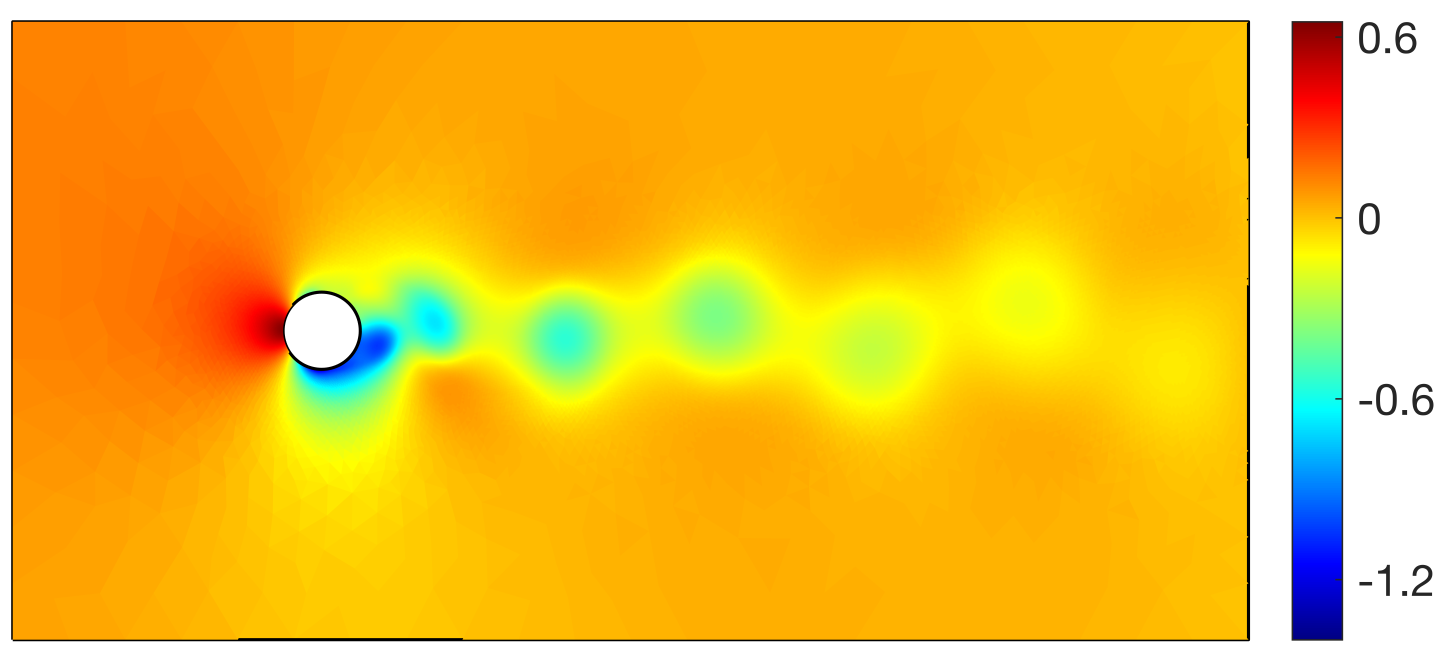}}
	\subfigure[]{\includegraphics[width=0.32\textwidth]{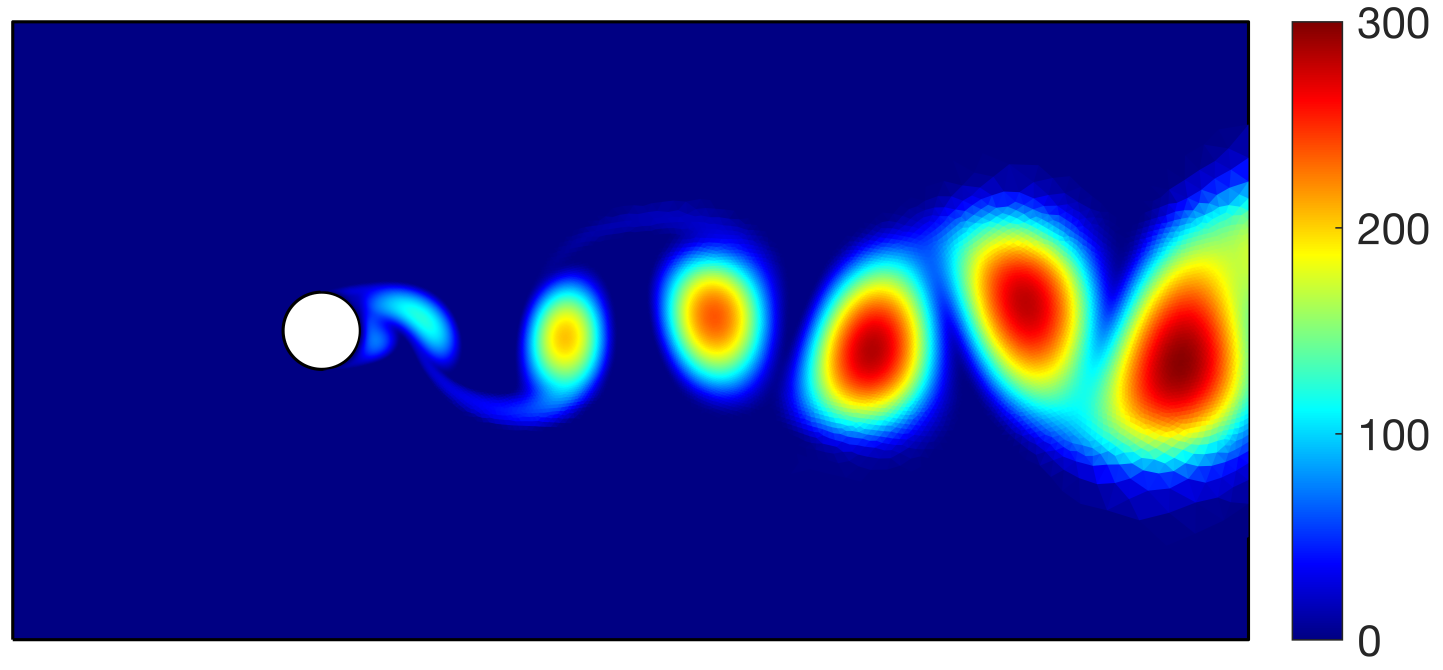}}
	\caption{Unsteady turbulent flow past a circular cylinder: Snapshots of the (a) magnitude of the velocity, (b) pressure and (c) turbulent viscosity at $t=61$.}
	\label{fig:turbulentCyl_Snapshots}
\end{figure}

A snapshot of the magnitude of the velocity, the pressure and the turbulent viscosity at time $t=61$ is presented in figure~\ref{fig:turbulentCyl_Snapshots}. 
The simulation corresponds to the FCFV solution in the second mesh using the HLL stabilisation. The results show a more complicated pattern of the vortices, compared to the laminar flow of section~\ref{sc:laminarCyl}, and, given the moderate Reynolds number employed, it can be observed that the turbulent effects are confined to the wake. 

As in the previous transient case, a tolerance of $10^{-6}$ is used for the nonlinear problems and the average number of Newton-Raphson iterations across all time steps is three in all cases. This is slightly higher than the number of iterations observed for the laminar flow, illustrating the extra difficulty in solving the coupled RANS-SA system. 

\begin{figure}[!tb]
	\centering
	\subfigure[]{\includegraphics[width=0.49\textwidth]{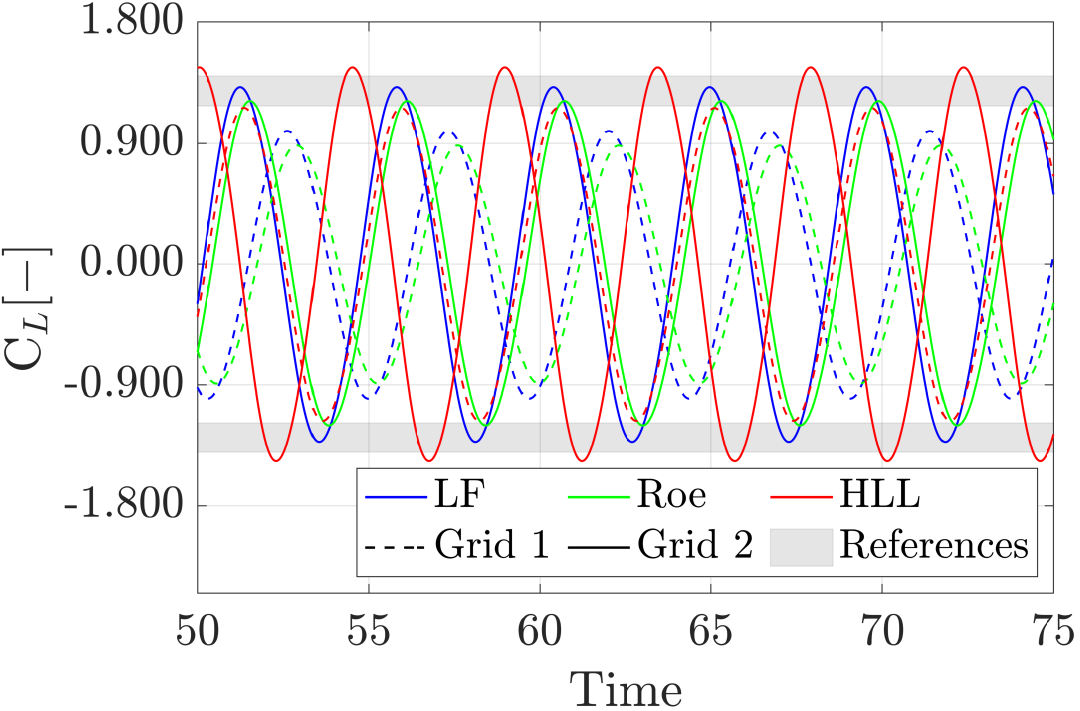}}
	\subfigure[]{\includegraphics[width=0.49\textwidth]{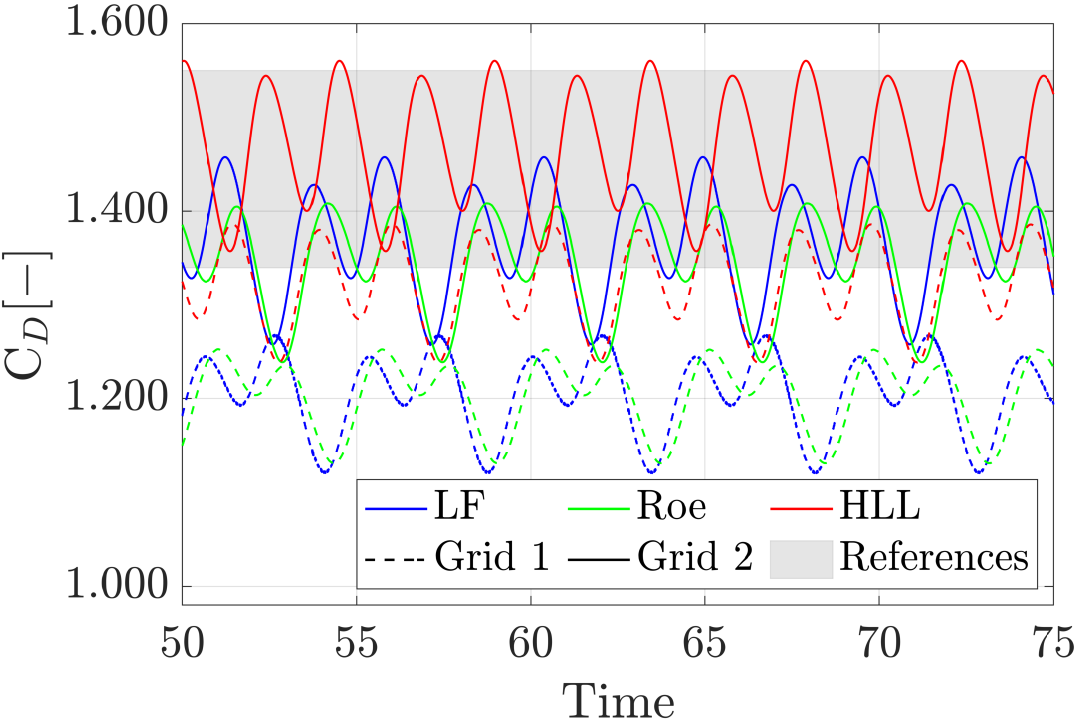}}
	\caption{Unsteady turbulent flow past a circular cylinder: (a) Lift and (b) drag coefficients as a function of time for the two meshes and the three convective stabilisations.}
	\label{fig:tubulentCyl_ClCd}
\end{figure}

The accuracy of the FCFV simulations is assessed by comparing the computed lift ($C_L$) and drag ($C_D$) coefficients and the Strouhal number ($S_t$) against values found in the literature. Figure~\ref{fig:tubulentCyl_ClCd} shows the lift and drag coefficients as a function of time for the two meshes and the three convective stabilisations, where the shadowed area represents the range of values found in the literature~\cite{Nithiarasu2006,Selvam1997,Saghafian2003}.

For the references considered, the number of elements varies between 2,000 and 47,000, but it is important to note that in these references higher order approximations are considered for the simulations. The results reported in the literature are obtained for a variety of numerical schemes including finite volumes and stabilised finite elements.

A more detailed comparison is provided in table~\ref{tab:turbulentCyl}, reporting the amplitude of the lift coefficient, the mean value of the drag and the Strouhal number using both meshes and the three different stabilisations. 
\begin{table}[!tb]
	\centering
	\begin{tabular}{|c|c|c|c|c|c|c|}
		\hline
		\multicolumn{7}{|c|}{FCFV results} \\  
		\hline
		& \multicolumn{2}{|c|}{$C_L$ amplitude} & \multicolumn{2}{|c|}{Mean $C_D$} & \multicolumn{2}{|c|}{$S_t$} \\
		\hline
		Stabilisation & Mesh 1 & Mesh 2 & Mesh 1 & Mesh 2 & Mesh 1 & Mesh 2\\
		\hline
		LF  & 1.00 & 1.32 & 1.21 & 1.37 & 0.213 & 0.219 \\
		Roe & 0.90 & 1.20 & 1.20 & 1.34 & 0.213 & 0.218 \\
		HLL & 1.18 & 1.46 & 1.32 & 1.46 & 0.218 & 0.224 \\
		\hline	
		\multicolumn{7}{|c|}{Literature results} \\  
		\hline
		Reference & \multicolumn{2}{|c|}{$C_L$ amplitude} & \multicolumn{2}{|c|}{Mean $C_D$} & \multicolumn{2}{|c|}{$S_t$} \\
		\hline
		\cite{Nithiarasu2006} & \multicolumn{2}{|c|}{1.40} & \multicolumn{2}{|c|}{1.35} & \multicolumn{2}{|c|}{0.167} \\
		\cite{Selvam1997}     & \multicolumn{2}{|c|}{1.18} & \multicolumn{2}{|c|}{1.34} & \multicolumn{2}{|c|}{[0.16, 0.19]} \\
		\cite{Saghafian2003}  & \multicolumn{2}{|c|}{--}   & \multicolumn{2}{|c|}{1.55} & \multicolumn{2}{|c|}{[0.217,0.244]} \\
		\hline		
	\end{tabular}
	\caption{Unsteady turbulent flow past a circular cylinder: Amplitude of the lift coefficient, mean value of the drag and Strouhal number using both meshes and the three different stabilisations and reference results.}
	\label{tab:turbulentCyl}
\end{table}
Results reported in the literature for the three quantities of interest are also included.

To conclude, the simulation using the first mesh and the HLL flux is repeated using the laminar FCFV solver, without the SA turbulence model. For this case, the $C_L$ amplitude obtained is 1.74, the mean $C_D$ is 1.71 and the Strouhal number is 0.230. These results substantially differ from the results obtained with the SA model, shown in Table~\ref{tab:turbulentCyl}, and are outside of the range of results reported in the references~\cite{Nithiarasu2006,Selvam1997,Saghafian2003}. This shows, again, the need to incorporate the SA turbulence model to accurately simulate the problems considered here, even if the FCFV method proposed is first-order and therefore introduces higher dissipation when compared to other, second-order, FV schemes.

%==========================================================================
\section{Concluding remarks}               \label{sc:conclusions}
%==========================================================================

The development and application of the FCFV method for laminar and turbulent viscous incompressible flows has been presented, including the introduction of three new stabilisations, inspired by classical Riemann solvers, for the convective part of the RANS and SA equations. 

The Couette and the lid-driven cavity flow problems were used to test the convergence of the error under mesh refinement. In all cases, the results confirmed the first-order convergence of the error for all the variables, namely the cell velocity, the face velocity, the velocity-gradient tensor and the pressure.  The performance of three convective stabilisations proposed in this work was also tested using the lid-driven cavity flow, showing a slightly superior performance of the HLL stabilisation, when compared to the LF and Roe stabilisation. 

For a set of laminar and turbulent benchmarks, severe mesh distortion and stretching was introduced to test the robustness of the proposed FCFV method. In all the examples, it was observed that mesh distortion induces very little difference in the results, even when cell distortion is introduced in the boundary layer region, for high Reynolds number flows.

Monolithic and staggered approaches to handle the RANS-SA system are also discussed and numerically compared.
Future work will focus on the parallel implementation of the FCFV method in three dimensions and the application to large scale problems. To this end, particular attention will be paid to the solution of the global system of equations and the potential of the staggered approach to reduce the memory footprint.

%==========================================================================
\section*{Acknowledgements}
%==========================================================================
The authors acknowledge the support of: H2020 MSCA ITN-EJD ProTechTion (Grant No. 764636) that partially funded the PhD scholarship of LMV; Spanish Ministry of Science, Innovation and Universities and Spanish State Research Agency MICIU/AEI/10.13039/501100011033 (Grants No. PID2020-113463RB-C33 to MG, PID2020-113463RB-C32 to AH and CEX2018-000797-S to MG and AH); Generalitat de Catalunya (Grant No.  2021-SGR-01049 to MG and AH); MG is Fellow of the Serra H\'unter Programme of the Generalitat de Catalunya.

%==========================================================================
%\bibliographystyle{plain}
\bibliographystyle{elsarticle-num}
\bibliography{FCFV} 
%==========================================================================

%==========================================================================
\appendix

%==========================================================================
\section{Convective stabilisation}\label{app:convectiveTau}
%==========================================================================

This appendix presents three new convective stabilisation tensors for the incompressible RANS equations and the SA turbulence model. The convective stabilisations $\btau^a$ and $\tauT^a$ are used to define the full stabilisation tensors $\btau$ and $\tauT$ in the local and global problems given by equations \eqref{eq:localDiscrete} and \eqref{eq:globalDiscrete}, respectively. They are used and compared in several examples presented in Section~\ref{sc:examples}.

The rationale is inspired by classical Lax-Friedrichs, Roe and Harten-Lax-van Leer Riemann solvers for compressible flows~\cite{Toro2009} and it is derived following the unified approach presented in~\cite{JVP_HDG-VGSH:20,vila2022non}. 

The convective part of the RANS momentum equation can be written as 
\begin{equation} \label{eq:NS-Conv}
\dt{\bu} + \frac{\partial \bF_k(\bu)}{\partial x_k}  = \bm{0}, 
\end{equation}
where $\bF_k(\bu) = u_k \bu$, or in quasi-linear form as
\begin{equation} \label{eq:NS-ConvQuasiLin}
\dt{\bu} + \bA_k \frac{\partial \bu}{\partial x_k}  = \bm{0},
\end{equation}
where $\bA_k = \partial \bF_k/ \partial \bu$.

To derive the stabilisation tensors $\btau^a$, the system of equations is restricted to an arbitrary direction $\bn$, which in the context of the FCFV will correspond to the normal direction to an edge/face of a cell. The resulting system is
\begin{equation} \label{eq:NS-ConvDiag}
	\dt{\bu} + \bAn \frac{\partial \bu}{\partial \bn}  = \bm{0},
\end{equation}
with $\bAn = (\bu\cdot\bn)\Insd +\bu\otimes\bn$. 

The spectral decomposition of $\bAn$ leads to $\bAn = \mat{R}\, \mat{\Lambda}\, \mat{R}^{-1}$, where $\mat{R}$ is the matrix whose columns are the right eigenvectors of $\bAn$ and   $\mat{\Lambda}=\diag\left(\lambda_1,\dotsc ,  \lambda_{\nsd}\right)$ with $\lambda_1 =\dotsb=  \lambda_{\nsd{-}1} =  \vn$, $\lambda_{\nsd} = 2\vn$, and $\vn= \bu\cdot\bn$. To ensure a minimum value of the stabilisation, the regularised eigenvalues are considered, namely $\hat{\lambda}_1  =\dotsb=  \hat{\lambda}_{\nsd{-}1} =  \max\left(\abs{\bu\cdot\bn},\varepsilon\right)$, 
$\hat{\lambda}_{\nsd} = \max\left(2\abs{\bu\cdot\bn},\varepsilon\right)$, where $\varepsilon>0$ is a real, user-defined numerical parameter.

Using the maximum eigenvalue of $\mat{\hat{\Lambda}}=\diag\left(\hat{\lambda}_1,\dotsc ,  \hat{\lambda}_{\nsd}\right)$, the stabilisation inspired by the Lax-Friedrichs (LF) Riemman solver is defined as
\begin{equation}\label{eq:LFConv}
\btauLF = \max\left\{2\abs{\bhu\cdot\bn},\varepsilon \right\}\Insd.
\end{equation}
The stabilisation inspired by the Roe Riemman solver is defined as
\begin{equation}\label{eq:RoeConv}
\btauRoe = \mat{R}\, \mat{\hat{\Lambda}}\, \mat{R}^{-1}.
\end{equation}
From an implementation point of view, it is convenient to rewrite the Roe stabilisation as
\begin{equation}\label{eq:RoeConv2}
	\btauRoe =
	\begin{cases}
		\varepsilon\,\Insd   &\text{for $\abs{\bhu\cdot\bn} \leq \varepsilon/2$,}\\
		\varepsilon\,\Insd + \sign(\bhu{\cdot}\bn)\left(2-(\varepsilon/\abs{\bhu{\cdot}\bn})\right)(\bhu\otimes\bn)  &\text{for $\abs{\bhu\cdot\bn} \in \left(\varepsilon/2,\varepsilon \right)$,}\\
		 \sign(\bhu{\cdot}\bn) \left[(\bhu{\cdot}\bn)\Insd+(\bhu\otimes\bn)\right]&\text{for $\abs{\bhu\cdot\bn} \geq \varepsilon$,}\\		
	\end{cases}
\end{equation}

The last stabilisation considered is inspired by the Harten-Lax-van Leer (HLL) Riemann solver. This stabilisation, defined as
\begin{equation}\label{eq:HLLConv}
\btauHLL = \max\{2(\bhu\cdot\bn),\varepsilon\}\Insd,
\end{equation}
employs an estimate of the largest wave speed of the Riemann problem.

For the SA turbulence model, the derivation of the convective stabilisation $\tauT^a$ is simpler due to the scalar nature of the equation. The stabilisation inspired from both LF and Roe Riemann solvers is simply
\begin{equation}\label{eq:LFRoeConvSA}
	\tauTLF = \tauTRoe= \max\left\{\abs{\bhu\cdot\bn},\varepsilon \right\}.
\end{equation}
whereas the stabilisation inspired by the HLL Riemann solver is 
\begin{equation}\label{eq:HLLConvSA}
	\tauTHLL= \max\left\{\bhu\cdot\bn,\varepsilon \right\}.
\end{equation}

It can be observed that both the Lax-Friedrichs and the Roe stabilisations provide a unique value of the stabilisation as seen from the two cells sharing an edge/face. In contrast, the HLL stabilisation provides different values for the two cells sharing an edge/face. 

The Roe Riemann solver of the RANS equations, given in equation~\eqref{eq:RoeConv2}, provides a continuous expression when $\abs{\bhu\cdot\bn}$ tends to $\varepsilon$ and $\varepsilon/2$, which numerical experiments have confirmed to be a key factor to guarantee convergence of the numerical scheme. 

In all the examples, the numerical parameter $\varepsilon$ is selected as $5 \times 10^{-2}$ for the LF and HLL stabilisations of RANS equations, whereas a slightly higher value of $10^{-1}$ is used for the Roe stabilisation. For the SA equation, the value of $\varepsilon$ is taken as $10^{-2}$ for all stabilisations. 

%==========================================================================
\section{Computational aspects}               \label{app:FCFV_computational}
%==========================================================================

This appendix briefly discusses some computational aspects to be considered when implementing the proposed FCFV for solving the incompressible RANS equations with the SA turbulence model.

%==========================================================================
\subsection{Solution of global and local problems}  \label{app:FCFV_compSystems}
%==========================================================================

The cost of each Newton-Raphson iteration is clearly dominated by the cost of solving the global problem. The size of the sparse non-symmetric linear system of equations~\eqref{eq:NRiterationReduced} is $\nequa =\numfa(\nsd+1) + \numel$, where $\numfa$ is the total number of edges/faces in the mesh. In two dimensions and for a mesh with $N$ nodes, $\nequa \simeq 11N$ for a mesh of triangular cells and $\nequa \simeq 7N$ for a quadrilateral mesh. In this work, the solution of the global system is performed using the parallel direct method implemented in the Harwell subroutine library code \texttt{ma41}~ \cite{amestoy1989vectorization,amestoy1993memory}.

The size of each local problem is only $(\nsd+1)^2$. The local problems can be trivially solved in parallel as there is no communication required between cells. Furthermore, the solution of a local problem does not actually require solving a linear system of equations of size $(\nsd+1)^2$. After solving the global system~\eqref{eq:NRiterationReduced}, the increments of $\bLe$ and $\bqTe$ are explicitly obtained from the increments of $\bhuV$ and $\bhnuV$, using the first and third equations in~\eqref{eq:localDiscrete}. With the updated values of $\bhuV$, $\bhnuV$ and $\bqTe$, the last equation of~\eqref{eq:localDiscrete} is used to obtain the increment of the turbulent variable $\nuTe$. Finally, the velocity can be updated using the second equation of~\eqref{eq:localDiscrete}. In summary, each local problem involves the solution of four independent equations, where three provide explicit expressions for the unknowns, whereas one requires the solution of a nonlinear problem with a scalar unknown.

%==========================================================================
\subsection{Convergence of the Newton-Raphson method}          \label{app:NRconvergence}
%==========================================================================

To check the convergence of the Newton-Raphson iterations, the following residual is employed in all the examples
\begin{equation} \label{eq:NRstoppingCriteria}
\norm{\mathcal{R}^n}_\infty := \max\left\{\frac{\norm{\bR^{n}_{\hat{u}}}_\infty}{\norm{\vect{f}^n_{u}}_{\infty}}, \frac{\norm{\bR^{n}_{\hat{\nu}}}_{\infty}}{\norm{	\vect{f}^n_{\nuT}}_{\infty}} \right\}
\end{equation}
where the normalising factors are defined as
\begin{equation}
\vect{f}^n_{e,u}\! := \volE \bse^n {+} \sum_{j \in \Dset}\!  \areaFj  \bigl( \btau_j^{n-1} {-} (\bu_{D,j}^n {\cdot} \bn_j)\Insd \bigr) \bu_{D,j}^n  
\end{equation}
and
\begin{equation}
\vect{f}^n_{e,\nuT}\! := \sum_{j \in \Dset}\!  \areaFj  ( \tauT_j^{n-1} {-} \bu_{D,j}^n {\cdot} \bn_j) \nuT_{D,j}^n
\end{equation}
for the RANS and SA equations respectively.

\begin{figure}[!tb]
	\centering
	\includegraphics[width=0.48\textwidth]{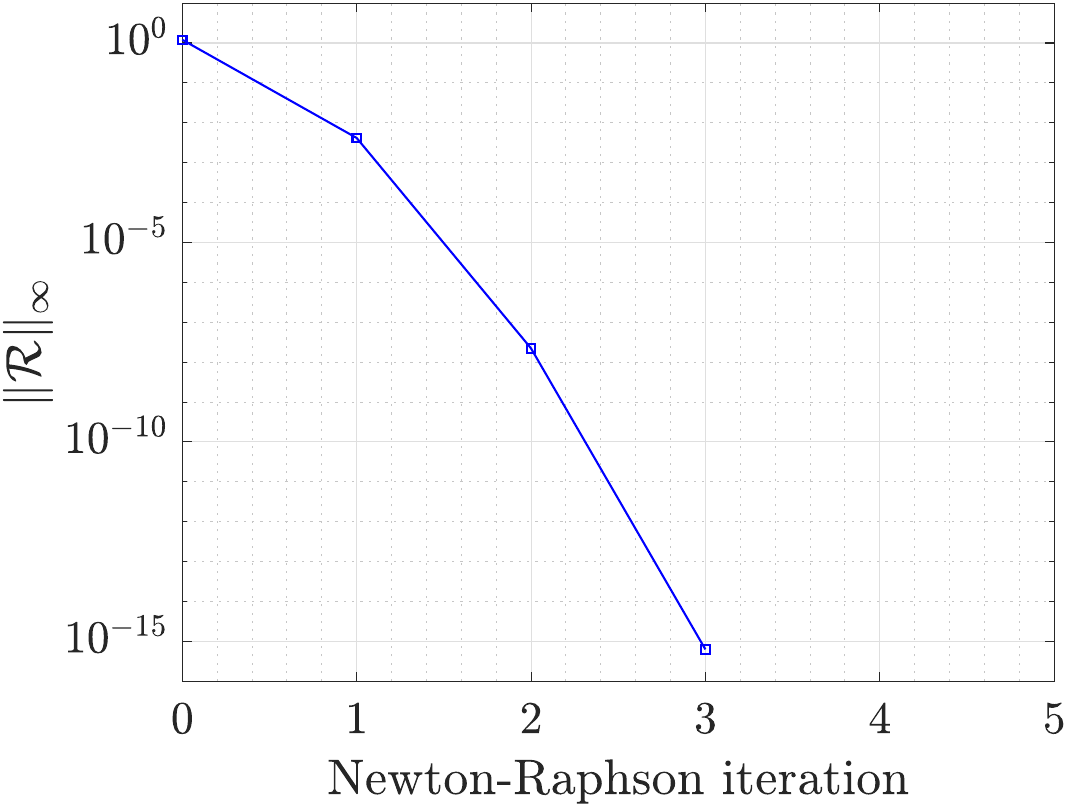}
	\caption{Residual of equation~\ref{eq:NRstoppingCriteria} as a function of the number of Newton-Raphson iterations for the Couette example.}
	\label{fig:couette_NRiterations}
\end{figure}

To illustrate the quadratic convergence attained, figure~\ref{fig:couette_NRiterations} shows the residual of equation~\ref{eq:NRstoppingCriteria} as a function of the number of Newton-Raphson iterations for the Couette example of Section~\ref{sc:couette}.

%==========================================================================
\subsection{Monolithic versus staggered solution}               \label{app:FCFV_computationalMonoStaggered}
%==========================================================================

The solution strategy used in sections~\ref{sc:plate} and~\ref{sc:turbulentCyl} employs a monolithic coupling of the RANS and SA equations. An alternative approach would be to solve the RANS and SA equations in a staggered manner. Such an approach requires solving a global system to obtain $\bhuV$ and $\brhoV$ followed by a global system to obtain $\bhnuV$. This is then repeated until convergence.

The staggered approach requires solving two global systems, associated with the RANS and SA equations, of size $\nequa^\texttt{RANS} =\numfa\nsd + \numel$ and $\nequa^\texttt{SA} =\numfa$, respectively. A priori, it is not feasible to compare the cost of the two approaches because it is not possible to know the number of iterations required by the staggered scheme to converge, but some comments can be made about the memory requirements of both strategies.

Assuming that the number of exterior edges/faces in the mesh is negligible, when compared to the number of interior edges/faces, the number of non-zero entries in the global system of the monolithic approach is
\begin{equation}
	\nnz \simeq	\sum_{e = 1}^{\numel} \numfa^e \left[ \numfa^e (\nsd^2 + 1) + 2 \nsd (\numfa^e + 1) \right] - \numfa (\nsd + 1)^2 .
\end{equation}
In contrast, the number of non-zero entries of the global problems associated to the RANS and SA equations in the staggered approach is
\begin{equation}
	\nnz^\texttt{RANS} \simeq \sum_{e = 1}^{\numel} \numfa^e \nsd (\numfa^e \nsd+2) - \numfa \nsd^2,
	\qquad
	\nnz^\texttt{SA} \simeq \sum_{e = 1}^{\numel} (\numfa^e)^2 - \numfa.
\end{equation}

In two dimensions and for a mesh with $N$ nodes, the global matrix of the monolithic approach has $\nnz \simeq 159N$ for a mesh of triangular cells and $\nnz \simeq 142N$ for a quadrilateral mesh. For the staggered approach, the global matrices have $\nnz^\texttt{RANS} \simeq 84N$ and $\nnz^\texttt{SA} \simeq 15N$ for triangular meshes and $\nnz^\texttt{RANS} \simeq 72N$ and $\nnz^\texttt{SA} \simeq 14N$ for quadrilateral meshes. This means that the staggered approach requires approximately 60\% of the memory required by the monolithic approach to store the global matrices. 

%==========================================================================
\subsection{Pseudo-time marching for steady-state computations}   \label{app:relaxation}
%==========================================================================

When a steady-state simulation is of interest, the solution of a pseudo-transient problem, where $t$ denotes the artificial or pseudo-time, is often preferred due to the difficulty to find an initial guess that guarantees convergence of the Newton-Raphson algorithm. In this scenario, time accuracy is not a requirement and, therefore, it is common to utilise a first-order time marching scheme to reach the steady state. Furthermore, strategies to vary the time step as the solution evolves are attractive because at the beginning of a simulation smaller time steps are required to guarantee convergence of the nonlinear problem, whereas as the solution approaches steady state it is possible to use extremely large time steps.

When time marching to a steady state, this work employs the low order BDF1 scheme. Several strategies have been proposed to automatically adjust the time step during steady state simulations, e.g. a simple exponential law~\cite{Kelley1998Convergence,Bucker2009CFL}. More elaborated approaches such as the so-called switched evolution relaxation (SER)~\cite{Mulder1985experiments,Bucker2009CFL,Kelley1998Convergence} approach use the ratio between the nonlinear residuals at two consecutive time steps to adjust the time step. This work adapts the SER approach to the proposed FCFV scheme.

At the beginning of a simulation, the CFL is initialised to a low value, namely $\CFL^0 = 0.1$. This is particularly important to avoid the potential negative impact of performing an impulsive start with an initial condition not satisfying the boundary conditions, e.g. using free-stream values for the initial condition. Following~\cite{Crivellini2013SADG,Crivellini2011matrixFree,Kelley1998Convergence,Ceze2013Pseudo,SU2tutorials}, after the first time step the CFL is updated according to the expression
\begin{equation} \label{eq:RelaxCFL}
	\CFL^{n+1} =\min \left\{ \CFL^n / f^\gamma, \CFLmax \right\},
\end{equation}
where 
\begin{equation} \label{eq:RelaxF}
	f = \max \biggl\{\frac{\norm{\bR^{n}_{u,s}}_{\infty}}{\norm{\bR^{n-1}_{u,s}}_{\infty}}, \frac{\norm{\bR^{n}_{\nuT,s}}_{\infty}}{\norm{\bR^{n-1}_{\nuT,s}}_{\infty}} \biggr\} 
\end{equation}
and the exponent $\gamma$ is taken as $\gamma_\text{max}>1$ if both residuals decrease with respect to the previous time step, i.e., if $f \leq 1$, or as $\gamma_\text{min}<1$ if at least one of the residuals increases, i.e., if $f>1$.

In equation~\eqref{eq:RelaxF} the residual vector $\bR_{u,s}$ and $\bR_{\nuT,s}$ correspond to the result of assembling the elemental contributions of the second and fourth residual in equation~\eqref{eq:localDiscrete}, respectively, but without including the term corresponding to the discrete time derivative. It is worth noting that only two residuals of the local problem~\eqref{eq:localDiscrete} are employed to define $f$ in equation~\eqref{eq:RelaxF}, because the other two residuals are linear. Furthermore, the residuals of the global problem~\eqref{eq:globalDiscrete} are also linear with respect to $\bu$, $\bhu$, $\bL$, $\rho$ and $\hnu$.

\begin{figure}[!tb]
	\centering
	\subfigure[]{\includegraphics[width=0.48\textwidth]{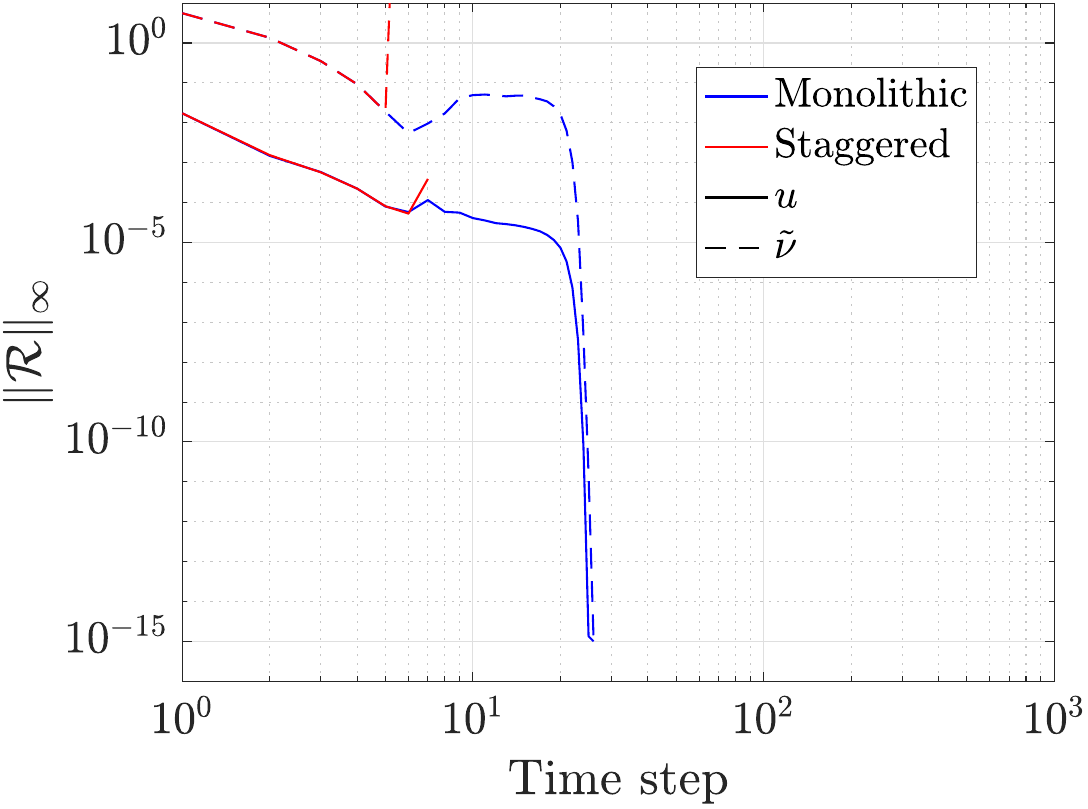}}
	\subfigure[]{\includegraphics[width=0.48\textwidth]{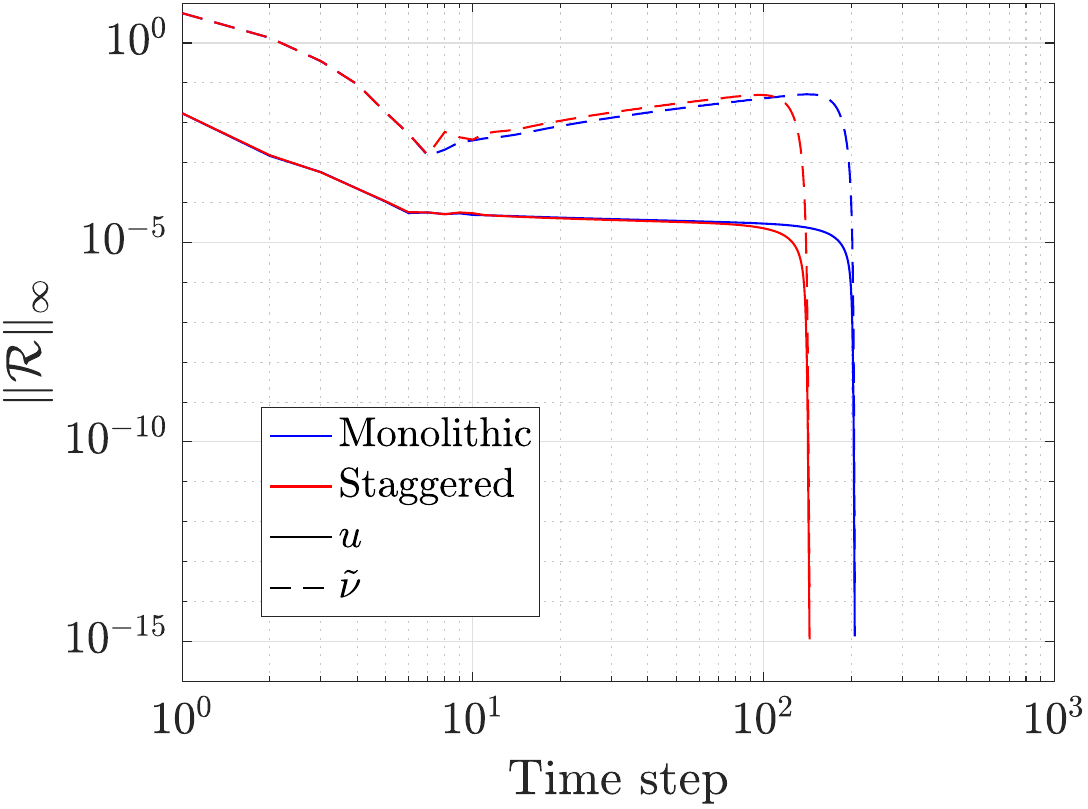}}
	\caption{Turbulent flow over a flat plate: Evolution of the residuals with (a) $\gamma_\text{max}=2$ and (b) $\gamma_\text{max}=1.2$.}
	\label{fig:plate_ResSER}
\end{figure}

Following~\cite{Crivellini2013SADG,Crivellini2011matrixFree,SU2tutorials}, in all the numerical examples the maximum allowed CFL is taken as $\CFLmax = 10^{20}$. When employing the monolithic FCFV approach, the parameters to define the exponent $\gamma$ are taken as $\gamma_\text{max}=2$ and $\gamma_\text{min}=0.1$. When using the staggered FCFV approach, numerical experiments show that it is necessary to employ a lower value for $\gamma_\text{max}$. To illustrate this phenomenon, let us consider the turbulent flow over a flat plate problem presented in section~\ref{sc:plate}. Figure~\ref{fig:plate_ResSER} shows the evolution of the residuals of the RANS and SA equations during the pseudo-time marching.
The results correspond to the second quadrilateral mesh and using the HLL stabilisation. When using the value of $\gamma_\text{max}=2$ the monolithic approach converges in less than 30 time steps, whereas the staggered approach diverges after six time steps, as displayed in figure~\ref{fig:plate_ResSER}(a). This illustrates a lack of stability induced by the staggered approach. A simple remedy consists of using a more conservative approach to increase the CFL number, for instance lowering the value of $\gamma_\text{max}$. The results with $\gamma_\text{max}=1.2$ in figure~\ref{fig:plate_ResSER}(b) show that the staggered approach converges. As expected, the more conservative increase of the CFL results in convergence, but with a substantially larger number of time steps.

\end{document}